\documentclass{aa}%[structabstract]{aa}
\usepackage{txfonts}
\usepackage{graphicx}
\usepackage{natbib}
\usepackage{longtable}
\usepackage{subeqnarray}
\usepackage{cases}
\usepackage{ulem}

\usepackage[colorlinks=true,citecolor=blue]{hyperref}

\begin{document}

\title{Observations of multiple NH$_3$ transitions in W33}
\author{K. Tursun \inst{1,2}
\and C. Henkel \inst{3,4,1}
\and J. Esimbek \inst{1,5}
\and X. D. Tang \inst{1,5}
\and T. L. Wilson \inst{3}
\and A. Malawi \inst{4}
\and E. Alkhuja \inst{4,6}
\and F. Wyrowski \inst{3}
\and R. Mauersberger \inst{3}
\and K. Immer \inst{3,7}
\and H. Asiri \inst{4}
\and J. J. Zhou \inst{1,5}
\and G. Wu \inst{3,1}
}
\institute{
Xinjiang Astronomical Observatory, Chinese Academy of Sciences, Urumqi 830011, P. R. China \\
e-mail: kadirya@xao.ac.cn, jarken@xao.ac.cn
\and
University of Chinese Academy of Sciences, Beijing 100080, P. R. China
\and
Max-Planck-Institut f\"ur Radioastronomie, Auf dem H\"ugel 69, D-53121 Bonn, Germany \\
e-mail: chenkel@mpifr-bonn.mpg.de
\and
Astronomy Department, King Abdulaziz University, P.O.
Box 80203, 21589 Jeddah, Saudi Arabia
\and
Key Laboratory of Radio Astronomy, Chinese Academy of Sciences, Urumqi 830011, P. R. China
\and
Institut f{\"u}r Physik und Astronomie, Universit{\"a}t Potsdam, Karl-Liebknecht-Str. 24/25, 14476-Golm, Germany
\and
Leiden Observatory, Leiden University, PO Box 9513, 2300 RA Leiden, The Netherlands
}

\abstract
{At a distance of 2.4\,kpc, W33 is an outstanding massive and luminous 10\,pc sized star forming complex containing quiescent infrared dark clouds as well as highly active infrared bright cloud cores heated by young massive stars. We report measurements of ammonia (NH$_3$) inversion lines in the frequency range 18--26\,GHz, obtained with the 40$\arcsec$ resolution of the 100\,m Effelsberg telescope. We have detected the ($J$,\,$K$)\,=\,(1,1), (2,2), (3,3), (4,4), (5,5), (6,6), (2,1) and (3,2) transitions. There is a maser line in the (3,3) transition towards W33\,Main. Brightness temperature and line shape indicate no significant variation during the last $\sim$36\,yr. We have determined kinetic temperatures, column densities and other physical properties of NH$_3$ and the molecular clouds in W33. For the total-NH$_3$ column density, we find for 40$\arcsec$ (0.5\,pc) sized regions 6.0\,($\pm$2.1)\,$\times$\,10$^{14}$, 3.5\,($\pm$0.1)\,$\times$\,10$^{15}$, 3.4\,($\pm$0.2)\,$\times$\,10$^{15}$, 3.1\,($\pm$0.2)\,$\times$\,10$^{15}$, 2.8\,($\pm$0.2)\,$\times$\,10$^{15}$ and 2.0\,($\pm$0.2)\,$\times$\,10$^{15}$\,cm$^{-2}$ at the peak positions of W33\,Main, W33\,A, W33\,B, W33\,Main1, W33\,A1 and W33\,B1, respectively. W33\,Main has a total-NH$_3$ fractional abundance of 1.3\,($\pm$0.1)\,$\times$\,10$^{-9}$ at the peak position. High values of 1.4\,($\pm$0.3)\,$\times$\,10$^{-8}$, 1.6\,($\pm$0.3)\,$\times$\,10$^{-8}$, 3.4\,($\pm$0.5)\,$\times$\,10$^{-8}$, 1.6\,($\pm$0.5)\,$\times$\,10$^{-8}$ and 4.0\,($\pm$1.2)\,$\times$\,10$^{-8}$ are obtained at the central positions of W33\,A, W33\,B, W33\,Main1, W33\,A1, and W33\,B1. From this, we confirm the already previously proposed different evolutionary stages of the six W33 clumps and find that there is no hot core in the region approaching the extreme conditions encountered in W51-IRS2 or Sgr\,B2. The ortho-to-para-NH$_3$ abundance ratios suggest that ammonia should have been formed in the gas phase or on dust grain mantles at kinetic temperatures of $\gtrsim$\,20\,K. We determine kinetic temperatures only using NH$_3$\,(1,1) and (2,2), and from this we provide gas volume densities for the six main sources in the W33 region. Using our new $T_{\rm kin}$ values shows that our volume densities are similar to those estimated by \cite{2014A&A...572A..63I}, suggesting that ammonia beam filling factors are close to unity.}


\maketitle

\section{Introduction}
\label{sect:Introduction}
The pyramidal ammonia (NH$_3$) molecule provides the
unique opportunity to trace molecular cloud excitation up to temperatures of $\sim$2000\,K by
observing its characteristic inversion transitions within a limited frequency interval
(20\,--\,50\,GHz; e.g., \citealt{1983ARA&A..21..239H,2006A&A...460..533W}). Most accessible lines are
covering an even smaller range, between 20 and 30\,GHz. The frequencies of the lines
connecting the two states of an inversion-doublet (arising from oscillations of the nitrogen
nucleus through the plane of the three hydrogen nuclei) depend on the total angular momentum $J$ and
its projection on the molecular axis, $K$, with $K$\,=\,0, 3, 6, 9... belonging to ortho-NH$_3$ and
$K$\,=\,1, 2, 4, 5, 7... representing para-NH$_3$.

Dozens of inversion lines can be detected, provided kinetic temperatures and ammonia column densities
are sufficient. These conditions prevail in "hot cores", dense molecular clumps near sites of very recent
massive star formation (e.g., \citealt{1986A&A...162..199M,1988A&A...201..123M,1987A&A...182..137H,
2013A&A...549A..90H,1988A&A...201..285H,1992A&A...256..618C,1993A&A...276..445H,1995A&A...294..667H,1993A&A...276L..29W,
1997ApJ...488..241Z,2011ApJ...739L..13G}). The enormous NH$_3$ column densities reaching up to\,$>$\,$10^{19}$\,cm$^{-2}$ are
believed to be caused by dust grain mantle evaporation (e.g., \citealt{1987A&A...182..137H,1987A&A...172..311W,
1988MNRAS.231..409B}). The warm dense clumps are characterized by temperatures $T_{\rm kin}$\,$>$\,100\,K,
$\chi$(para-NH$_3$)\,=\,$N$(NH$_3$)/$N$(H$_2$)\,$\sim$\,$10^{-5...-6}$ and source averaged ammonia column
densities regularly surpassing $10^{18}$\,cm$^{-2}$, while star forming regions in earlier or later evolutionary
stages are characterized by lower values.

Ortho- and para-NH$_3$ can be considered as almost independent molecules, with mixing rates of order $10^{-6}$\,yr$^{-1}$
\citep{1969ApJ...157L.113C}. Dipole transitions between $K$-ladders are forbidden. Within each $K$-ladder, states
with ($J$\,>\,$K$) are non-metastable; these can decay rapidly (10-100\,s) via far-infrared (FIR) $\triangle$$J$\,=\,1
transitions to the $J$\,=\,$K$ metastable levels. The metastable inversion doublets provide the bulk of the emission outside hot
cores and decay via the much slower ($\sim$$10^{9}$\,s)\,$\triangle K$\,=\,$\pm$3 transitions. Metastable transitions have very
similar excitation temperatures in case of para-NH$_3$ and hence the assumption of equal $T_{\rm ex}$ used to derive
rotation temperatures between different $K$-ladders is justified \citep{1973ApJ...186..501M,1988MNRAS.235..229D}.
The situation is different for ortho-NH$_3$, where due to peculiarities of the $K$\,=\,0 ladder (single states instead of inversion
doublets; e.g. \citealt{1983A&A...124..322G}), the behavior of the excitation temperatures of the metastable levels is not as simple.
In particular, one finds that the (3,3) line should be inverted in a fairly relevant density range (e.g.,
$\sim10^{4}-10^{5}$\,cm$^{-3}$ for $T_{\rm kin}$\,$\sim$\,50\,K). This effect is caused
by the forbidden nature of collisions between the lower (3,3) level, (3,3\,-), and the ground (0,0) state
\citep{1983A&A...122..164W}. Because of its large number of transitions sensitive to a wide range of excitation conditions and the fact
that it can be detected in a great variety of regions, NH$_3$ is perhaps second only to CO in importance and
an absolutely essential tracer to be studied wherever molecular gas is prominent.

\begin{figure}[t]
\vspace*{0.2mm}
\begin{center}
\includegraphics[trim={16.1cm 2.5cm 17.1cm 1.8cm}, clip, width=0.52\textwidth]{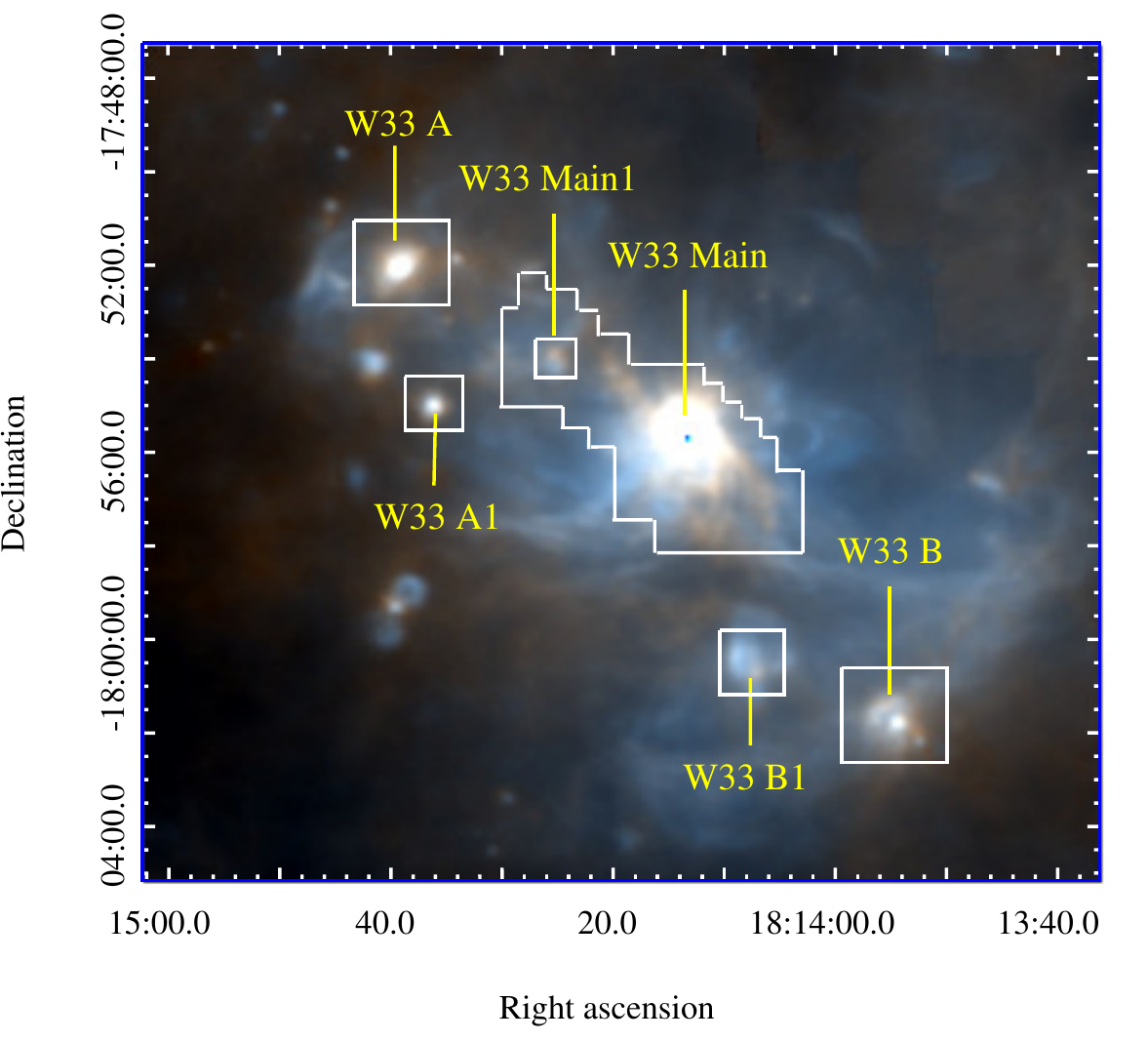}
\end{center}
\caption[]{Color image of the high-mass star forming complex W33 and its surroundings (blue for 70\,$\mu$m, red for 160\,$\mu$m, all derived from $Herschel$ data). The six boxes indicate the regions observed by us in NH$_3$.}
\label{Map1}
\end{figure}

With its relatively small distance of $\sim$2.4\,kpc \citep{2013A&A...553A.117I}, W33 is an outstanding massive
($\gtrsim$\,$10^{5}$\,${\rm M}_{\odot}$) and luminous ($10^{6}$\,${\rm L}_{\odot}$) 10\,pc sized star forming
complex (Fig.\,\ref{Map1}) that contains regions ranging from quiescent infrared dark clouds to highly active infrared bright
hotspots, associated with young massive stars (e.g., \citealt{2014A&A...572A..63I,2015ApJ...805..110M}).
Masing transitions of water and methanol have been detected in W33\,A and W33\,B at the edges, and in W33\,Main
at the center of the complex (e.g. \citealt{1977A&AS...30..145G,1981ApJ...250..621J,1986A&A...157..318M,
1990ApJ...354..556H,2013A&A...553A.117I}), while OH masers reside in W33\,A and W33\,B \citep{1974ApJ...187...41W,
1998MNRAS.297..215C}.  A cluster of three IR sources, located in W33\,Main, was detected by \citet{1977ApJ...211..421D},
while W33\,A contains an IR source with deep absorption features at 3 and 10\,$\mu$m \citep{1977ApJ...211..421D,
1978ApJ...226..863C,2010A&A...515A..45D}. Young stellar clusters are associated with W33\,A, W33\,B, and W33\,Main. W33\,Main
is also exhibiting strong radio continuum emission \citep{1984ApJ...283..573S,2012PASP..124..939H}.

These sources provide targets at quite different evolutionary stages (see Table\,\ref{table:1}).
W33\,Main1, W33\,A1, and W33\,B1 are high-mass protostellar objects, with W33\,Main1 being in
a particularly early stage devoid of any substantial heating source. The other two objects appear to be
warmer and thus more evolved. W33\,A and W33\,B have been considered as hot cores \citep{2014A&A...572A..63I},
with their complex chemistry being greatly influenced by recently evaporated dust grain mantles. W33\,Main is even more
evolved, also hosting an H\,{\scriptsize II} region giving rise to strong radio continuum emission
(e.g. \citealt{1984ApJ...283..573S}). And finally, there are the stellar clusters which have already successfully dispersed most or
all of their ambient molecular material (e.g. \citealt{2011ApJ...733...41M,2015ApJ...805..110M}).
In this survey, we mainly study W33\,Main, W33\,A, W33\,B, W33\,Main1, W33\,A1, and W33\,B1 (see Fig.\,\ref{Map1}).
The regions observed by us in NH$_3$ are marked by five different boxes in Fig.\,\ref{Map1}.
Furthermore, detailed source parameters of the six W33 clumps are given in Table\,\ref{table:1}.

W33 has so far not systematically been studied in NH$_3$. Published are the 2.5$\arcmin$ resolution NH$_3$\,(1,1) map by \citet{2012MNRAS.426.1972P},
the 5$\arcsec$ resolution NH$_3$\,(1,1) channel maps by \citet{1989ApJ...347..349K}, (1,1) and (2,2) emission
from W33\,A by \citet{2010ApJ...725...17G}, and the four principal metastable lines (($J, K$)\,=\,(1,1) to (4,4))
toward W33\,Main \citep{1982A&A...110L..20W}, but these data are far short of what can be reached nowadays with
new K-band broadband receivers ranging from 18-26\,GHz.

In this paper we provide a systematic study of the spectral characteristics of W33 in the
$\lambda$\,$\sim$\,1.3\,cm band with the Effelsberg-100\,m telescope ranging between 18\,GHz to 26\,GHz.
The article is organized as follows: In Sect.\,\ref{sect:Observation} we introduce
our observations and data reduction. Results are highlighted in Sect.\,\ref{sect:Results}.
The discussion is presented in Sect.\,\ref{sect:discussion} and our main conclusions are
summarized in Sect.\,\ref{sect:summary}.

%-----------------Figure 1

%-------------------------- Table 1.0
\begin{table*}[t]
\centering
\begin{footnotesize}
\setlength{\tabcolsep}{5 pt}
\caption{Source parameters of the six W33 clumps.}
\begin{tabular}{lccccccccc}
\hline \hline
\vspace{4pt}
Source & R.A.\,(J2000) & Dec.\,(J2000) &   $L_{\rm bol}$  & $M_{\rm source}$ & $T_{\rm dust}$ & $N(\rm H_{2})$ & Evol.\,Stage & Obs. Size \\
\vspace{2pt}
&($^{\rm h}$\,$^{\rm m}$\,$^{\rm s}$)&($^{\circ}$\,$^{\arcmin}$\,$^{\arcsec}$)&($10^{3}$\,${L}_{\odot}$) & ($10^{3}$\,${M}_{\odot}$)& (K) & ($10^{23}$\,cm$^{-2}$) &   &    \\
\hline
W33\,Main&18:14:13.50&-17:55:47.0&449&4.0 $\pm$ 2.5  & 42.5 $\pm$ 12.6 &  4.6 $\pm$ 1.6 &Hot Core with H\,{\scriptsize II} region  &$380\arcsec\,\times\,320\arcsec$ \\
W33\,A    & 18:14:39.10 & -17:52:03.0   & 41   & 3.4 $\pm$ 2.3 & 28.6 $\pm$ 5.3 & 2.5 $\pm$ 0.6 &Hot Core  &$40\arcsec\,\times\,40\arcsec$  \\
W33\,B    & 18:13:54.40 & -18:01:52.0   & 22   & 1.9 $\pm$ 1.1 & 26.5 $\pm$ 3.9  &  2.1 $\pm$ 0.5  &Hot Core  &$120\arcsec\,\times\,120\arcsec$  \\
W33\,Main1&18:14:25:00&-17:53:58.0&11&0.5 $\pm$ 0.3 & 28.6 $\pm$ 5.6  &  0.9 $\pm$ 0.2  &High-mass protostellar  & 1 position \\
W33\,A1&18:14:36.10&-17:55:05.0&6&0.4 $\pm$ 0.5 & 25.0 $\pm$ 7.3  &  1.7 $\pm$ 0.6  & High-mass protostellar  &  1 position  \\
W33\,B1&18:14:07.10&-18:00:45.0&16&0.2 $\pm$ 0.1 & 38.6 $\pm$ 11.4  & 0.5 $\pm$ 0.2 & High-mass protostellar  &$40\arcsec\,\times\,40\arcsec$ \\
\hline
\end{tabular}
\label{table:1}
\tablefoot{Column\,1: source name; Cols.\,2,\,3: reference position; Col.\,4: bolometric luminosity; Col.\,5: total interstellar gas mass of the clumps; Col.\,6: temperature of the dust; Col.\,7: H$_{2}$ column density; Col.\,8: evolutionary stage of star forming regions related to W33; values or information given in Cols.\,4\,--\,8 are taken from \cite{2014A&A...572A..63I}; Col.\,9: observed area in this survey (see Fig.\,\ref{Map1}). Here and elsewhere we arrange the order of sources in the way that we start with the largest and most evolved object to then go to more compact targets representing earlier evolutionary stages.}
\end{footnotesize}
\end{table*}

\section{Observations and data reduction}
\label{sect:Observation}
\subsection{NH$_3$ observations}
\label{sect:Observation1}
The data were taken in January 2018, with the 100-m Effelsberg telescope\footnote{The 100-m telescope
at Effelsberg is operated by the Max-Plank-Institut f\"ur Radioastronomie (MPIFR) on behalf of
the Max-Plank-Gesellschaft (MPG).} near Bonn/Germany. Measurements were carried out with a
dual channel (LCP/RCP) K-band (17.9\,GHz-26.2\,GHz) HEMT receiver. With $T_{\rm sys}$\,$\sim$\,60\,K on a
$T^\ast_{A}$ scale the 5$\sigma$ noise level is $\sim$30\,mK for 1\,km\,s$^{-1}$ wide channels.
Four subbands, WFF4\,(17.9\,--\,20.4\,GHz), WFF3\,(20.0\,--\,22.5\,GHz),
WFF2\,(21.6\,--\,24.1\,GHz) and WFF1\,(23.7\,--\,26.2\,GHz), covered simultaneously the entire
frequency range with an overlap of at least 300\,MHz between adjacent subbands.
Over this whole frequency range, the FWHM (Full Width Half Maximum) beamsize varied
from 35$\arcsec$ (0.4\,pc) to 50$\arcsec$ (0.6\,pc) ($\sim40\arcsec$ (0.5\,pc) at 23\,GHz).

The survey encompasses a total of $\sim$20 observing hours. The focus was checked every few hours,
in particular after sunrise and sunset. Pointing was obtained every hour toward a nearby pointing
source (mostly towards PKS\,1830--211) and was found to be accurate to about 5$\arcsec$. The strong
continuum source 3C\,286 was used to calibrate the spectral line flux, assuming a standard flux density
of 2.5\,Jy at 22\,GHz \citep{1994A&A...284..331O}. The conversion factor from Jy on a flux density
scale ($S_{\nu}$) to K on a main beam brightness temperature scale ($T_{\rm mb}$) is
$T_{\rm mb}$/$S_{\nu}$\,$\sim$\,1.7\,K\,Jy$^{-1}$ at 18.5\,GHz, 1.5\,K\,Jy$^{-1}$ at 22\,GHz, and 1.4\,K\,Jy$^{-1}$
at 23.7\,GHz. All velocities are with respect to the Local Standard of Rest (LSR). Specific
observational details of the eight detected transitions of NH$_3$ are listed in Table\,\ref{table:2}.

For the six W33 sources (Fig.\,\ref{Map1}), a total of 218 positions were measured. 182 positions belong to the region
containing W33\,Main and W33\,Main1, 9 positions were observed to map W33\,A, 17 positions are related to W33\,B and
9 positions to W33\,B1. Only the central position was observed towards W33\,A1. The sizes of the different maps shown
in Fig.\,\ref{Map1} are listed in Table\,\ref{table:1}. In most cases a total of four minutes of on+off source integration
time were spent on individual positions. However, to control potential variations in pointing and calibration, the central
continuum position of W33\,Main (see Table\,\ref{table:1}) was frequently measured. We fitted the NH$_3$\,(1,1) main lines (the central group
of NH$_3$\,(1,1) hyperfine components) to study the stability of the system (see Sects.\,\ref{sect-3-1} and \ref{maser-line} for details).
All data were taken in a position switching mode with offset positions $900\arcsec$ in azimuth, alternating between right and left. The spacing of the maps is $20\arcsec$. Thus the maps are fully sampled.

\subsection{Data reduction}
\label{sect-2-2}
The CLASS packages of GILDAS\footnote{http://www.iram.fr/IRAMFR/GILDAS/}
were used for all the data reduction. For ammonia, we chose
two fitting methods, `GAUSS' fit and NH$_3$\,(1,1) fit. In order to convert
hyperfine blended line widths to intrinsic line widths in the ammonia
inversion spectrum (e.g., \citealt{1998ApJ...504..207B}), we fitted
the averaged spectra using the GILDAS routine `NH$_3$\,(1,1)' fitting method
which can fit all 18 hyperfine components simultaneously. From the
NH$_3$\,(1,1) fit we obtain integrated intensity $\int$$T_{\rm MB} {\rm d}\upsilon$,
the Local Standard of Rest line center velocity $V_{\rm LSR}$, intrinsic line widths of individual hyperfine
structure (hfs) components $\Delta v$, and optical depth $\tau$ (see Table\,\ref{table:A.1}).
The `GAUSS' fit was used to obtain integrated intensities
$\int$$T_{\rm MB} {\rm d}\upsilon$, line center velocities $V_{\rm LSR}$, line widths $\Delta v$
of the other seven observed NH$_3$ transitions (see Table\,\ref{table:2} and, in the Appendix, e.g. Tables\,\ref{table:A.2} to \ref{table:A.4}
for the (2,2) to (4,4) lines). Main beam brightness temperatures $T_{\rm MB}$ of all the detected eight transitions
are also obtained from `GAUSS' fit (see Tables\,\ref{table:A.1} to \ref{table:A.4}).

\section{Results and analysis}
\label{sect:Results}
\subsection{The ammonia peak positions}
\label{sect-3-1}
We provide the molecular lines and line parameters obtained by Gaussian or
hyperfine structure fits toward the peak positions in our survey. We have detected the NH$_3$\,(1,1), (2,2), (3,3), (4,4), (5,5) and
(6,6) metastable lines in W33\,Main, W33\,A, and W33\,B (see Figs.\,\ref{W33Main,A} and \ref{W33B,non.}).
The non-metastable (2,1) and (3,2) transitions were also detected toward these molecular hotspots in
the W33 region (see Fig.\,\ref{W33B,non.} right panel). The (1,1), (2,2), (4,4), (5,5) and (6,6) inversion
lines were detected in absorption against the radio continuum in W33\,Main (see Fig.\,\ref{W33Main,A} left
panel). For W33\,Main,  the $(J,K)$\,=\,(1,1), (2,2), (4,4), (5,5) and (6,6) absorption lines exhibit a two component
velocity structure with $V_{\rm LSR}$\,$\sim$\,33\,km\,s$^{-1}$ and $V_{\rm LSR}$\,$\sim$\,38\,km\,s$^{-1}$ at
offsets $(\Delta\alpha, \Delta\delta)$\,=\,($0\arcsec, 0\arcsec$) and ($+20\arcsec, 0\arcsec$)
with respect to the reference position R.A.\,: 18:14:13.50, DEC.\,: -17:55:47.0 (J2000; see also Table\,\ref{table:1}). We have also
obtained a tentative (7,7) absorption line, but its signal-to-noise ratio, 2.1, is low. Other offset positions present
emission lines, only showing a single velocity component (Fig.\,\ref{W33Main,A} left panel, Figs.\,\ref{FgA.1} to \ref{FgA.6}, and Tables\,\ref{Table-absorption}, \ref{Table-emission}, \ref{table:A.1} to \ref{table:A.4}). The (3,3) line shows significant emission near the central
($0\arcsec, 0\arcsec$) offset positions (see Fig.\,\ref{FgA.3}), only showing one velocity component, which is intermediate between the above mentioned 33\,km\,s$^{-1}$ and 38\,km\,s$^{-1}$ features. The single emission component is wide enough to cover both features seen in the absorption lines of the other transitions. The absorption in the other lines appears to originate from a region smaller than our beam because the NH$_3$\,(4,4) line shows weak emission at offsets $(\Delta\alpha, \Delta\delta)$\,=\,($-20\arcsec, 0\arcsec$) and ($-20\arcsec, +20\arcsec$), the NH$_3$\,(5,5) line shows weak emission at offsets ($-20\arcsec, +20\arcsec$) and ($-20\arcsec, -20\arcsec$), and the (6,6) line also shows emission at the offset ($-20\arcsec, 0\arcsec$) (see Figs.\,\ref{FgA.4}, \ref{FgA.5}, and \ref{FgA.6}).

The peak position of W33\,Main was frequently measured during the observing sessions (because of W33's low declination, each session only lasted for a few hours). Using this data set, we calculated the values and standard deviations of the mean of the resulting flux densities for the NH$_3$\,(1,1), (2,2) and (3,3) lines (see Appendix\,\ref{Appendix B}). Figs.\,\ref{FgB.1} and \ref{FgB.2} show resulting values as a function of elevation for the NH$_3$\,(1,1) and (3,3) lines and as a function of time also for the (3,3) transition. Note that due to the low elevations of the source, differences in our corrections for elevation dependent gain variations of the telescope are minimal. These mainly play a role at high elevations. The standard deviations of the mean of the flux densities for the NH$_3$\,(1,1) and (2,2) metastable transitions (the latter is not shown in Appendix\,\ref{Appendix B}) are $4.2\%$ and $4.0\%$, respectively, with flux densities of $-$1.33\,Jy and $-$0.95\,Jy, corresponding to $-$1.86\,K and $-$1.33\,K on a main beam brightness temperature scale. For the NH$_3$\,(3,3) line, the most intense in this analysis, the scatter is slightly larger, with a standard deviation of the mean of $4.4\%$, while line shapes remain undistinguishable. Also in this case, the fluctuations are well below the level that would convincingly indicate source variability.

Towards W33\,A and W33\,B, the NH$_3$\,(1,1) to (6,6) lines are in emission (see Fig.\,\ref{W33Main,A},
right panel, and Fig.\,\ref{W33B,non.}, left panel). Here, the NH$_3$\,(1,1) to (3,3) lines exhibit the
strongest emission at all offsets (see Figs.\,\ref{FgA.7} and \ref{FgA.8}). The (4,4) to (6,6) lines show
weak emission at the central $(\Delta\alpha, \Delta\delta)$\,=\,($0\arcsec, 0\arcsec$) offsets while the
signals at the ($+20\arcsec, 0\arcsec$) and ($-20\arcsec, 0\arcsec$) offsets are even fainter (see Table\,\ref{table:1} and Figs.\,\ref{FgA.7}
and \ref{FgA.8}). Whether we have also detected the (7,7) line is not clear. Here the signal-to-noise ratio is only 1.8.
The two non-metastable lines ($J$\,>\,$K$,\,$K$)\,=\,(2,1) and (3,2) have been detected toward the central positions of
W33\,Main, W33\,A, and W33\,B (see the right panel of Fig.\,\ref{W33B,non.} as well as Figs.\,\ref{FgA.7} and \ref{FgA.8}).

We have also detected the NH$_3$\,(1,1), (2,2), (3,3) and (4,4) metastable emission lines in W33\,Main1 and W33\,A1.
Towards W33\,B1 we only see the NH$_3$\,(1,1), (2,2), and (3,3) emission lines. The non-metastable NH$_3$\,(2,1) and (3,2)
transitions were not detected in these W33 regions (see Figs.\,\ref{W33Main1,A1} and \ref{W33B1,non.}).
W33\,Main1, showing extended emission, is also tentatively detected in the (6,6) transition of ortho-NH$_3$
(Fig.\,\ref{W33Main1,A1}). W33\,B1 shows extended emission in the (1,1), (2,2) and (3,3) lines, which are strongest toward the
($0\arcsec, 0\arcsec$) and ($0\arcsec, +20\arcsec$) offset positions. Weak emission is also seen at offsets ($0\arcsec, -20\arcsec$),
($-20\arcsec, 0\arcsec$), ($-20\arcsec, +20\arcsec$) and the three positions with a right ascension offset of $+20\arcsec$ (see Fig.\,\ref{FgA.9}).

%-------------------------- Table 2
\begin{table}[t]
\centering
\caption{Measured NH$_3$ transition parameters.}
\begin{tabular}{ccccc}
\hline \hline
Line & Frequency & Resolution & $E_{\rm low}$/k$^{a}$ \\
 & MHz  & km\,s$^{-1}$ & K \\
\hline
NH$_3$\,(1,1) & 23694.4955    & 0.48   & 23.2  \\
NH$_3$\,(2,2) & 23722.6333    & 0.48   & 64.1  \\
NH$_3$\,(3,3) & 23870.1292    & 0.48   & 122.9 \\
NH$_3$\,(4,4) & 24139.4163    & 0.47   & 199.3  \\
NH$_3$\,(5,5) & 24532.9887    & 0.47   & 293.6  \\
NH$_3$\,(6,6) & 25056.0250    & 0.46   & 405.6  \\
NH$_3$\,(2,1) & 23098.8190    & 0.49   & 80.4   \\
NH$_3$\,(3,2) & 22834.1820    & 0.50   & 149.9   \\
\hline
\end{tabular}
\label{table:2}
\tablefoot{$^{a}$ Energy of the lower level above the ground state.}
\end{table}

\subsection{NH$_3$ distribution}
\label{sect-3-2}
\begin{figure*}
\centerline{\hbox{
\includegraphics[width=6.5cm,height=6.2cm,angle=0]{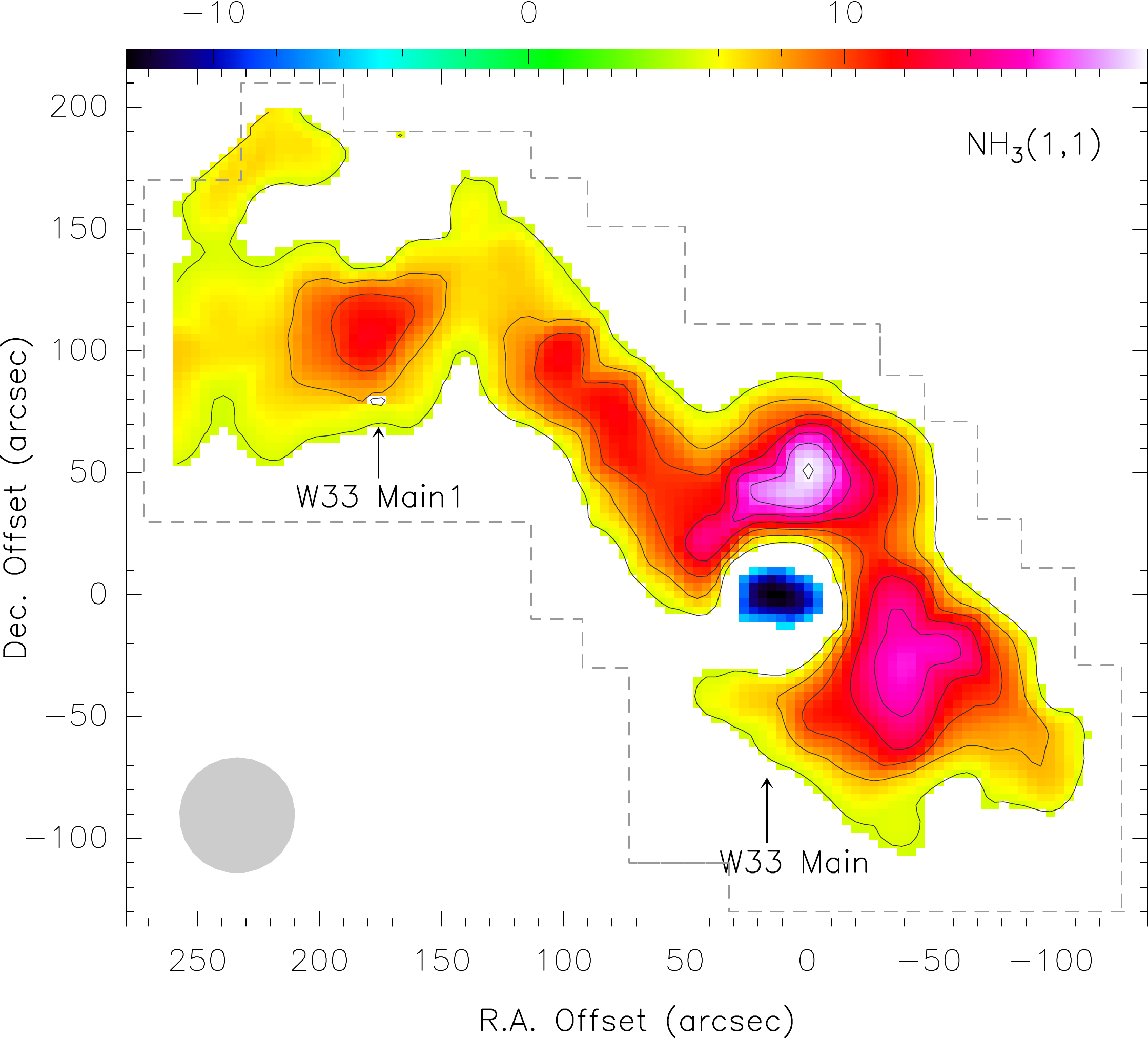}
\hspace{0.08cm}
\includegraphics[width=6.3cm,height=6.2cm,angle=0]{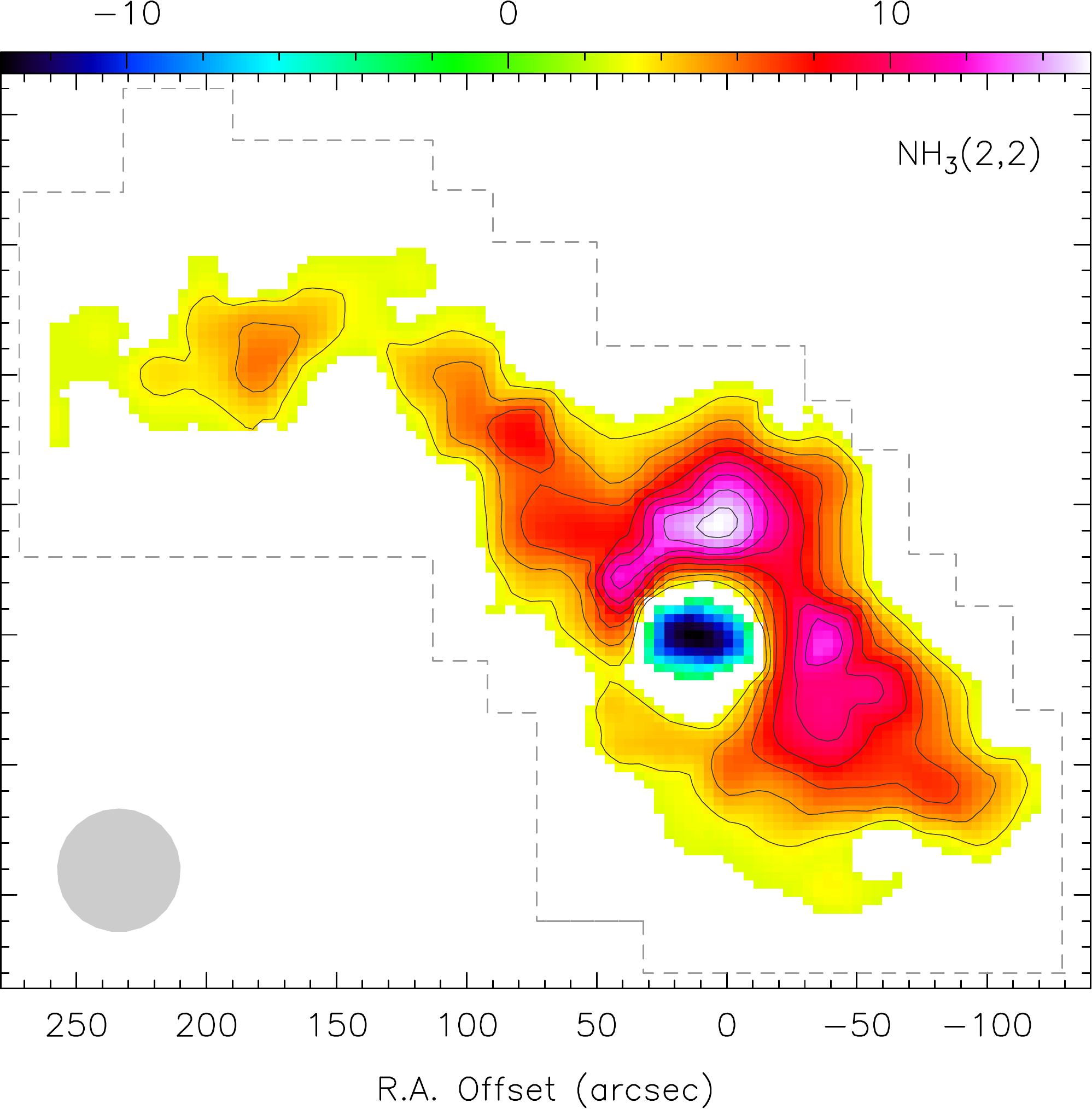}
\hspace{0.08cm}
\includegraphics[width=6.3cm,height=6.2cm,angle=0]{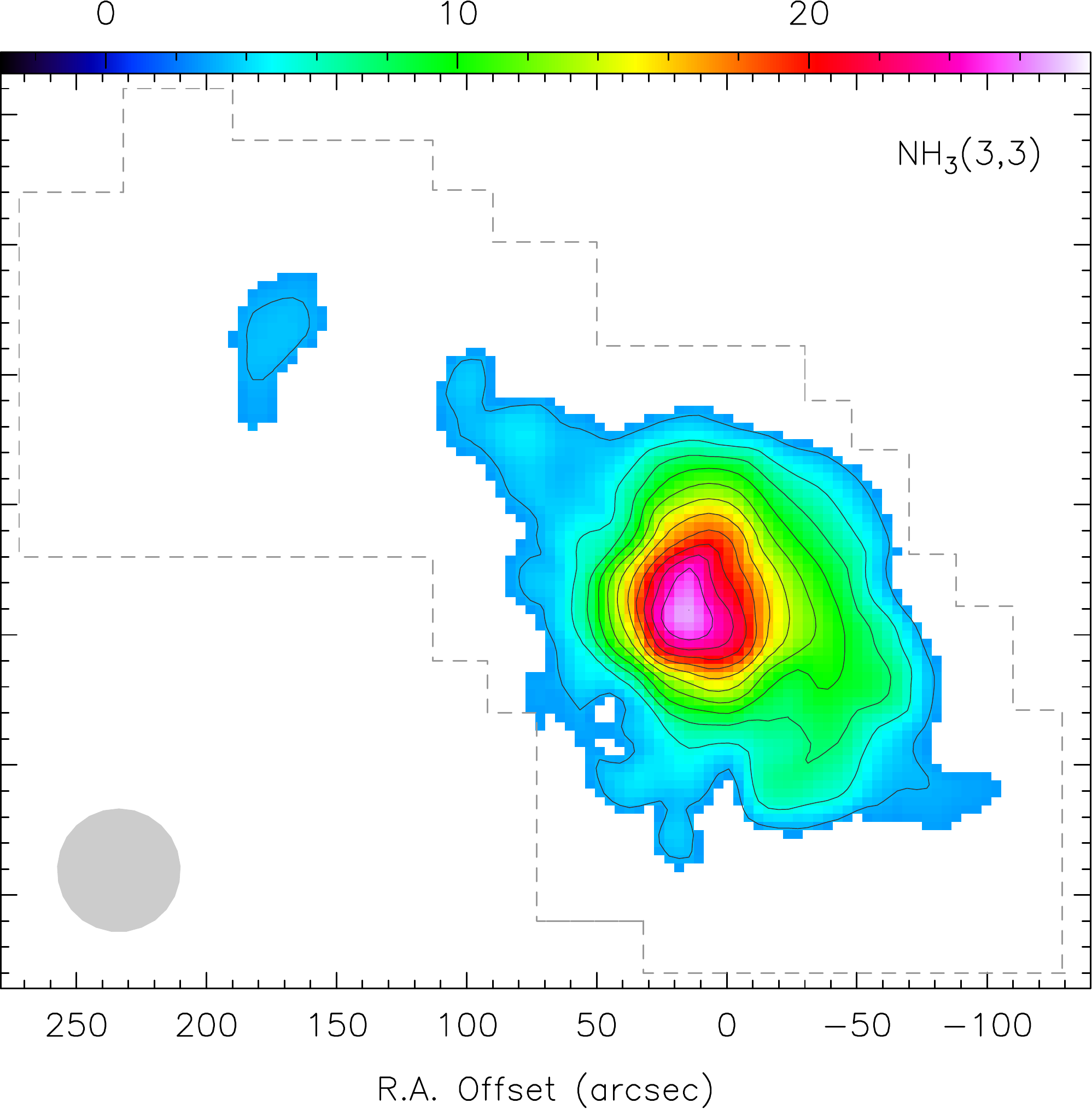}}}
\caption[]{Integrated intensity maps of NH$_3$\,(1,1) (\textit{left}), (2,2) (\textit{middle}), and (3,3) (\textit{right}) for the W33\,Main and W33\,Main1 regions. The reference position is R.A.\,: 18:14:13.50, DEC.\,: -17:55:47.0 (J2000). The integration range is 32 to 40 \,km\,s$^{-1}$. Contours start at 3.14\,K\,km\,s$^{-1}$ (3$\sigma$) on a main beam brightness temperature scale and go up in steps of 3.14\,K\,km\,s$^{-1}$. The unit of the color bar is K\,km\,s$^{-1}$. The limits of the mapped region are indicated with gray dashed lines. While the NH$_3$\,(1,1) and (2,2) lines show absorption near the reference position, the (3,3) line emission indicates a peak in this region. The half-power beam width is illustrated as a gray filled circle in the lower left corners of the images.}
\label{integrated-intencity}
\end{figure*}

A total of 218 positions were measured (see Sect.\,\ref{sect:Observation1}). 168, 130, and 86 of these were detected at a 3$\sigma$ level in case of the NH$_{3}$\,(1,1), (2,2), and (3,3) lines, respectively. This sigma ($\sigma$) value is the
peak/rms value multiplied by the square root of the number of channels contributing significantly to the line. NH$_3$\,(1,1), (2,2),
and (3,3) velocity-integrated intensity maps of the main groups of hyperfine components for the W33\,Main and W33\,Main\,1 regions are
presented in Fig.\,\ref{integrated-intencity}. Intensities were integrated over the Local Standard of Rest velocity ($V_{\rm LSR}$)
range of 32 to 40 \,km\,s$^{-1}$. The NH$_3$\,(1,1) emission shows an extended distribution and traces the denser molecular structure. NH$_3$\,(2,2) is detected in a slightly less extended region, while the NH$_3$\,(3,3) distribution is even more compact. In each panel, the half-power beam width is illustrated as a gray circle in the lower left corners of the images. The limits of the mapped region are
indicated with gray dashed lines.

All our metastable and non-metastable transitions exhibit the well known comparatively broad
$V_{\rm LSR}$\,$\sim$\,36\,km\,s$^{-1}$ to $V_{\rm LSR}$\,$\sim$\,58\,km\,s$^{-1}$ total velocity range
(e.g., \citealt{2013A&A...553A.117I,2014A&A...572A..63I}). W33\,Main, W33\,A, W33\,Main1, W33\,A1,  W33\,B1
absorption and emission lines have an observed radial velocity range of 32 to 40\,km\,s$^{-1}$, while the
W33\,B lines show a different central radial velocity of $\sim$58\,km\,s$^{-1}$ (see Figs.\,\ref{W33Main,A},
\ref{W33B,non.}, \ref{W33Main1,A1}, \ref{W33B1,non.} ). These two radial velocities are consistent with
those found in the CO observations of \cite{1983ApJ...265..791G}.

%-----------------Figure 2
\begin{figure*}[t]
\centering
\includegraphics[width=0.45\textwidth]{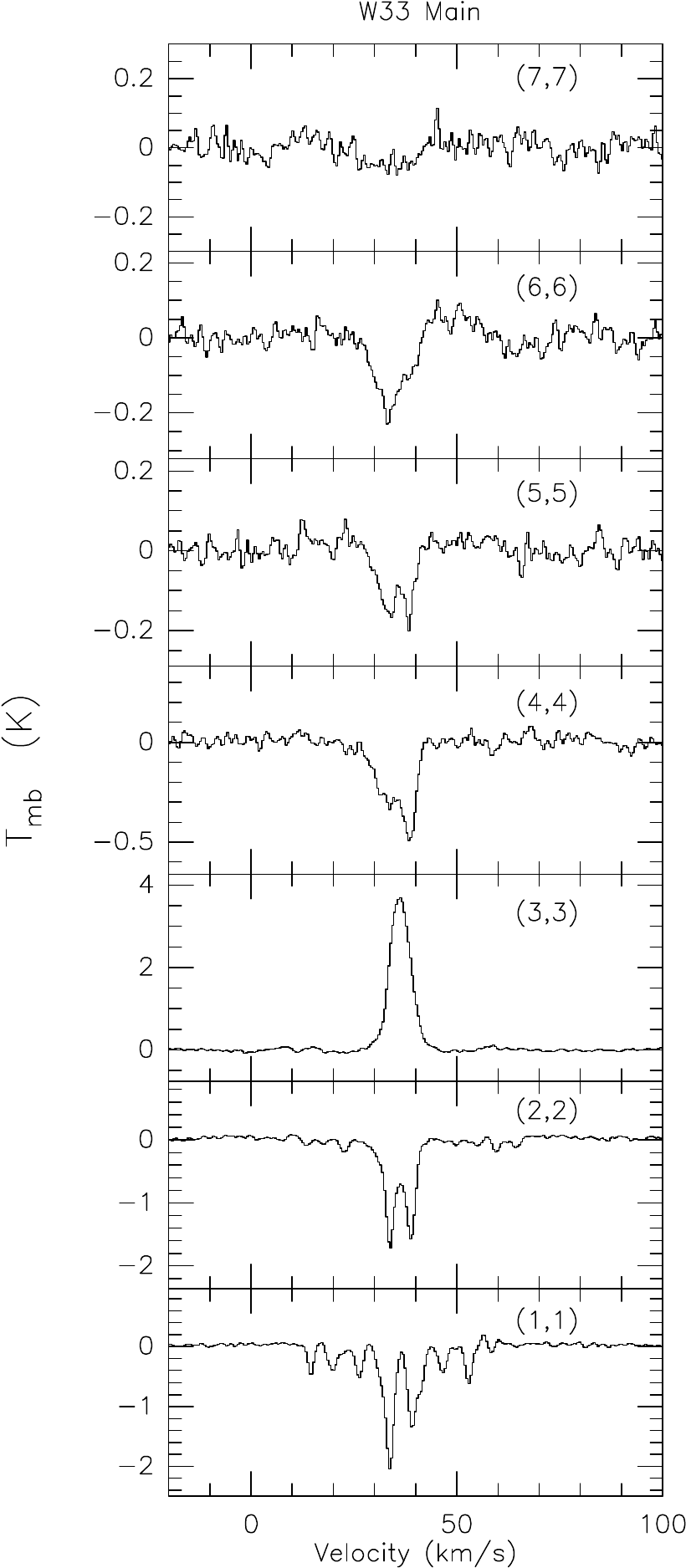}
\includegraphics[width=0.388\textwidth]{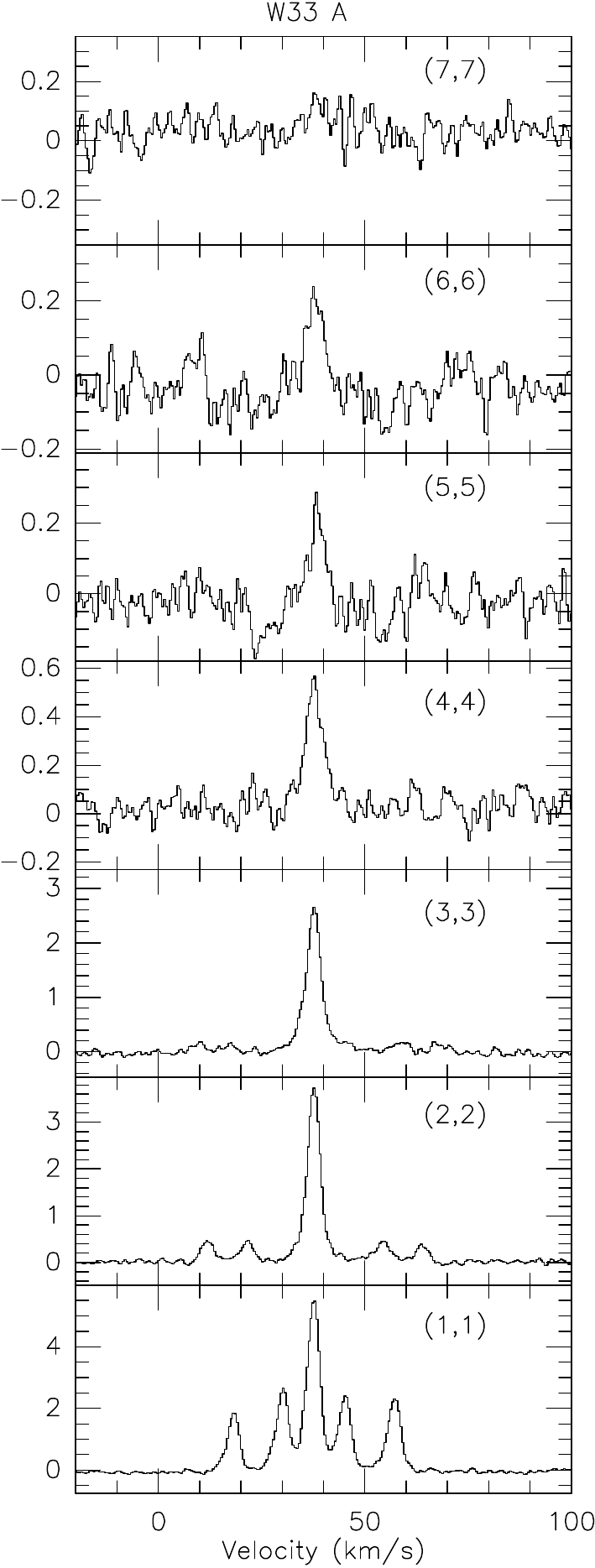}
\caption{NH$_3$ spectra of metastable lines at offset ($0\arcsec, 0\arcsec$) with respect to the reference positions given in Table\,\ref{table:1} for W33\,Main (\textit{left panel}) and W33\,A (\textit{right panel}). The channel widths are 0.48, 0.48, 0.48, 0.47, 0.47, 0.46, and 0.44\,km\,s$^{-1}$ for the NH$_3$\,(1,1) to (7,7) lines, respectively. The velocity scale is Local Standard of Rest, here and elsewhere.}
\label{W33Main,A}
\end{figure*}

%-----------------Figure 3
\begin{figure*}[t]
\centering
\includegraphics[width=0.45\textwidth]{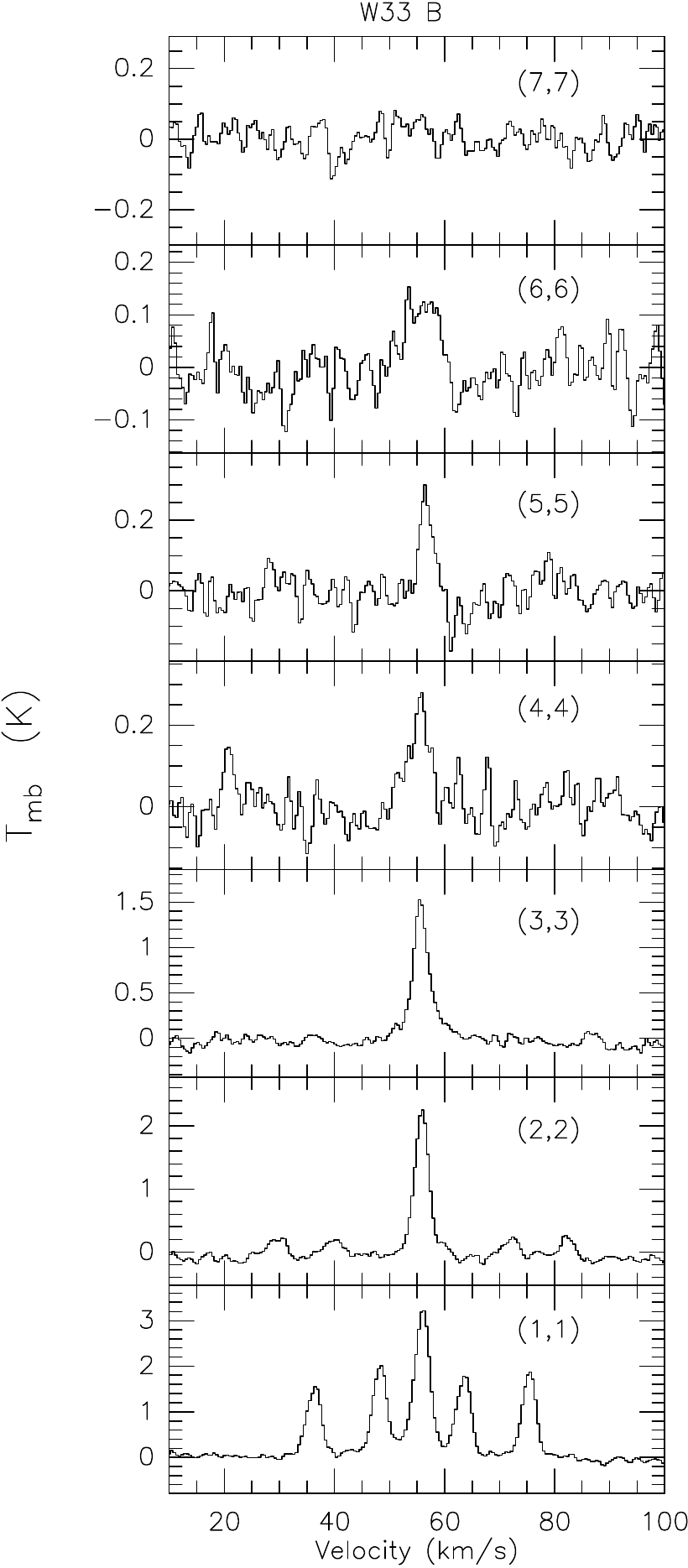}
\includegraphics[width=0.39\textwidth]{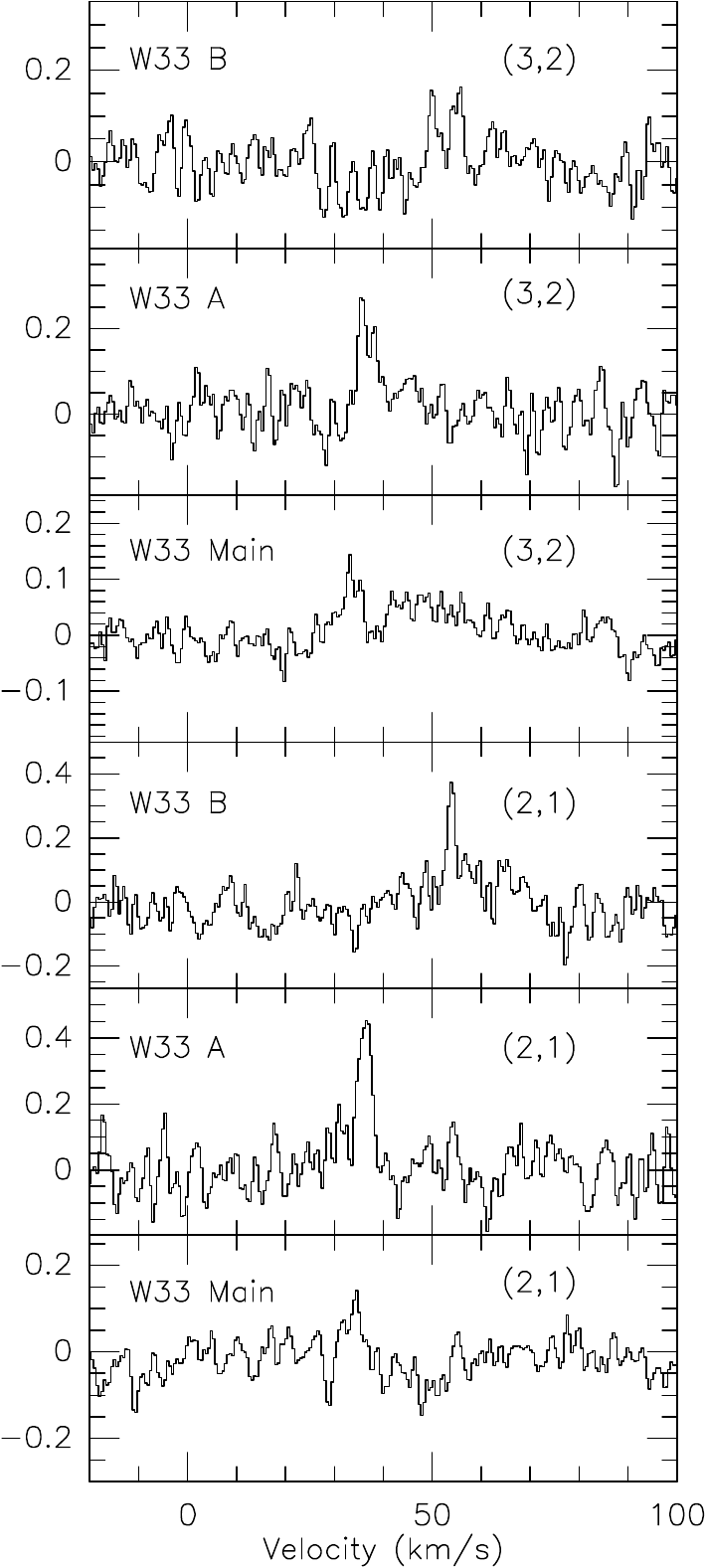}
\caption{NH$_3$ spectra of metastable lines at offset ($0\arcsec, 0\arcsec$) with respect to the reference positions given in Table\,\ref{table:1} for W33\,B (\textit{left panel}). Non-metastable lines from W33\,Main, W33\,A, and W33\,B are shown in the \textit{right panel}. Channel widths are 0.48, 0.48, 0.48, 0.47, 0.47, 0.46, and 0.44\,km\,s$^{-1}$ for the NH$_3$\,(1,1) to (7,7) lines, respectively, while the corresponding values for the NH$_3$\,(2,1) and (3,2) lines are 0.49 and 0.50\,km\,s$^{-1}$.}
\label{W33B,non.}
\end{figure*}

%-----------------Figure 4
\begin{figure*}[t]
\centering
\includegraphics[width=0.45\textwidth]{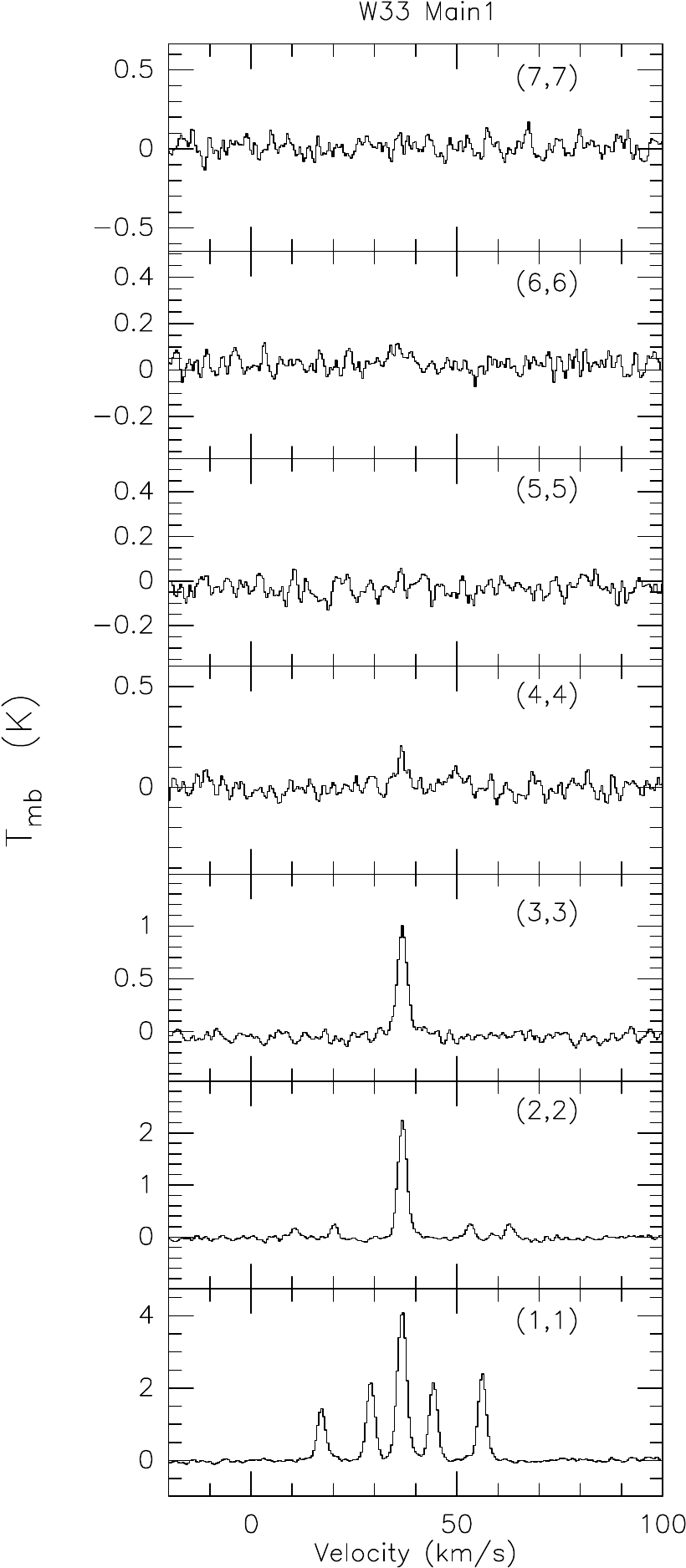}
\includegraphics[width=0.388\textwidth]{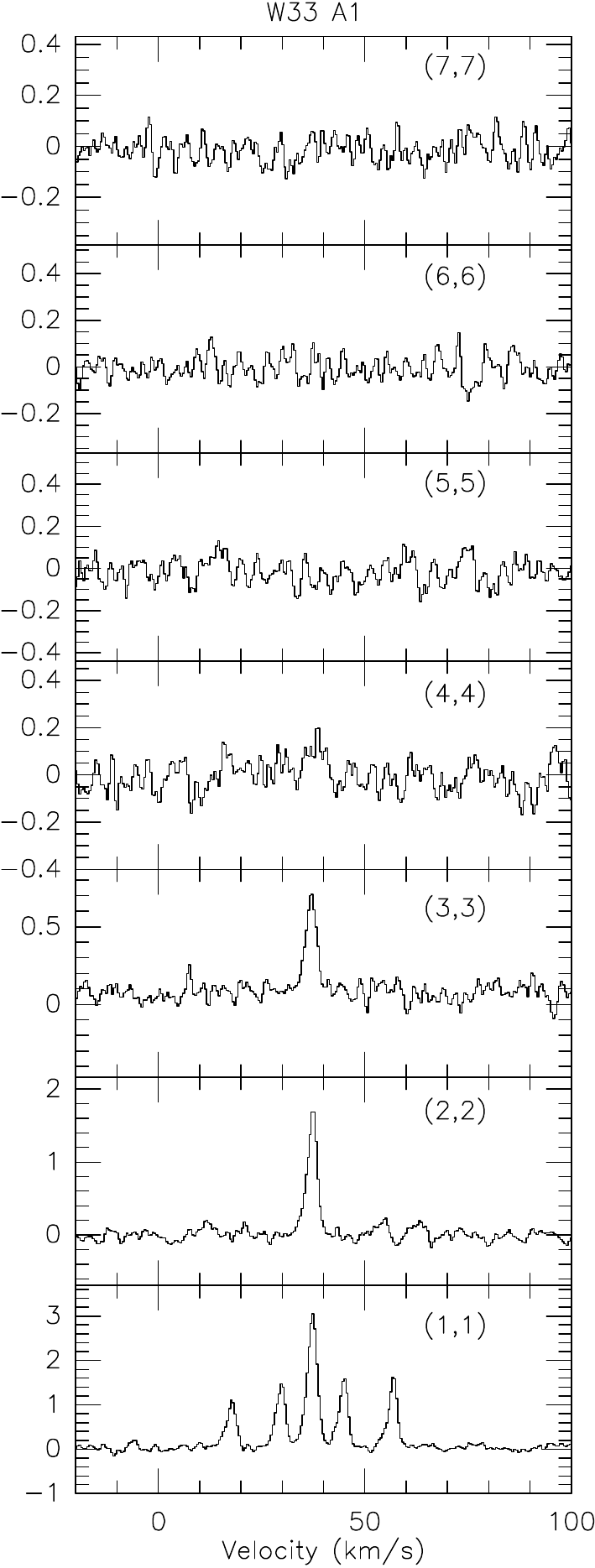}
\caption{NH$_3$ spectra of metastable lines at offset ($0\arcsec, 0\arcsec$) with respect to the reference positions given in Table\,\ref{table:1} for W33\,Main1 (\textit{left panel}) and W33\,A1 (\textit{right panel}). The channel widths are 0.48, 0.48, 0.48, 0.47, 0.47, 0.46, and 0.44\,km\,s$^{-1}$ for the NH$_3$\,(1,1) to (7,7) lines, respectively.}
\label{W33Main1,A1}
\end{figure*}

%-----------------Figure 5
\begin{figure*}[t]
\centering
\includegraphics[width=0.45\textwidth]{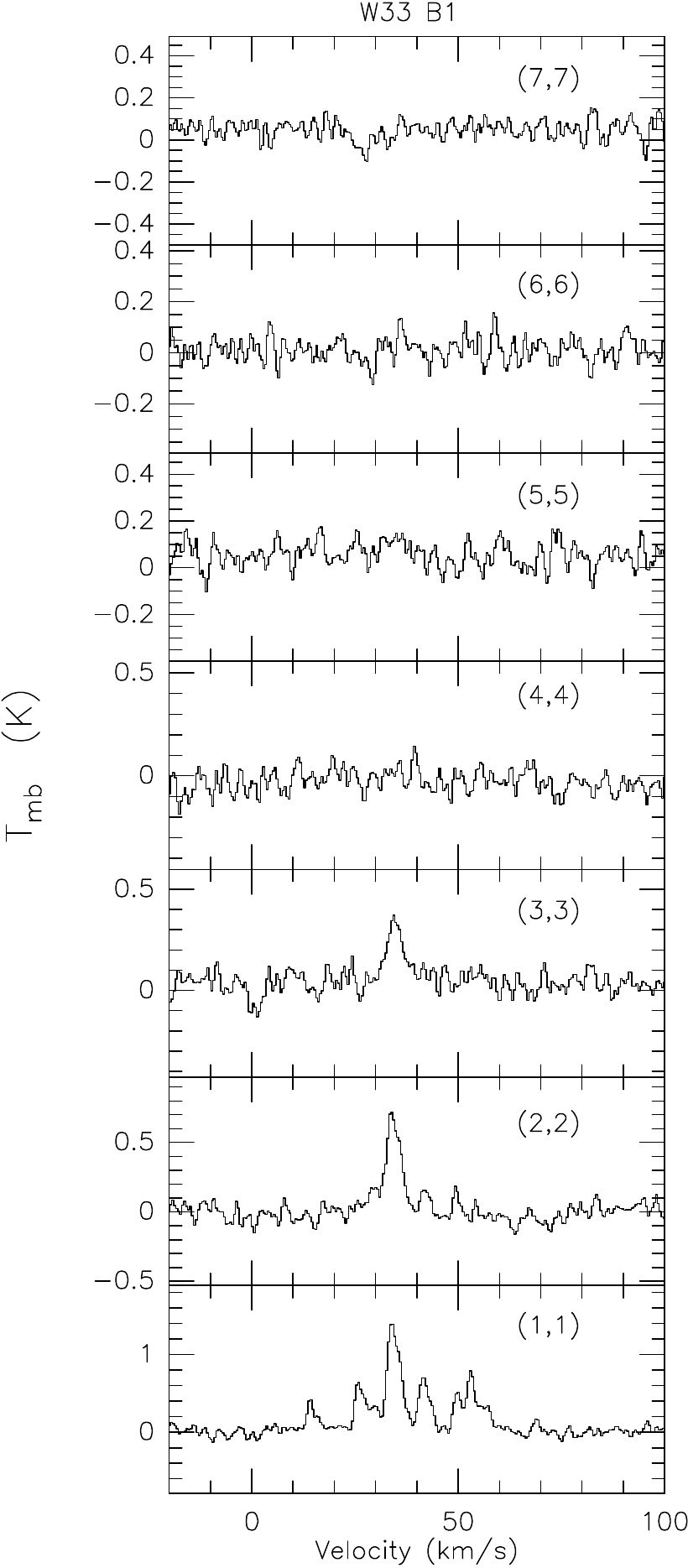}
\includegraphics[width=0.39\textwidth]{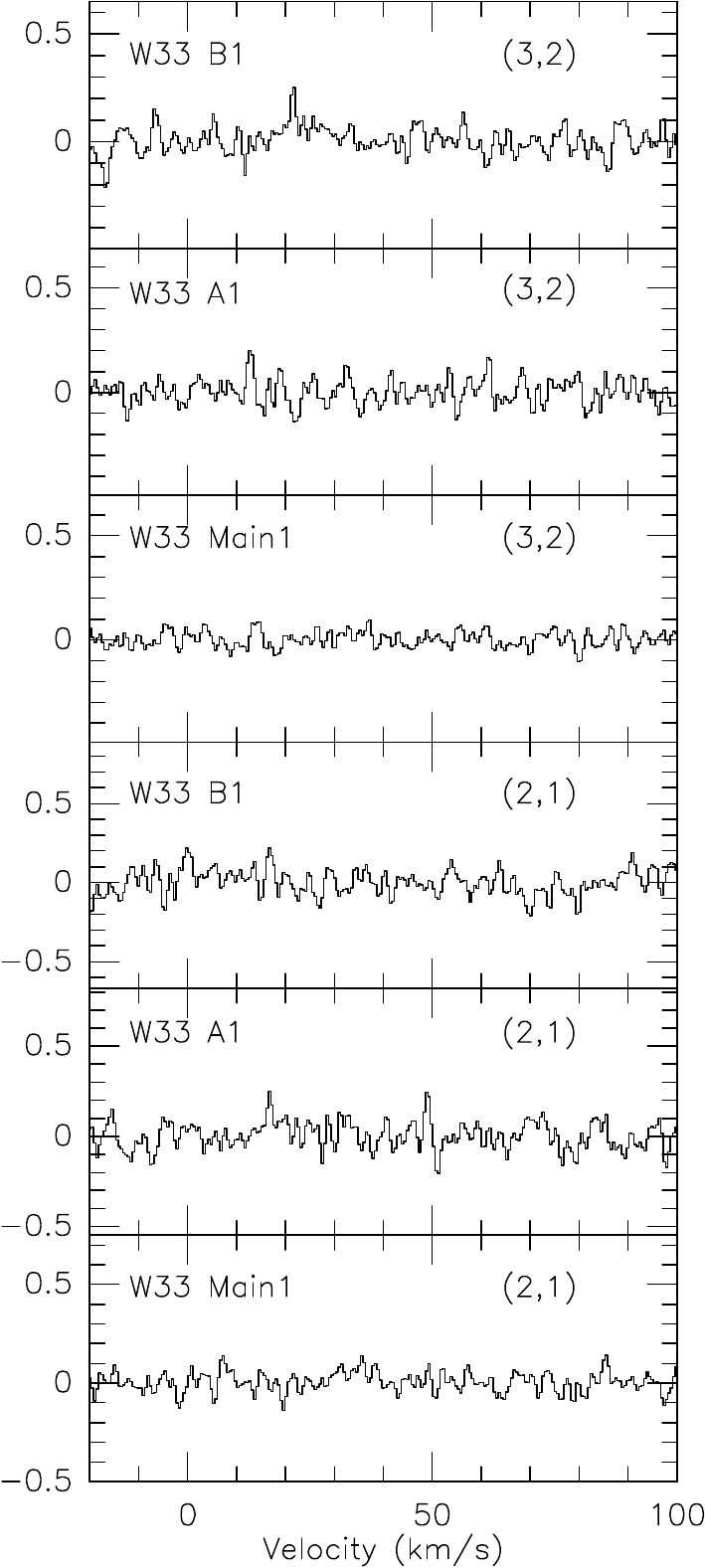}
\caption{NH$_3$ spectra of metastable lines at offset ($0\arcsec, 0\arcsec$) with respect to the reference positions given in Table\,\ref{table:1} for W33\,B1 (\textit{left panel}). Non-metastable lines from W33\,Main1, W33\,A1, and W33\,B1 are shown in the \textit{right panel}. Channel widths are 0.48, 0.48, 0.48, 0.47, 0.47, 0.46, and 0.44\,km\,s$^{-1}$ for the NH$_3$\,(1,1) to (7,7) lines, respectively, while the corresponding values for the NH$_3$\,(2,1) and (3,2) lines are 0.49 and 0.50\,km\,s$^{-1}$.}
\label{W33B1,non.}
\end{figure*}

%-----------------Figure 6

%-----------------Figure 7
\begin{figure*}
\centering
\includegraphics[width=16.0cm,height=20.0cm,angle=0]{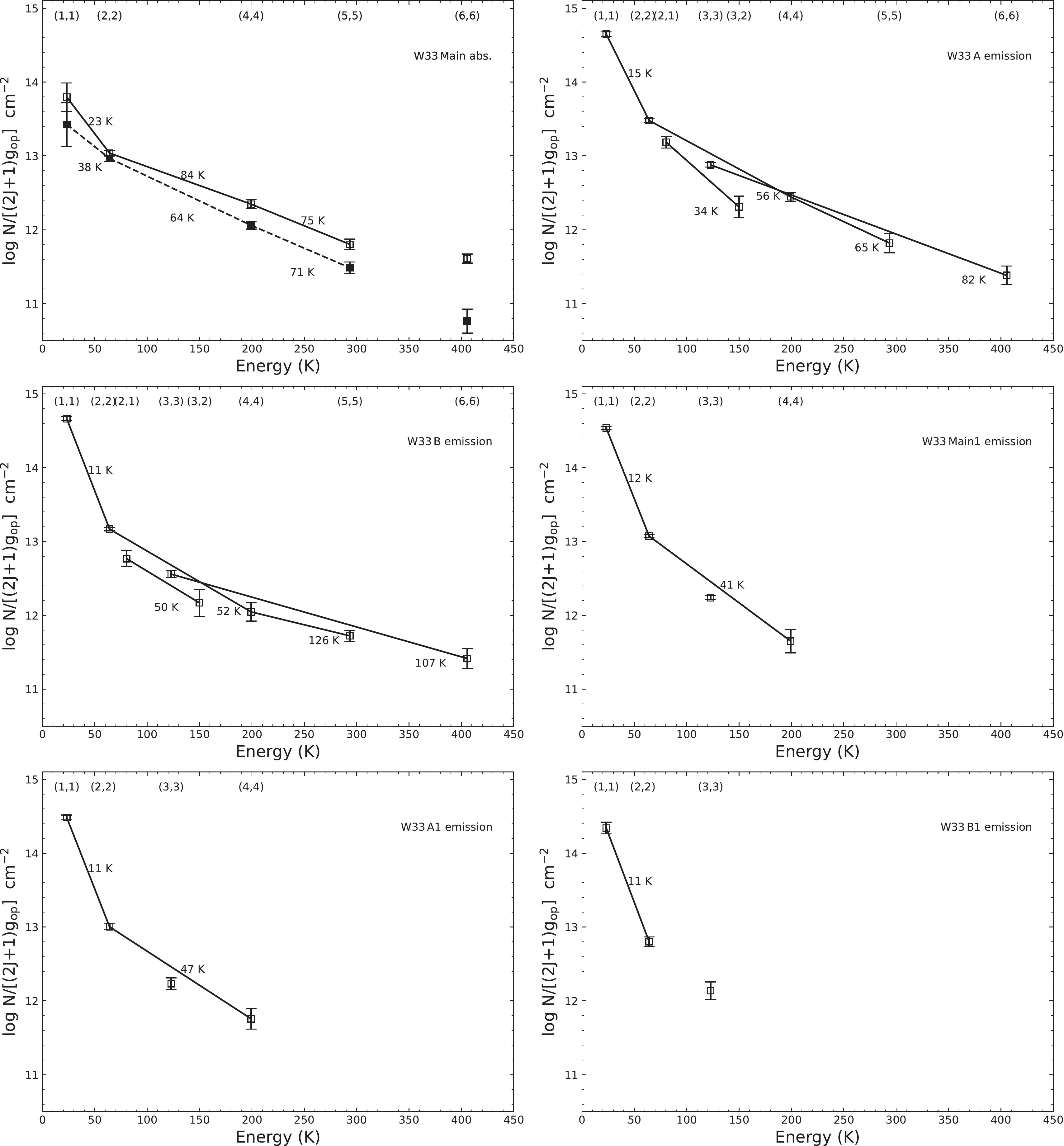}
\caption{The Boltzmann plots (rotation diagrams) for the W33\,Main absorption lines (\textit{top left}), W33\,A emission lines (\textit{top right}), W33\,B emission lines (\textit{second row left}), W33\,Main1 emission lines (\textit{second row right}), W33\,A1 emission lines (\textit{third row left}), and W33\,B1 emission lines (\textit{third row right}). The taken positions are those of Table\,\ref{table:1}. The solid and dashed lines in the top left panel represent the first velocity component at $V_{\rm LSR}$\,$\sim$\,33\,km\,s$^{-1}$ and second velocity component at $V_{\rm LSR}$\,$\sim$\,38\,km\,s$^{-1}$, respectively. For para-NH$_3$, $g_{\rm op}$\,$=$\,1. For ortho-NH$_3$, here the (3,3) and (6,6) levels, $g_{\rm op}$\,$=$\,2. The rotational temperatures are obtained from the corresponding slopes. The numbers mark the rotational temperatures in\,K.}
\label{rotationdiagram}
\end{figure*}

\subsection{NH$_3$ column density}
\label{ammonia-column-density}
Tables\,\ref{Table-absorption} and \ref{Table-emission} list main beam brightness temperatures, velocities,
FWHM line widths, optical depths, and column densities for all measured transitions toward the six reference positions (Table\,\ref{table:1}). Towards the reference position of W33\,Main, opacities of the NH$_3$\,(1,1), (2,2), (4,4), (5,5) and (6,6) absorption lines (Table\,\ref{Table-absorption}) were calculated using
\begin{eqnarray}
\tau = -{\rm ln} \left(1 - \frac{|T_{\rm L}|}{T_{\rm c}}\right).
\label{Eq-1}
\end{eqnarray}
Here $T_{\rm L}$ is the observed line temperature assuming full continuum source coverage and $T_{\rm c}$ is the corresponding temperature of the continuum source.

In case of lines showing significant saturation effects, so that the optical depth can be determined by the
GILDAS `NH$_3$\,(1,1)' fitting method (see Sect.\,\ref{sect-2-2}), the column density in
the $(J,K)$ state is obtained with the optical depth $\tau_{\rm tot}$, following \citet{1986A&A...162..199M}, by
\begin{eqnarray}
N(J,K)=\frac{1.65\times10^{14}}{\nu} \,\frac{J(J+1)}{K^{2}}\,\Delta v\,\tau_{\rm tot}\,T_{\rm ex}\ \ {\rm cm^{-2},}
\label{Eq-2}
\end{eqnarray}
where $N$ is in cm$^{-2}$, the FWHM line width $\Delta v$ is in \,km\,s$^{-1}$, the line frequency $\nu$
is in GHz, and the excitation temperature $T_{\rm ex}$ is in K. In the optically thick case,
the excitation temperature $T_{\rm ex}$ is derived from the main beam brightness temperatures $T_{\rm MB}$
and the optical depth $\tau$ by
\begin{eqnarray}
T_{\rm MB} = (T_{\rm ex}-2.7)(1-exp(-\tau))
\label{Eq-3}
\end{eqnarray}
In the optically thin case, the main beam brightness temperatures $T_{\rm MB}$ can be approximated by $T_{\rm ex}$\,$\tau$, so that Eq.(\ref{Eq-2}) can also be used in these instances.

The logarithm of normalized column densities, $\log$\,$N/[(2J+1)g_{\rm op}]$, as a function of the energy of the involved states above the ground
state is shown in Fig.\,\ref{rotationdiagram}. Normalization is obtained by dividing $N(J, K)$ by the statistical
weight of the respective transition, $(2J+1)$$g_{\rm op}$, with $g_{\rm op}$\,=\,1 for
para-NH$_3$ ($K$\,=\,1, 2, 4, 5, 7, 8, 10) and $g_{\rm op}$\,=\,2 for ortho-NH$_3$ ($K$\,=\,3, 6, 9).

\subsection{Rotation temperature}
\label{rotation-temperature}
We calculated the rotation temperature between different energy levels with $T_{\rm rot}$\,=\,--log$e/a$\,$\approx$\,--0.434/$a$ \citep{2000A&A...361L..45H}, where the slope $a$ is obtained by linear fitting the Boltzmann plot, relating normalized intensity
to excitation above the ground state. In Fig.\,\ref{rotationdiagram}, we show the rotation diagrams for the six metastable and
two non--metastable NH$_3$ absorption lines for W33\,Main and all the clearly detected W33\,A, W33\,B, W33\,Main1, W33\,A1, and W33\,B1 emission lines, respectively. In the six main W33 sources, all our NH$_{3}$ lines with the exception of the ($J,K$)\,=\,(1,1) transition are optically thin. The NH$_3$\,(1,1) line is always optically thick (see Tables\,\ref{Table-absorption}, \ref{Table-emission}). In this case, the derived optical depth affects the determination of the column density in the (1,1) state. This effect is estimated using Eqs.\,\ref{Eq-2} and \ref{Eq-3} in Sect.\,\ref{ammonia-column-density}.

The upper left panel of Fig.\,\ref{rotationdiagram} shows the Boltzmann plots of the two velocity components of the W33\,Main absorption lines (for the position, see Table\,\ref{table:1}), where the first velocity component ($V_{\rm LSR}$\,$\sim$\,33\,km\,s$^{-1}$) of the two inversion transition lines NH$_3$\,$(J,K)$\,=\,(1,1) and (2,2) gives the value $T_{\rm rot}$\,=\,23\,$\pm$\,5\,K. For the (2,2) and (4,4) lines we obtain $T_{\rm rot}$\,=\,84\,$\pm$\,3\,K, while the (4,4) and (5,5) lines give $T_{\rm rot}$\,=\,75\,$\pm$\,2\,K. The rotational temperature of the para-NH$_3$ species obtained by fitting all four absorption lines is $T_{\rm rot}$\,=\,64\,$\pm$\,11\,K. The second velocity component ($V_{\rm LSR}$\,$\sim$\,38\,km\,s$^{-1}$) has $T_{\rm rot}$\,=\,38\,$\pm$\,6\,K for the NH$_3$\,$(J,K)$\,=\,(1,1) and (2,2) transitions. For the NH$_3$\,(2,2) and (4,4) lines, the corresponding value becomes 64\,$\pm$\,1\,K. The NH$_3$\,(4,4) and (5,5) lines give 71\,$\pm$\,3\,K and a fit to all four para-NH$_3$ transitions indicates 62\,$\pm$\,4\,K (see Fig.\,\ref{rotationdiagram} top left panel and Table\,\ref{Table-Trot}). No $T_{\rm rot}$ value can be derived from ortho-NH$_3$, because the (3,3) line is seen in emission and shows quite a different line shape.

For the W33\,A emission lines (position in Table\,\ref{table:1}), we obtain $T_{\rm rot}$\,=\,45\,$\pm$\,9\,K by
fitting the para-NH$_3$\,(1,1), (2,2), (4,4) and (5,5) lines, and $T_{\rm rot}$\,=\,82\,$\pm$\,5\,K for the (3,3) and (6,6)
ortho-NH$_3$ transitions. For para-NH$_3$ the (1,1) and (2,2) lines give $T_{\rm rot}$\,=\,15\,$\pm$\,1\,K, while the (2,2)
and (4,4) lines give 56\,$\pm$\,2\,K, the (4,4) and (5,5) lines give 65\,$\pm$\,7\,K, and the (2,1) and (3,2) lines yield 34\,$\pm$\,3\,K (see Fig.\,\ref{rotationdiagram} top right panel, and Table\,\ref{Table-Trot}). For W33\,B (see Table\,\ref{table:1} for the position), the four para-NH$_3$ emission lines, i.e. the (1,1), (2,2), (4,4), (5,5) transitions, give $T_{\rm rot}$\,=\,44\,$\pm$\,12\,K. In contrast, the (3,3) and (6,6) ortho-NH$_3$ lines yield $T_{\rm rot}$\,=\,107\,$\pm$\,8\,K. The rotation temperature between the lowest inversion doublets of para-NH$_3$, the (1,1) and (2,2) lines, is $T_{\rm rot}$\,=\,11\,$\pm$\,1\,K, while the (2,2) and (4,4) lines give 52\,$\pm$\,5\,K. The other two para-NH$_3$ transitions, (4,4) and (5,5), give $T_{\rm rot}$\,=\,126\,$\pm$\,19\,K. The two non-metastable lines (2,1) and (3,2) lines yield $T_{\rm rot}$\,=\,50\,$\pm$\,6\,K (see Fig.\,\ref{rotationdiagram}, second row, left panel and Table\,\ref{Table-Trot}). For W33\,Main\,1 (see Table\,\ref{table:1} for the reference position), we get $T_{\rm rot}$\,=\,29\,$\pm$\,9\,K by fitting the three para-NH$_3$ species, i.e. the (1,1), (2,2) and (4,4) transitions. The NH$_3$\,(1,1) and (2,2) lines show $T_{\rm rot}$\,=\,12\,$\pm$\,1\,K, and the (2,2) and (4,4) lines give 41\,$\pm$\,4\,K (see Fig.\,\ref{rotationdiagram}, second row, right panel and Table\,\ref{Table-Trot}). For W33\,A\,1 the three para-NH$_3$ transition give $T_{\rm rot}$\,=\,31\,$\pm$\,11\,K. For the rotational temperature, (1,1) and (2,2), (2,2) and (4,4), we obtain 11\,$\pm$\,1\,K and 47\,$\pm$\,4\,K, respectively (Fig.\,\ref{rotationdiagram} third row, left panel and Table\,\ref{Table-Trot}). For the W33\,B1 emission lines the (1,1) and (2,2) para-NH$_3$ transitions indicate $T_{\rm rot}$\,=\,11\,$\pm$\,1\,K (see Fig.\,\ref{rotationdiagram} third row, right panel and Table\,\ref{Table-Trot}).

\subsection{Kinetic temperature}
\label{Kinetic temperature}
To obtain kinetic temperatures, $T_{\rm kin}$, of all observed regions of our map, the NH$_3$\,(1,1) and (2,2) lines are the best choice because their emission is quite extended (see Fig.\,\ref{integrated-intencity}). We obtained the rotation temperature of NH$_3$\,(1,1) and (2,2) using the same method described in Sect.\,\ref{rotation-temperature}, which is equivalent to
\begin{eqnarray}
\label{Eq1}
T_{\mbox{\tiny rot}}(1,2)= \frac{-41.5}{{\rm ln}({N}_{22}/5) - {\rm ln}({N}_{11}/3)}\ \  {\rm K,}
 \label{equation: Trot}
\end{eqnarray}
where $N_{\rm 11}$ and $N_{\rm 22}$ are the column densities of NH$_3$\,(1,1) and (2,2) lines from Eqs.\,\ref{Eq-2} and \ref{Eq-3}. The optically thick NH$_3$\,(1,1) line is not only relevant when deriving column densities of the (1,1) state but also when deriving the rotation temperatures of the gas traced by both the NH$_3$\,(1,1) and (2,2) transitions. We account for this effect by using Eq.\,\ref{equation: Trot} in this section. Then we calculate the $T_{\rm rot}$ values, which are presented in Table\,\ref{table:A.5}. To better visualize our $T_{\rm rot}$ values, in Fig.\,\ref{Fig:C.1} left panel, we add the rotational temperatures $T_{\rm rot}$ map of all observed regions, where Eq.\,\ref{equation: Trot} has been used to calculate the corresponding $T_{\rm rot}$ values.

Following \citet{2004A&A...416..191T} to connect rotational with kinetic temperatures, we used
\begin{eqnarray}
\label{Eq2}
T_{\mbox{\tiny kin}}=\frac{T_{\mbox{\tiny rot}}(1,2)}{1-\frac{T_{\mbox{\tiny rot}}(1,2)}
{41.5}\ln\left( 1+1.1 \, {\rm exp}\left(\frac{-16}{T_{\mbox{\tiny rot}}(1,2)}\right)\right)}\ \ {\rm K,}
 \label{equation: Tkin}
\end{eqnarray}
where the energy gap between the (1,1) and (2,2) states is $\Delta E_{12}$\,=\,41.5\,K. \citet{2004A&A...416..191T} ran different Monte Carlo models involving the NH$_3$\,($J,K$)\,=\,(1,1), (2,1), and (2,2) inversion doublets and an $n(r)$ = $n_0 / (1 + (r/r_{o})^{2.5})$ density distribution to compare their observationally determined approximately constant rotational temperatures with modelled kinetic temperatures in dense quiescent molecular clouds. Eq.\,(\ref{equation: Tkin}) is derived from fitting $T_{\rm kin}$ in the range of 5 to 20\,K. So there is a caveat in using it for higher temperatures.

\begin{figure}[t]
\vspace*{0.2mm}
\begin{center}
\includegraphics[width=0.5\textwidth]{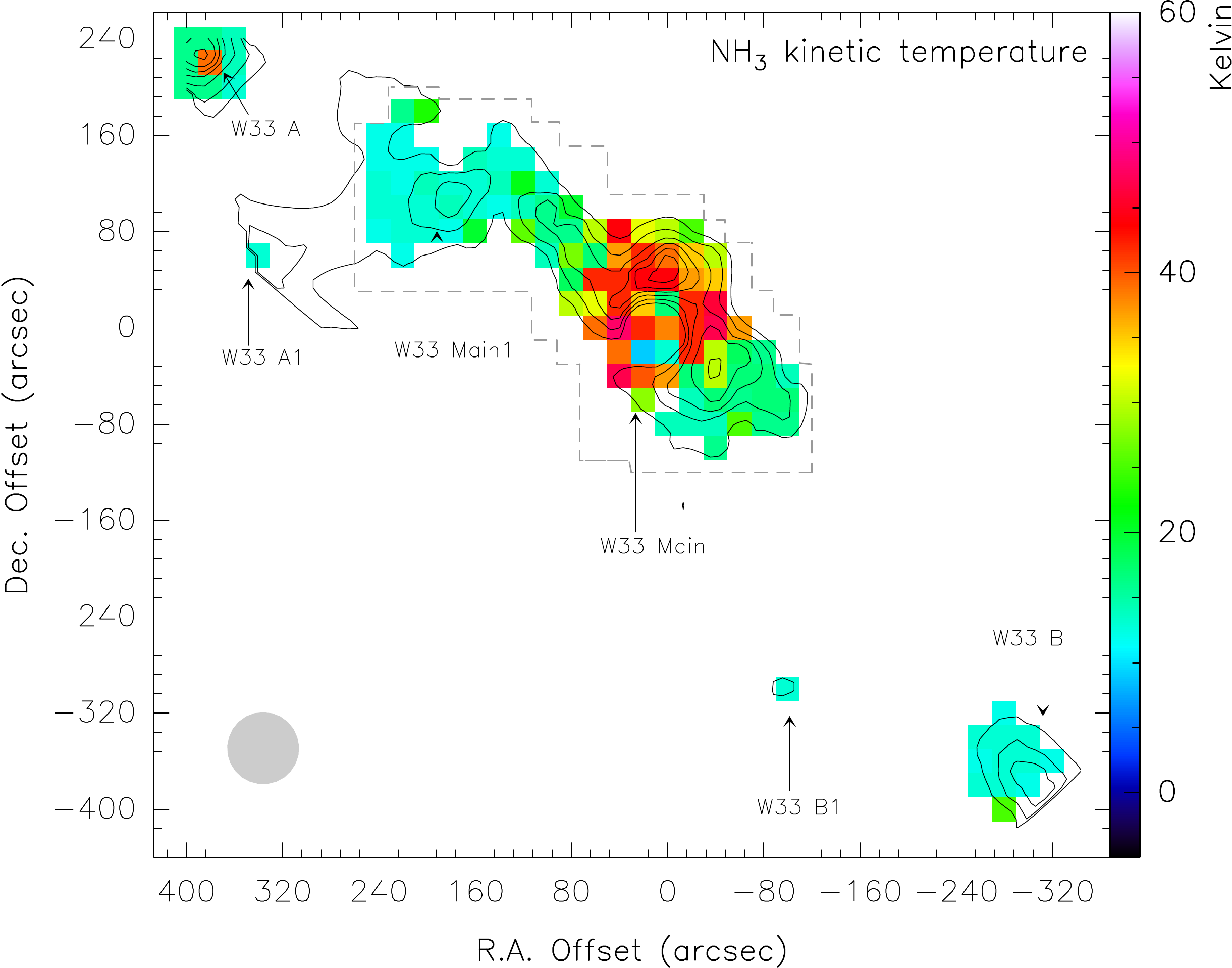}
\end{center}
\caption[]{Map of the kinetic temperature in units of Kelvin for the W33 regions, obtained from the para-NH$_3$\,(2,2)/(1,1) lines. The reference position is R.A.\,: 18:14:13.50, DEC.\,: -17:55:47.0 (J2000). The integration range is 32 to 40\,km\,s$^{-1}$. Contours are the same as in the left panel of Fig.\,\ref{integrated-intencity}. The limits of the mapped region over significant parts of our map are indicated by gray dashed lines. The half-power beam width is illustrated as a gray filled circle in the lower left corner of the image.}
\label{Tkin-map}
\end{figure}

The gas kinetic temperatures derived from the NH$_3$\,(2,2)/(1,1) map for the six main W33 sources are shown in Fig.\,\ref{Tkin-map}.
The kinetic temperatures of the dense gas in W33\,Main derived from the NH$_3$\,(2,2)/(1,1) ratios range from 9 to 49\,K with an average of
21\,$\pm$\,11\,K (errors are standard deviations of the mean). We find that the kinetic temperatures in the dense gas
around the central ($0\arcsec, 0\arcsec$) offset positions of our W33\,Main mapped region are high ($\sim$38\,K; see Fig.\,\ref{Tkin-map}).
This region contains a young stellar cluster associated with an H\,{\scriptsize II} region. The gas kinetic temperatures
from para-NH$_3$ in the other five W33 regions are cooler ranging from 13 to 39\,K with an average of
18\,$\pm$\,8\,K in W33\,A. Lower gas temperatures associate with W33\,B ranging from 8 to 39\,K with an average 15\,$\pm$\,7\,K, while dense gas in the W33\,B1 region shows kinetic temperatures ranging from 7 to 39\,K with an average of 21\,$\pm$\,12\,K. Furthermore, with the three para- and one ortho-NH$_3$ lines detected towards W33\,Main1 and W33\,A1, we use the NH$_3$\,(1,1) and (2,2) temperatures determined as $T_{\rm kin}$\,=\,13\,K for both sources (see Table\,\ref{table:A.5}).

\subsection{Total NH$_3$ column density and H$_2$ volume density}
\label{column-volume-density}
Using Eq.\,(\ref{Eq-2}), we calculate NH$_3$ column densities for the observed metastable and non-metastable
inversion doublets (see Tables\,\ref{Table-absorption} and \ref{Table-emission}). We have also determined the column
density of the $(J,K)$\,=\,(0,0) ground state, using the method described in \cite{2017ApJ...850...77K}, which is
\begin{eqnarray}
{N}_{00}=\frac{1}{3} \exp\left(\frac{23.2\,{\rm K}}{{T}_{\rm kin,\,12}}\right) {N}_{11}\,,
\label{equation: n00 extrapolation}
\end{eqnarray}
where the energy difference between the NH$_3$\,(0,0) and NH$_3$\,(1,1) is 23.2\,K, $T_{\rm kin,\,12}$ is the gas kinetic temperature derived from the (1,1) and (2,2) doublets (see Sect.\,\ref{Kinetic temperature} and Table\,\ref{table:A.5}). We obtained total column densities of ammonia by adding the column densities of all convincingly detected metastable NH$_{3}$ lines and the $(J,K)$\,=\,(0,0) column density from the ortho-NH$_{3}$ ground state, following \cite{2017ApJ...850...77K}

\begin{eqnarray}
{N}_{\rm tot}=\left[\frac{1}{3} \exp \left( \frac{23.2\,{\rm K}}{{T}_{12,\,\rm kin}}\right) +1 \right] {N}_{11} + {N}_{22} + {N}_{33} + {N}_{44} \\ \nonumber + {N}_{55} + {N}_{66}\,.
\label{equation: total column density}
\end{eqnarray}
The column densities of the (0,0) state are given in Tables\,\ref{Table-absorption} and \ref{Table-emission} for W33\,Main,
W33\,A, W33\,B, W33\,Main1, W33A1, and W33\,B1, respectively. In addition, the column densities of para-NH$_3$ and
ortho-NH$_3$, and total-NH$_3$\,(para+ortho) column densities for the six main W33 sources are calculated,
and listed in Table\,\ref{Table-abundance}. The W33 complex shows a broad distribution of total-NH$_3$ column densities.
Among the six main sources, we obtain at the reference positions (Table\,\ref{table:1}) the lowest value for W33\,Main, $N$(total-NH$_3$)\,=\,6.0\,($\pm$2.1)\,$\times$\,10$^{14}$\,cm$^{-2}$ (for the detailed calculation of this value, see Sect.\,\ref{maser-line}). The highest value is derived for W33\,B, $N$(total-NH$_3$)\,=\,3.4\,($\pm$0.2)\,$\times$\,10$^{15}$\,cm$^{-2}$ (see Table\,\ref{Table-abundance}). As can be seen in the central panel of Fig.\,\ref{Fig:C.1}, we provide a total column density map of NH$_3$ for the six W33 sources. In addition, we determined the total NH$_3$ mass in the area covered by the Effelsberg data. The total NH$_3$ masses for each source are obtained by integrating the total-$N$(NH$_3$) values over the covered regions. For the total NH$_3$ masses, we find 13, 1.9, 3.4, 0.2, 0.2 and 1.3\,$\times$\,10$^{-4}$\,${M}_{\odot}$ for W33\,Main, W33\,A, W33\,B, W33\,Main1, W33\,A1 and W33\,B1, respectively.

The volume density of H$_{2}$ molecules has been obtained from the (1,1) line, using the method described in Equation (2) of \citet{1983ARA&A..21..239H}, which is:
\begin{equation}
n({\rm H}_2) = \frac{A}{C}  \left[ \frac{ J_\nu(T_{\rm ex}) - J_\nu(T_{\rm bg}) }
{ J_\nu(T_{\rm kin}) - J_\nu(T_{\rm ex}) } \right]  \left[ 1 +
\frac{J_\nu(T_{\rm kin})}{h\nu/k_{\scriptscriptstyle\rm B}} \right].
\label{Eq-4}
\end{equation}

$A$ is the Einstein coefficient for spontaneous emission (=\,1.71\,$\times$\,10$^{-7}$\,s$^{-1}$) and $C$ is
the collisional de-excitation rate ($\sim$8.5\,$\times$\,10$^{-11}$\,cm$^{-3}$\,s$^{-1}$) for the (1,1) line in
\citet{1988MNRAS.235..229D}. $T_{\rm bg}\,=\,2.73$\,K is the black body background radiation temperature,
and $J_\nu(T)$ is defined by
\begin{equation}
J_\nu(T) = \frac{h\nu}{k_{\scriptscriptstyle\rm B}}\left( {\rm e}^{h\nu/k_{\scriptscriptstyle\rm B}T} -1 \right)^{-1}.
\label{Eq-5}
\end{equation}
The relation between the gas kinetic temperature (here we rely exclusively on the (1,1) and (2,2) lines; see Sec.\,\ref{Kinetic temperature}),
and the excitation temperature can provide a reliable estimate of the gas volume density. However, the gas density calculated using Eq.(\ref{Eq-4}) may be significantly underestimated if the beam is not filled uniformly, i.e. if the sizes of our W33 sources are smaller than the beam size of 40$\arcsec$. We use Eq.(\ref{Eq-4}) to set a lower bound on the gas density, $n$(H$_{2}$), adopting a beam filling factor of $\eta$\,=\,1. Note that if $T_{\rm ex}\,=\,T_{\rm kin}$, Eq.(\ref{Eq-4}) is invalid and $n$(H$_{2}$) has to be calculated in a different way (see, e.g., \citealt{1983QJRAS..24..267H,2012ApJ...753...50P}). However, this problem did not occur in our case (see Table\,\ref{table:A.5}). In Table\,\ref{table:A.5}, we also present the obtained volume densities of the six main sources.

\section{Discussion}
\label{sect:discussion}
\subsection{Comparison with previously obtained volume densities}
\label{Comparison}
We calculated the volume density of hydrogen molecules $n({\rm H}_2)$ using the column densities $N_{\rm H_{2},\rm source}$ from Table\,5 of \cite{2014A&A...572A..63I}, i.e., $n({\rm H}_2)$\,=\,$N_{{\rm H_2},{\rm source}}$\,/\,2$r_{\rm source}$. $r_{\rm source}$ is the size of the respective clumps, taken from Table\,5 of \cite{2014A&A...572A..63I}. The volume densities of hydrogen molecules obtained from this method are 4.1\,$\times$\,10$^{3}$, 3.2\,$\times$\,10$^{3}$, 0.6\,$\times$\,10$^{4}$, 0.6\,$\times$\,10$^{4}$, 2.1\,$\times$\,10$^{4}$, 0.3\,$\times$\,10$^{4}$\,cm$^{-3}$ for W33\,Main, W33\,A, W33\,B, W33\,Main1, W33\,A1, and W33\,B1, respectively. The volume densities of \cite{2014A&A...572A..63I} and our results, using the new $T_{\rm kin}$ values (see Table\,\ref{table:A.5} and Sect.\,\ref{column-volume-density}) show that our volume densities are $\sim$3.0, $\sim$8.1, $\sim$1.8, and $\sim$3.0 times higher in W33\,Main, W33\,A, W33\,B, and W33\,Main1, respectively, $\sim$1.9 time lower in W33\,A1, and equal in W33\,B1. The good agreement indicates that beam filling factors in both the \cite{2014A&A...572A..63I} and our data are similar and possibly close to unity. Since spatial distributions of different species tend to differ, we consider the latter as a viable possibility.

\subsection{Variations of the NH$_3$ abundance}
\label{ammonia abundance}
Total-NH$_3$ column densities $N({\rm NH_{3}})$ are compared with the column densities of H$_{2}$ derived from
the Atacama Pathfinder Experiment (APEX) telescope using its 870\,$\mu$m continuum data \citep{2014A&A...572A..63I},
where the H$_{2}$ peak column densities (Table\,5 of \citealt{2014A&A...572A..63I}) is the best choice with respect to our
beam size of $40\arcsec$.

The fractional total-NH$_3$ abundances ($\chi$\,(total-NH$_3$)\,=\,(total-$N$(NH$_3$))/$N$(H$_2$)), calculated
for the peak positions of our six W33 sources, are listed in Table\,\ref{Table-abundance}. Therefore, the NH$_3$ abundances relative to those of molecular hydrogen are calculated to be 1.3\,($\pm$0.1)\,$\times$\,10$^{-9}$, 1.4\,($\pm$0.3)\,$\times$\,10$^{-8}$,
1.6\,($\pm$0.3)\,$\times$\,10$^{-8}$, 3.4\,($\pm$0.5)\,$\times$\,10$^{-8}$, 1.6\,($\pm$0.5)\,$\times$\,10$^{-8}$ and
4.0\,($\pm$1.2)\,$\times$\,10$^{-8}$ for the peak positions of W33\,Main, W33\,A, W33\,B, W33\,Main1, W33\,A1, and W33\,B1, respectively. The errors shown in parentheses are obtained using error propagation. The fractional total-NH$_3$ abundance map ($\chi$\,(total-NH$_3$)\,=\,(total-$N$\,(NH$_3$))/$N$(H$_2$)) is shown in the right panel of Fig.\,\ref{Fig:C.1}. The NH$_3$ abundances in the peak position of the six W33 sources are consistent with those in other Galactic
sources. The fractional NH$_3$ abundances are 2\,$\times$\,10$^{-8}$ in the cyanopolyyne peak of TMC-1 \citep{1987ASSL..134..561I}, 2\,$\times$\,10$^{-7}$ in the Orion ridge \citep{1987ASSL..134..561I}, (1--10)\,$\times$\,10$^{-8}$ \citep{1987ASSL..134..561I} and 8\,$\times$\,10$^{-8}$\,--\,10$^{-4}$ \citep{1993A&A...276..445H} in Sgr\,B2, the latter also including hot cores. In addition, averaged fractional ammonia abundance values of 1.2\,$\times$\,$10^{-7}$, 4.6\,$\times$\,$10^{-8}$, and 1.5\,$\times$\,$10^{-8}$, were obtained by \citet{2011ApJ...741..110D}, \citet{2012A&A...544A.146W}, and \citet{2019MNRAS.483.5355M} in clumps of the Bolocam Galactic Plane Survey (BGPS), the APEX Telescope Large Area Survey of the GALaxy (ATLASGAL), and the Hi-GAL survey, respectively.
Fractional abundances of $\sim$2--3\,$\times$\,$10^{-8}$ were derived for protostellar and starless cores
in the Perseus and Taurus-Auriga dark clouds as well as in infrared dark clouds \citep{2006A&A...455..577T,2009ApJ...696..298F,
2013A&A...552A..40C}.

The fractional NH$_3$ abundance varies among star-forming regions \citep{1983ApJ...270..589B}. In quiescent clouds, ammonia
should have a fractional abundance of $\sim$10$^{-7...-9}$, while in the hot cores its abundance could be two or three orders
of magnitude higher \citep{2013A&A...549A..90H}. The difference in the total-NH$_3$ abundances among the three
W33 sources at the peak position is that W33\,Main encounters the lowest value with 1.3\,($\pm$0.1)\,$\times$\,10$^{-9}$, while the total-NH$_3$ abundance of W33\,B 1.6\,($\pm$0.3)\,$\times$\,10$^{-8}$ is slightly higher than that of W33\,A 1.4\,($\pm$0.3)\,$\times$\,10$^{-8}$. As mentioned before, the obtained total NH$_3$ abundances at the peak positions are 3.4\,($\pm$0.5)\,$\times$\,10$^{-8}$, 1.6\,($\pm$0.5)\,$\times$\,10$^{-8}$, and 4.0\,($\pm$1.2)\,$\times$\,10$^{-8}$ for W33\,Main1, W33\,A1, W33\,B1 respectively. According to \citet{2014A&A...572A..63I}, W33\,Main is more evolved, also hosting an H\,{\scriptsize II} region, W33\,A and W33\,B can be considered as hot cores, W33\,B is rich in nitrogen \citep{2014A&A...572A..63I}. From our fractional total NH$_3$ abundances calculation, we can confirm the different evolutionary stages proposed by \citet{2014A&A...572A..63I} and find that there is no hot core in the region approaching the extreme conditions encountered in W51-IRS2 or Sgr\,B2. Figure\,\ref{Fig:C.2}, top left and top right panels, indicates that total-$N$(NH$_3$) and total fractional NH$_3$ abundance versus to the evolutionary sequence of the six W33 sources. The lower total-NH$_3$ fractional abundance in W33\,Main compared to the other five main W33 sources is likely due to the fact that W33\,Main is strongly affected by FUV photons originating from its H\,{\scriptsize II} region. Ammonia is a particularly sensitive molecular species with respect to FUV radiation (e.g., \citealt{2001ApJ...554L.143W}).

%-----------------Figure 8

\subsection{Ortho-to-para NH$_3$ ratio}
\label{ratio}
In massive star forming dense cores, outflow induced shock waves and rising levels of stellar radiation can liberate
NH$_3$ molecules confined to dust grains (e.g., \citealt{1990MNRAS.246..183N,1995A&A...294..815F}) and increase
the NH$_3$ abundance. Ortho-to-para abundance ratios of NH$_3$ can tell us about the contribution of liberated NH$_3$
molecules with respect to those formed in the gas phase \citep{1999ApJ...525L.105U}.

The ortho-to-para ratio depends on its origin either in the gas or in the dust phase. As \citet{1999ApJ...525L.105U} described,
if NH$_3$ is formed by gas phase reactions, the ortho-to-para ratios will be close to unity. On the other hand, a formation that occurred on dust grains released to the interstellar medium could raise the ortho-to-para ratio above unity. In the latter case, the ortho-to-para ratio also depends on the NH$_3$ formation temperature. Here, the ortho-to-para ratio is inversely proportional to the kinetic temperature of the gas at the time of the formation of the NH$_3$ molecules. The ratio is about 3 at formation temperature 10\,K \citep{2002PASJ...54..195T}. The ortho-to-para ratio in our case (($N_{00}$+$N_{33}$+$N_{66}$)\,/($N_{11}$+$N_{22}$+$N_{44}$+$N_{55}$)), calculated for the peak positions of our six main W33 sources, are listed in Table\,\ref{Table-abundance}. With the four para- and two ortho-NH$_3$ lines observed, we estimate the ortho-to-para abundance ratios to be 0.5\,($\pm$0.1), 1.3\,($\pm$0.1), 1.3\,($\pm$0.1) for W33\,Main, W33\,A and W33\,B, respectively (see Table\,\ref{Table-abundance}). The very low value toward W33\,Main is likely caused by the inclusion of the (3,3) maser line, which may introduce systematic errors which are difficult to quantify (see Sect.\,\ref{maser-line}). In addition, with three para-NH$_3$ transitions and one ortho-NH$_3$ line detected in W33\,Main1 and W33\,A1, we can determine the ortho-to-para abundance ratios to be 1.8\,($\pm$0.1), and 1.9\,($\pm$0.1), respectively (see Table\,\ref{Table-abundance}). For W33\,B1, the detected two para-NH$_3$ transitions and one ortho-NH$_3$ line provide an ortho-to-para abundance ratio 1.9\,($\pm$0.1). The errors in parentheses are calculated using error propagation.

As we discussed in Sects.\,\ref{rotation-temperature} and \ref{Kinetic temperature}, the NH$_3$\,(1,1) line is optically thick in all the W33 sources. The effect of an optically thick NH$_3$\,(1,1) line on the ortho-to-para ratio is estimated using equation ($N_{00}$+$N_{33}$+$N_{66}$)\,/($N_{11}$+$N_{22}$+$N_{44}$+$N_{55}$) in this section. Plotting the ortho-to-para ratios against the evolutionary stage (Fig.\,\ref{Fig:C.2} bottom left panel) of our six targets, we may see a trend of decreasing ortho-to-para ratios with evolutionary stage. \citet{2002PASJ...54..195T} suggests that an ortho-to-para ratio of $\sim$1.5 corresponds to a cool formation temperature of $\sim$20\,K. According to this, from the ortho-to-para abundance ratios of 0.5\,($\pm$0.1), 1.3\,($\pm$0.1), 1.3\,($\pm$0.1), 1.8\,($\pm$0.1), 1.9\,($\pm$0.1) and 1.9\,($\pm$0.1), we believe that ammonia has either been formed in the gas-phase or has been formed on dust grains in a medium with $\sim$\,20\,K or more to be then released into the interstellar medium.

\subsection{A comparison of kinetic temperatures with previously obtained data}
\label{temperature}
\cite{2014A&A...572A..63I} presented rotation diagrams of H$_{2}$CO, CH$_{3}$OH and CH$_{3}$CCH for W33\,Main, of H$_{2}$CO for W33\,A, of H$_{2}$CO, HNCO, CH$_{3}$CN and CH$_{3}$OH for W33\,B, and of H$_{2}$CO for W33\,Main1, W33\,A1 and W33\,B1 and obtained a different rotational temperature for each of the six W33 sources (see Table\,6, Figs.\,10 and 12 of \citealt{2014A&A...572A..63I}). We have compared the NH$_3$ temperatures with those
derived from the rotation diagrams of \cite{2014A&A...572A..63I}, and find that these are quite similar (Table\,6 of \citealt{2014A&A...572A..63I} and our Table\,\ref{Table-Trot}). However, the $T_{\rm rot}$ values of the \cite{2014A&A...572A..63I} study are affected by both kinetic temperature and density, while NH$_3$ allows for a determination of the kinetic temperature alone. So there is a degeneracy in the \cite{2014A&A...572A..63I} data, which is one of the main motivations for this study.

Nevertheless, our $T_{\rm kin}$ estimates from NH$_3$ are not uniform. The Boltzmann diagrams in Fig.\,\ref{rotationdiagram} clearly
show rising rotation temperatures with increasing excitation above the ground states. Similar results from Galactic sources (e.g., \citealt{1993A&A...276L..29W,2006A&A...460..533W}) were interpreted in terms of the presence of gradients in $T_{\rm kin}$, which may
indicate the existence of dense post-shock gas that is gradually cooling with increasing distance from the shock front \citep{2008A&A...485..451H}.
However, in a warm environment radiative transfer calculations show that higher metastable states favor higher rotation temperatures that
gradually approach the kinetic temperature even if the $T_{\rm kin}$ of the gas has only a single value \citep{1983A&A...122..164W,1988MNRAS.235..229D,1995A&A...294..815F,2008A&A...485..451H}.

From Tables\,\ref{Table-absorption} and \ref{Table-emission} it is clear that most of the NH$_3$ column densities reside in the (1,1) and (2,2) states. In addition to widespread emission from these inversion doublets, the relatively high column densities are the second main motivation why we chose the (1,1) and (2,2) transitions for our kinetic temperature estimates. Nevertheless, the presence of higher-$J$ metastable ammonia transitions indicates that higher excited gas is also present (see Fig.\,\ref{rotationdiagram}). Our detection of the non-metatable (2,1) and (3,2) lines in W33\,Main, W33\,A and W33\,B (Fig.\,\ref{W33B,non.}) may indicate the existence of gas components with high volume densities ($>$\,10$^{5}$\,cm$^{-3}$) and/or intense infrared radiation fields (e.g. \citealt{1985A&A...146..168M}). Again considering W33\,Main, W33\,A and W33\,B, it is clear from Fig.\,\ref{rotationdiagram} and Sect.\,\ref{rotation-temperature} that the rotation temperatures of para-NH$_3$ are highest in W33\,Main and lowest in W33\,B. Thus we find a hierarchy of kinetic temperatures with W33\,Main containing the warmest and W33\,B the coolest gas, while conditions in W33\,A are intermediate. These kinetic temperatures are compatible with the stages of evolution outlined by \citet{2014A&A...572A..63I} and are also indicative of significant temperature gradients within the dense gas of W33\,B. We found clear trends as a function of evolutionary stage in the gas kinetic temperatures (Fig.\,\ref{Fig:C.2} bottom right panel). From our ammonia $T_{\rm kin}$ determinations we thus conclude that large temperature gradients may be present in these three W33 clumps.

\subsection{Maser emission in the NH$_3$\,(3,3) line}
\label{maser-line}
Molecular masers associated with ongoing massive star formation have been detected in a large number of studies (OH: e.g. \citealt{1983ApJ...265..295H};  H$_{2}$O: e.g. \citealt{1996A&AS..120..283H}; CH$_{3}$OH: e.g. \citealt{1998MNRAS.301..640W}; and
NH$_{3}$\,(3,3): e.g. \citealt{2013A&A...549A..90H}). Compared to other masers, NH$_{3}$\,(3,3) masers are rare and most of the known
NH$_{3}$ maser lines are from non-metastable ($J$\,$>$\,$K$) inversion transitions \citep{2013A&A...549A..90H}. \cite{1982A&A...110L..20W} first
detected (3,3) maser emission, in the massive star forming region of W33. To date, NH$_3$\,(3,3) maser emission has been detected in more than a dozen of star forming clouds (e.g., NGC\,7538-IRS1, DR21(OH), NGC\,6334\,V, NGC\,6334\,I, W51, IRAS\,20126+4104, G5.89-0.39, G20.08-0.14N, G23.33\,-\,0.30, G30.7206-00.0826, G35.03+0.35, G28.34+0.06, W51C, W44, G5.7-0.0, G1.4-0.1; \citealt{1986A&A...160L..13M}; \citealt{1994ApJ...428L..33M}; \citealt{1995ApJ...439L...9K}; \citealt{1995ApJ...450L..63Z}; \citealt{1999ApJ...527L.117Z}; \citealt{2008ApJ...680.1271H}; \citealt{2009ApJ...706.1036G}; \citealt{2011MNRAS.416.1764W}; \citealt{2011MNRAS.418.1689U}; \citealt{2011ApJ...739L..16B}; \citealt{2012ApJ...745L..30W}; \citealt{2016ApJ...826..189M}; \citealt{2016ApJ...826..189M}; \citealt{2016ApJ...826..189M}; \citealt{2016ApJ...826..189M}). The NH$_{3}$\,(3,3) maser emission can occur at densities between 10$^{3.5}$\,$\lesssim$\,$n$(H$_{2}$)\,$\lesssim$\,10$^{7.3}$\,cm$^{-3}$, kinetic temperatures larger than about 20\,K, and column densities less than $N$(ortho-NH$_{3}$)\,$\lesssim$\,10$^{16.8}$\,cm$^{-2}$ \citep{1995ApJ...439L...9K}. \citet{1999ApJ...527L.117Z} propose that NH$_{3}$\,(3,3) masers are excited in shocked regions of molecular outflows.

In Figs.\,\ref{W33Main,A} (left panel) and \ref{FgA.3}, we can see that the absolute main beam brightness temperature of the NH$_{3}$\,(3,3) emission line is about twice as large as those of the (1,1) and (2,2) absorption lines and peaks at 36\,km\,s$^{-1}$. It indicates that this line is a weak NH$_3$ maser as described in \cite{1982A&A...110L..20W}. Because of a radial velocity of 36\,km\,s$^{-1}$, which is compatible with the velocities derived from other NH$_{3}$ transitions seen in absorption against the continuum, because of the low opacities of all our NH$_{3}$ absorption lines with the exception of the ($J,K$)\,=\,(1,1) transition (Table\,\ref{Table-absorption}) and in view of the moderate strength of the (3,3) emission, it is plausible that the (3,3) emission is based on inverted populations amplifying the background continuum. From our Effelsberg data, we obtain a continuum flux density of 18.7\,$\pm$\,0.2\,Jy or 26.2\,$\pm$\,0.3\,K on a main beam brightness temperature scale. The corresponding main beam brightness temperature of the (3,3) line is 3.7\,$\pm$\,0.1\,K, so that the line to continuum ratio becomes 0.141\,$\pm$\,0.002 (relative calibration
uncertainties are not included in this error budget). This may indicate that the maser line is unsaturated and optically thin so that in this case, with negative $T_{\rm ex}$ and negative $\tau$, the product $T_{\rm ex}$\,$\times$\,$\tau$ almost matches the value in case of quasi-thermal emission (see, for example, \citealt{1986A&A...160L..13M} and \citealt{1991A&A...247..516S}). Under these circumstances, we can also derive physical parameters, including the (3,3) line, and obtain a NH$_{3}$\,(3,3) column density of 2.0\,($\pm$0.1)\,$\times$\,10$^{14}$\,cm$^{-2}$ (see Table\,\ref{Table-emission}). However, this value strongly depends on the correctness of our approach. If, for example, the maser only amplifies a part of the background continuum, it's absolute opacity will be higher than estimated, leading to a higher ortho-NH$_{3}$ column density and ortho-to-para ratio than derived in Sect.\,\ref{ratio}.

In general, an important feature of maser lines is their variability. Such variability has also been seen in non-metastable ($J$\,>\,$K$)
ammonia maser lines (e.g. \citealt{2013A&A...549A..90H}). So we searched for variability of the maser line within the few days of observations of
our own data (see Sect.\,\ref{sect-3-1} and Appendix\,\ref{Appendix B}). From Fig.\,\ref{FgB.2}, we can clearly see that the mean
beam brightness temperatures of this NH$_{3}$\,(3,3) line indicate no significant variation during our observations.
Furthermore, we make the comparison between our data and previous studies (e.g. w.\,r.\,t. the NH$_{3}$\,(3,3) maser reported by \citealt{1982A&A...110L..20W}, also obtained with the Effelsberg 100m telescope). We use the `GAUSS' fit to obtain an NH$_{3}$\,(3,3) line main beam brightness temperature of $T_{\rm MB}$\,=\,3.7(0.1)\,K and radial velocity of $V_{\rm LSR}$\,=\,36.2(0.1)\,km\,s$^{-1}$ (the errors shown in parentheses are fitting uncertainties) at the ($0\arcsec, 0\arcsec$) offset position. \cite{1982A&A...110L..20W} obtain $T_{\rm MB}$\,=\,3.7\,K and $V_{\rm LSR}$\,=\,35.4\,km\,s$^{-1}$. While we cannot explain the difference in velocity, we can nevertheless conclude that in view of peak intensity and line shape no significant variations have occurred during the past $\sim$36\,yr. This is consistent with the fact that the $^{15}\rm NH_{3}$\,(3,3) maser of \cite{1986A&A...160L..13M} and \cite{1991A&A...247..516S} did not show variability over a timescale of several years. The lack of variability for this type of maser suggests that the region with inverted populations giving rise to (3,3) maser emission may be larger than those of most other maser transitions, of ammonia as well as other molecular species.

\section{Summary}
\label{sect:summary}
Using the 100-m telescope at Effelsberg, we have searched for NH$_3$ absorption and emission lines in the prominent massive star forming regions of the W33 complex. Our ammonia observations of the W33 region reveal the following main results:

\begin{enumerate}
\item
We have detected the NH$_3$\,(1,1), (2,2), (3,3), (4,4), (5,5) and (6,6) metastable lines in W33\,Main,
W33\,A, and W33\,B. The non-metastable NH$_3$\,(2,1) and (3,2) transitions were also measured towards these three
molecular hotspots in the W33 region. There is an already previously reported maser line observed in the NH$_{3}$\,(3,3) transition towards W33\,Main. The NH$_{3}$\,(1,1), (2,2), (4,4), (5,5) and (6,6) inversion lines are detected in absorption against
the radio continuum in W33\,Main, while all other mapped regions provide ammonia emission lines. We have detected the
NH$_3$\,(1,1), (2,2), (3,3) and (4,4) metastable inversion lines, all in emission, in W33\,Main1 and W33\,A1.
Towards W33\,B1 we detected only the NH$_3$\,(1,1), (2,2) and (3,3) emission lines. The non-metastable NH$_{3}$\,(2,1) and (3,2)
transitions were not detected in these regions.

\item
For the total-NH$_3$ column density, we find 6.0\,($\pm$2.1)\,$\times$\,10$^{14}$, 3.5\,($\pm$0.1)\,$\times$\,10$^{15}$, 3.4\,($\pm$0.2)\,$\times$\,10$^{15}$, 3.1\,($\pm$0.2)\,$\times$\,10$^{15}$, 2.8\,($\pm$0.2)\,$\times$\,10$^{15}$ and 2.0\,($\pm$0.2)\,$\times$\,10$^{15}$\,cm$^{-2}$ at the peak positions of W33\,Main, W33\,A, W33\,B, W33\,Main1, W33\,A1 and W33\,B1, respectively.

\item
We determine kinetic temperatures only using NH$_3$\,(1,1) and (2,2), and from this we provide estimates of gas volume densities for the six main sources in the W33 region. Using our new $T_{\rm kin}$ values shows that our volume densities are similar to those estimated by \cite{2014A&A...572A..63I}, suggesting that ammonia beam filling factors are close to unity.

\item
W33\,Main has total-NH$_3$ fractional abundances ($\chi$\,(total-NH$_3$)\,=\,(total-$N$(NH$_3$))/$N$(H$_2$)) of 1.3\,($\pm$0.1)\,$\times$\,10$^{-9}$ at the peak position. High values of 1.4\,($\pm$0.3)\,$\times$\,10$^{-8}$, 1.6\,($\pm$0.3)\,$\times$\,10$^{-8}$, 3.4\,($\pm$0.5)\,$\times$\,10$^{-8}$, 1.6\,($\pm$0.5)\,$\times$\,10$^{-8}$ and 4.0\,($\pm$1.2)\,$\times$\,10$^{-8}$ characterize the central positions of W33\,A, W33\,B, W33\,Main1, W33\,A1, and W33\,B1, respectively. From this we confirm the difference evolutionary stages proposed by \citet{2014A&A...572A..63I} and find that there is no hot core in the region approaching the extreme conditions encountered in W51-IRS2 or Sgr\,B2.

\item
Ortho-to-para-NH$_3$ abundance ratios are 0.5\,($\pm$0.1), 1.3\,($\pm$0.1), 1.3\,($\pm$0.1), 1.8\,($\pm$0.1), 1.9\,($\pm$0.1) and 1.9\,($\pm$0.1) for W33\,Main, W33\,A, W33\,B, W33\,Main1, W33\,A1, and W33\,B1, respectively. The low value for W33\,Main may be affected by unknwon systematic errors. The other values indicate that ammonia has either been formed in the gas-phase or has been formed on dust grains in a medium with $\sim$\,20\,K or more to be then released into the interstellar medium.

\item
From our ammonia $T_{\rm kin}$ determinations we suggest that large temperature gradients may be present in the dense molecular gas of W33\,Main, W33\,A, and W33\,B. Kinetic temperatures towards the our six W33 targets are compatible with the different stages of evolution outlined by \citet{2014A&A...572A..63I}.

\item
The maser emission in the NH$_{3}$\,(3,3) line towards W33 Main shows no significant variability during the course of our observations. More importantly, brightness temperature and line shape of this line also indicate no significant change during the last $\sim$36\,yr.

\end{enumerate}

\begin{acknowledgements}
We like to thank the anonymous referee for the useful suggestions that improved this study.
This work is based on observations made with the 100\,m Effelsberg telescope, which is
operated by the Max-Planck-Institut f\"ur Radioastronomie. We thank the staff of the
Effelsberg 100-m radio telescope for their assistance during the observations. This work was funded by
the National Natural Science foundation of China under grant 11433008, 11903070, 11973076, and 12173075,
the Heaven Lake Hundred-Talent Program of Xinjiang Uygur Autonomous Region of China, and the
CAS "Light of West China" Program under Grant 2018-XBQNXZ-B-024 and 2020-XBQNXZ-017. C.\,H. acknowledges
support by a Chinese Academy of Sciences President's International Fellowship Initiative for visiting scientists
(2021VMA0009 and 2022VMA0018). This research has used NASA's Astrophysical Data System (ADS).
\end{acknowledgements}

%-------------------------- Table 3
\begin{table*}[h]
\centering
\caption{Line parameters for the NH$_{3}$ absorption component.}
\begin{tabular}{ccccccccccl}
\hline \hline
Source & Offset & NH$_{3}$\,$(J,K)$ & $T_{\rm mb}$ &rms & $V_{\rm LSR}$ &  $\Delta v$ & $T_{\rm L}/T_{\rm C}$ & $\tau$ &  $N{(\rm NH_3)}$ & \\
 & (\arcsec, \arcsec) &  & K & K & km\,s$^{-1}$  & km\,s$^{-1}$ &  & &  cm$^{-2}$&  \\
\hline
W33\,Main  & 0, 0 & (1,1)&-1.89&0.15&33.7(0.1)&2.2(0.1)&     &1.1(0.5) &1.9($\pm$0.8)\,$\times$\,10$^{14}$&  \\
           &      &      &-1.25&0.15&39.0(0.1)&2.2(0.1)&     &0.6(0.4) &7.9($\pm$5.4)\,$\times$\,10$^{13}$&  \\
           &      & (2,2)&-1.54&0.05&33.8(0.1)&3.4(0.1)&0.06 &0.1(0.01) &5.5($\pm$0.5)\,$\times$\,10$^{13}$&   \\
           &      &      &-1.53&0.05&38.7(0.1)&2.9(0.1)&0.06 &0.1(0.01) &4.6($\pm$0.4)\,$\times$\,10$^{13}$&  \\
           &      & (4,4)&-0.30&0.03&33.7(0.3)&7.8(0.5)&0.01 &$<$0.1(0.01) &2.0($\pm$0.3)\,$\times$\,10$^{13}$&   \\
           &      &      &-0.39&0.03&38.7(0.1)&3.1(0.2)&0.01 &$<$0.1(0.01) &1.0($\pm$0.1)\,$\times$\,10$^{13}$&  \\
           &      & (5,5)&-0.16&0.04&33.4(0.2)&5.4(0.5)&0.01 &$<$0.1(0.01) &6.9($\pm$1.1)\,$\times$\,10$^{12}$&  \\
           &      &      &-0.16&0.04&38.4(0.1)&2.6(0.3)&0.01 &$<$0.1(0.01) &3.4($\pm$0.6)\,$\times$\,10$^{12}$&  \\
           &      & (6,6)&-0.20&0.04&33.4(0.3)&6.9(0.6)&0.01 &$<$0.1(0.01) &1.1($\pm$0.2)\,$\times$\,10$^{13}$&  \\
           &      &      &-0.07&0.04&39.1(0.4)&2.8(0.6)&0.01 &$<$0.1(0.01) &1.5($\pm$0.6)\,$\times$\,10$^{12}$&  \\
           &...   & (0,0)&     &    &         &        &     &          &1.4($\pm$0.3)\,$\times$\,10$^{14}$ &  \\
           &...   &      &     &    &         &        &     &          &4.4($\pm$2.6)\,$\times$\,10$^{13}$ &  \\
\hline
\end{tabular}
\tablefoot{The reference position is R.A.\,: 18:14:13.50, DEC.\,: -17:55:47.0 (J2000) for W33\,Main.
The errors in parentheses for $V_{\rm LSR}$, $\Delta v$ (full width to half maximum) and the opacity $\tau$ of the NH$_{3}$(1,1) line are fitting
uncertainties, while other errors in parentheses are calculated uncertainties (see Appendix\,C of \citealt{2020A&A...643A.178T}).
For the calculation method of the (0,0) column density, see Sect.\,\ref{column-volume-density}.}
\label{Table-absorption}
\end{table*}

%-------------------------- Table 4
\begin{table*}[t]
\centering
\caption{Line parameters for the NH$_{3}$ emitting component.}
\begin{tabular}{lccccccccc}
\hline \hline
Source & Offset & NH$_{3}$\,$(J,K)$ & $T_{\rm mb}$ & rms & $V_{\rm LSR}$ &  $\Delta v$  & $\tau$ & $N{(\rm NH_3)}$  \\
 & (\arcsec, \arcsec) &  & K & K & km\,s$^{-1}$ & km\,s$^{-1}$ &  & cm$^{-2}$  \\
\hline
W33\,Main  & 0, 0  & (3,3)$^{\dag}$&3.74&0.11&36.2(0.1)&5.8(0.1)   &$<$0.3(0.4)  &2.0($\pm$0.1)\,$\times$\,10$^{14}$   \\
           &        & (2,1)&0.11&0.04&33.9(0.4)&3.4(0.8)  &$<$0.4(0.1)  &1.6($\pm$0.8)\,$\times$\,10$^{13}$   \\
           &        & (3,2)&0.11&0.04&33.5(0.3)&4.4(0.8)   &$<$0.2(0.1)  &1.0($\pm$0.4)\,$\times$\,10$^{13}$   \\
W33\,A     & 0, 0 & (1,1)&5.18&0.28&37.5(0.1)& 3.3(0.1)   &3.6(0.1)    &1.3($\pm$0.1)\,$\times$\,10$^{15}$     \\
           &      & (2,2)&3.62&0.15&37.6(0.1)& 4.0(0.1)  &$<$0.4(0.2)  &1.5($\pm$0.1)\,$\times$\,10$^{14}$  \\
           &      & (3,3)&2.48&0.13&37.6(0.0)&4.6(0.1)  &$<$0.4(0.1)  &1.1($\pm$0.1)\,$\times$\,10$^{14}$   \\
           &      & (4,4)&0.50&0.06&37.7(0.1)&5.9(0.4)   &$<$0.3(0.1)  &2.5($\pm$0.3)\,$\times$\,10$^{13}$   \\
           &      & (5,5)&0.24&0.06&38.5(0.2)& 3.8(0.5)   &$<$0.3(0.1)  &7.3($\pm$2.2)\,$\times$\,10$^{12}$     \\
           &      & (6,6)&0.21&0.06&37.9(0.3)&3.9(0.6)  &$<$0.4(0.1)  &6.3($\pm$1.8)\,$\times$\,10$^{12}$  \\
           &      & (2,1)&0.48&0.07&36.1(0.2)&3.7(0.5)  &$<$0.2(0.1)  &7.7($\pm$1.4)\,$\times$\,10$^{13}$   \\
           &      & (3,2)&0.18&0.07&36.4(0.4)&3.7(0.6)   &$<$0.4(0.1)  &1.4($\pm$0.5)\,$\times$\,10$^{13}$   \\
           & ...  & (0,0)&    &    &         &           &             &1.8($\pm$0.1)\,$\times$\,10$^{15}$    \\
W33\,B     &0, 0  & (1,1)&3.08&0.13&55.9(0.1)&2.5(0.1)   &6.8(0.2)   &1.4($\pm$0.1)\,$\times$\,10$^{15}$     \\
           &      & (2,2)&2.27&0.11&55.8(0.1)&3.1(0.1)  &$<$0.2(0.1)  &7.4($\pm$0.4)\,$\times$\,10$^{13}$  \\
           &      & (3,3)&1.43&0.12&55.7(0.0)&3.8(0.1)  &$<$0.3(0.1)  &5.1($\pm$0.5)\,$\times$\,10$^{13}$   \\
           &      & (4,4)&0.24&0.05&55.4(0.2)&4.9(0.6)   &$<$0.4(0.1)  &1.0($\pm$0.3)\,$\times$\,10$^{13}$   \\
           &      & (5,5)&0.27&0.05&56.6(0.1)&2.7(0.3)   &$<$0.3(0.1)  &5.8($\pm$0.9)\,$\times$\,10$^{12}$    \\
           &      & (6,6)&0.14&0.05&55.7(0.4)&6.3(0.6)  &$<$0.4(0.1)  &6.7($\pm$2.1)\,$\times$\,10$^{12}$  \\
           &      & (2,1)&0.35&0.05&53.8(0.1)&1.9(0.4)   &$<$0.2(0.1)  &2.9($\pm$0.7)\,$\times$\,10$^{13}$   \\
           &      & (3,2)&0.17&0.05&55.1(0.3)&2.8(0.7)   &$<$0.2(0.1)  &1.0($\pm$0.4)\,$\times$\,10$^{13}$    \\
           & ...  & (0,0)&    &    &         &           &             &2.8($\pm$0.1)\,$\times$\,10$^{15}$    \\
W33\,Main1 & 0,0  & (1,1)&4.11&0.08&36.6(0.1)&2.1(0.1)&5.2(0.2)&1.0($\pm$0.1)\,$\times$\,10$^{15}$ \\
           &      & (2,2)&2.21&0.06&36.7(0.1)& 2.6(0.1)&$<$0.3(0.1)&5.9($\pm$0.2)\,$\times$\,10$^{13}$\\
           &      & (3,3)&0.97&0.06&36.7(0.1)& 2.7(0.1)&$<$0.4(0.1)&2.4($\pm$0.2)\,$\times$\,10$^{13}$ \\
           &      & (4,4)&0.16&0.04&36.5(0.2)& 2.9(0.5)& $<$0.3(0.1)&4.0($\pm$1.5)\,$\times$\,10$^{12}$      \\
           & ...  & (0,0)&    &    &         &           &         & 1.9($\pm$0.1)\,$\times$\,10$^{15}$ \\
W33\,A1    & 0,0  & (1,1)&2.92&0.12&37.2(0.1)&2.4(0.1)&4.8(0.2)&9.1($\pm$0.7)\,$\times$\,10$^{14}$   \\
           &      & (2,2)&1.63&0.10&37.2(0.1)& 2.9(0.1)&$<$0.4(0.1)&5.0($\pm$0.5)\,$\times$\,10$^{13}$   \\
           &      & (3,3)&0.65&0.09&36.9(0.1)&4.0(0.4)&$<$0.3(0.1)&2.4($\pm$0.4)\,$\times$\,10$^{13}$  \\
           &      & (4,4)&0.14&0.04&38.2(0.4)& 4.3(0.8)& $<$0.2(0.1)&5.1($\pm$1.6)\,$\times$\,10$^{12}$      \\
           & ...  & (0,0)&    &    &         &        &           &1.8($\pm$0.1)\,$\times$\,10$^{15}$  \\
W33\,B1    & 0,0  & (1,1)&1.28&0.13&34.2(0.1)&3.6(0.1)&3.2(0.4)&6.6($\pm$1.2)\,$\times$\,10$^{14}$   \\
           &      & (2,2)&0.66&0.06&34.2(0.1)&4.6(0.3)&$<$0.3(0.1)&3.2($\pm$0.5)\,$\times$\,10$^{13}$    \\
           &      & (3,3)&0.30&0.06&34.8(0.3)&6.9(1.0)&$<$0.2(0.1)&1.9($\pm$0.5)\,$\times$\,10$^{13}$  \\
           & ...  & (0,0)&    &    &         &        &           & 1.3($\pm$0.1)\,$\times$\,10$^{15}$ \\
\hline
\end{tabular}
\tablefoot{For the reference position of each W33 source, see in Table\,\ref{table:1}. $^{\dag}$(3,3) is the NH$_{3}$\,(3,3) maser line in W33\,Main. The errors are derived in the same way as in Table\,\ref{Table-absorption}. For the calculation method of the (0,0) column densities, see Sect.\,\ref{column-volume-density}.}
\label{Table-emission}
\end{table*}

%-------------------------- Table 5
\begin{table*}[h]
\centering
\begin{footnotesize}
\setlength{\tabcolsep}{2.7pt}
\caption{NH$_3$ rotation temperatures obtained from our main six W33 sources (see Sect.\,\ref{rotation-temperature} and Fig.\,\ref{rotationdiagram}).}
\begin{tabular}{lccclcccc}
\hline \hline
 & W33\,Main & W33\,Main  & W33\,A & W33\,B &  W33\,Main1 & W33\,A1 & W33\,B1 &  \\
 &  absorption lines & absorption lines     &   &  &  &       \\
 &  $V_{\rm LSR}$\,$\sim$\,33\,km\,s$^{-1}$ & $V_{\rm LSR}$\,$\sim$\,38\,km\,s$^{-1}$ &      &   &  &  &       \\
\hline
$T_{\rm rot,\rm 12}$/K                         &23\,$\pm$\,5&38\,$\pm$\,6& 15\,$\pm$\,1& 11\,$\pm$\,1  &12\,$\pm$\,1&11\,$\pm$\,1&11\,$\pm$\,1&  \\
$T_{\rm rot,\rm 24}$/K                         &84\,$\pm$\,3&64\,$\pm$\,1& 56\,$\pm$\,2& 52\,$\pm$\,5  &41\,$\pm$\,4&47\,$\pm$\,4&            &  \\
$T_{\rm rot,\rm 36}$/K                         &            &            & 82\,$\pm$\,5& 107\,$\pm$\,8 &            &            &            &  \\
$T_{\rm rot,\rm 45}$/K                         &75\,$\pm$\,2&71\,$\pm$\,3& 65\,$\pm$\,7& 126\,$\pm$\,19&            &            &            &  \\
$T_{\rm rot,\rm \, non-metastable\,lines}$/K   &            &            & 34\,$\pm$\,3& 50\,$\pm$\,6  &           &             &            &  \\
$T_{\rm rot,\rm \, all\,para-NH_3\,species}$/K &64\,$\pm$\,11&62\,$\pm$\,4& 45\,$\pm$\,9&44\,$\pm$\,12 & 29\,$\pm$\,9&31\,$\pm$\,11&            &   \\
\hline
\end{tabular}
\tablefoot{For the chosen positions, see Table\,\ref{table:1}.}
\label{Table-Trot}
\end{footnotesize}
\end{table*}

%-------------------------- Table 6
\begin{table*}[h]
\begin{footnotesize}
\setlength{\tabcolsep}{2.7pt}
\caption{Total (para+ortho) ammonia column densities, total-NH$_3$\,(para+ortho) fractional abundances, and ortho-to-para abundance ratios at the peak positions of the W33 region.}
\begin{tabular*}{\textwidth}{@{\hspace{\tabcolsep}
\extracolsep{\fill}}lccccc}
\hline\hline
Source &            & $N{(\rm NH_3)}$&              &  Fractional     & Ortho/para \\
       & Ortho-$N{(\rm NH_3)}$ & Para-$N{(\rm NH_3)}$ & Total-$N{(\rm NH_3)}$       &  $\chi$(total-NH$_3$) & ratio \\
\cline{2-4}
      &            & cm$^{-2}$    &              &                      &       \\
%\vspace{2pt} \\
\hline
W33\,Main absorption lines $V_{\rm LSR}$\,$\sim$\,33\,km\,s$^{-1}$ &1.5($\pm$0.3)\,$\times$\,10$^{14}$  &
2.7($\pm$0.9)\,$\times$\,10$^{14}$  &4.2($\pm$1.2)\,$\times$\,10$^{14}$
    &0.9($\pm$0.1)\,$\times$\,10$^{-9}$  & 0.5($\pm$0.1) \\

W33\,Main absorption lines $V_{\rm LSR}$\,$\sim$\,38\,km\,s$^{-1}$ &0.4($\pm$0.3)\,$\times$\,10$^{14}$  &
1.4($\pm$0.7)\,$\times$\,10$^{14}$  &1.8($\pm$1.0)\,$\times$\,10$^{14}$
    &0.4($\pm$0.1)\,$\times$\,10$^{-9}$  & 0.3($\pm$0.1) \\

W33\,A &1.9($\pm$0.1)\,$\times$\,10$^{15}$  &
1.5($\pm$0.1)\,$\times$\,10$^{15}$  &3.5($\pm$0.1)\,$\times$\,10$^{15}$
     &1.4($\pm$0.3)\,$\times$\,10$^{-8}$  & 1.3($\pm$0.1) \\
W33\,B &1.9($\pm$0.1)\,$\times$\,10$^{15}$   &
1.5($\pm$0.1)\,$\times$\,10$^{15}$    &3.4($\pm$0.2)\,$\times$\,10$^{15}$
   &1.6($\pm$0.3)\,$\times$\,10$^{-8}$  & 1.3($\pm$0.1) \\
W33\,Main1  &2.0($\pm$0.1)\,$\times$\,10$^{15}$   &
1.1($\pm$0.1)\,$\times$\,10$^{15}$    &3.1($\pm$0.2)\,$\times$\,10$^{15}$
   &3.4($\pm$0.5)\,$\times$\,10$^{-8}$  & 1.8($\pm$0.1) \\
W33\,A1    &1.8($\pm$0.1)\,$\times$\,10$^{15}$   &
9.7($\pm$0.8)\,$\times$\,10$^{14}$    &2.8($\pm$0.2)\,$\times$\,10$^{15}$
   &1.6($\pm$0.5)\,$\times$\,10$^{-8}$  & 1.9($\pm$0.1) \\
W33\,B1    &1.3($\pm$0.1)\,$\times$\,10$^{15}$   &
6.9($\pm$1.2)\,$\times$\,10$^{14}$    &2.0$(\pm$0.2)\,$\times$\,10$^{15}$
   &4.0($\pm$1.2)\,$\times$\,10$^{-8}$  & 1.9($\pm$0.1) \\
\hline
\end{tabular*}
\tablefoot{The column densities of H$_{2}$ are taken from \cite{2014A&A...572A..63I}. The errors in parentheses are calculated uncertainties (see Appendix\,C of \citealt{2020A&A...643A.178T}).}
\label{Table-abundance}
\end{footnotesize}
\end{table*}

%----------------------------------Appendix A
%\Online
\begin{appendix}
\onecolumn
\section{Ammonia spectra toward W33 and derived physical parameters}
\label{Appendix A}
%-------------------------- Table A.1
\begin{table*}[h]
\centering
\caption{Observed parameters of the NH$_3$\,(1,1) lines detected in the W33. }
\begin{tabular}{lccccccccc}
\hline \hline
Source & Offset & $\int$$T_{\rm MB}$d$v$ &$V_{\rm LSR}$ & $\Delta v$ & $T_{\rm MB}$ & rms & $\tau$ \\
 & (\arcsec, \arcsec) & K\,km\,s$^{-1}$ & km\,s$^{-1}$ & km\,s$^{-1}$ & K & K &\\
\hline
W33\,Main  & 0, 0 & -2.13(0.02)& 33.7(0.5) &2.2(0.1)& -1.89&0.15&0.5(0.2) \\
           &      & -1.43(0.02)& 39.0(0.5) &2.2(0.1)& -1.25&0.15&0.2(0.1) \\
           & 20, 0 &-7.15(0.19)&33.6(0.5)&2.9(0.5)& -2.34&0.12&0.4(0.2) \\
           &       &-8.15(0.21)&39.5(0.5)&4.7(0.5)& -1.63&0.09&0.1(0.1) \\
           & -20, 0 &2.47(0.15) &36.4(0.1) &3.9(0.1)& 2.18&0.15 &0.3(0.1) \\
           & 0, 20 & 3.63(0.15)&35.9(0.1)&2.1(0.1)&2.18&0.14 &1.2(0.2) \\
           & 0, -20 &2.33(0.14)&36.6(0.1)&1.9(0.1) &1.30&0.15&1.3(0.3) \\
           & 20, 20 &3.05(0.15)&35.7(0.1)&2.2(0.1)&2.18&0.14&0.8(0.2) \\
           & -20, 20 & 2.76(0.14) &36.2(0.1)&3.4(0.1)&2.47&0.15&0.2(0.1) \\
           & -20, -20 & 2.90(0.15)&36.1(0.1)&4.0(0.1)&2.33&0.14&0.5(0.1) \\
           & 20, -20 &1.59(0.29)&36.8(0.1)&1.6(0.1)&0.87&0.14&1.4(0.6) \\
           &$\vdots$&$\vdots$ &$\vdots$&$\vdots$&$\vdots$&$\vdots$ &$\vdots$        \\
W33\,A     & 0, 0 &11.00(0.13)&37.5(0.1)&3.3(0.1)&5.18&0.28&1.8(0.1) \\
           & 20, 0 &7.70(0.44)&37.0(0.1)&3.5(0.1)&4.07&0.14&1.5(0.2) \\
           & -20, 0 &12.21(0.58)&37.8(0.1)&2.4(0.1)&4.94&0.15&2.4(0.2) \\
           & 0, 20 &9.16(0.44)&37.3(0.1)&3.1(0.1)&4.51&0.15&1.7(0.2) \\
           & 0, -20 &6.39(0.29) &37.6(0.1)&2.8(0.1)&3.64&0.15&1.3(0.2) \\
           & 20, 20 &7.56(0.44)&36.7(0.1)&3.4(0.1)&3.92&0.14&1.6(0.2)  \\
           & -20, 20 &12.65(0.58)&37.6(0.1)&2.4(0.1)&5.23&0.15&2.4(0.2) \\
           & -20, -20 &5.23(0.29)&37.6(0.1)&2.5(0.1)&2.62&0.14&1.7(0.2)  \\
           & 20, -20&4.07(0.15)&36.9(0.1)&4.3(0.1)&2.18&0.15&1.0(0.2)  \\
W33\,B     & 0, 0 &10.10(0.27)&55.9(0.1)&2.5(0.1)& 3.08&0.13&3.4(0.1) \\
           &20, 0 &5.23(0.44)&56.4(0.1)&2.1(0.1)&1.59&0.01&3.1(0.4) \\
           &-20, 0 &10.03(0.73)&55.5(0.1)&2.3(0.1)&2.76&0.14&3.7(0.4) \\
           &0, 20 &6.11(0.44)&56.1(0.1)&2.3(0.1)&2.18&0.01&2.7(0.4) \\
           &0, -20&6.25(0.44)&56.2(0.1)&2.5(0.1)& 2.47&0.15&2.4(0.3) \\
           &20, 20&6.25(0.44)&56.7(0.1)&1.8(0.1)& 2.18&0.02&3.0(0.3) \\
           &-20, 20&5.09(0.43)&55.6(0.1)&2.2(0.1)& 2.04&0.14&2.4(0.4) \\
           & -20, -20 &10.03(0.73)&55.4(0.1)&2.5(0.1)&2.76&0.15&3.8(0.4) \\
           &20, -20&3.05(0.29)&56.7(0.1)&2.2(0.1)& 1.31&0.15&2.0(0.4) \\
           &40, 0&2.33(0.29)&56.6(0.1)&2.1(0.2)& 1.02&0.02&1.8(0.4) \\
           & 0, 40&4.07(0.28) &56.0(0.1)&2.1(0.1)&1.59&0.02&2.3(0.4) \\
           &-40, 0&6.07(0.65)&55.1(0.1)&2.3(0.1)&2.00&0.14&3.1(0.5)  \\
           &0, -40&2.19(0.29)& 56.7(0.1)&2.7(0.2)&1.50&0.31&0.8(0.4) \\
           &60, 0 &2.61(0.26)&56.4(0.2)&1.5(0.4)&0.40&0.07&8.2(6.8) \\
           &-60, 0&2.13(0.29)&54.7(0.1)&3.1(0.2)&0.96&0.11&1.9(0.6)        \\
           &0, 60 &0.82(0.21)&55.9(0.2)&3.7(0.4)&0.40&0.19&0.4(0.2)         \\
           &0, -60&1.11(0.09)&56.8(0.1)&2.8(0.3)&1.13&0.09&0.1(0.1)         \\
W33\,Main1 & 0, 0 &11.0(0.14)&36.6(0.1)&2.1(0.1)&4.11&0.08&2.6(0.1) \\
W33\,A1    & 0, 0 &7.26(0.19)&37.2(0.1)&2.4(0.1)&2.92&0.12&2.4(0.1)       \\
W33\,B1    & 0, 0 &2.55(0.15)&34.2(0.1)&3.6(0.1)&1.28&0.13&1.6(0.2)        \\
           &20, 0 &1.10(0.29)&34.3(0.1)&2.4(0.3)&0.98&0.07&0.4(0.2)         \\
           &-20, 0&1.13(0.19)&34.7(0.2)&4.2(0.3)&0.86&0.13&0.5(0.4)         \\
           & 0, 20&2.09(0.33)&34.1(0.1)&2.9(0.2)&0.69&0.16&2.9(0.8)         \\
           &0, -20&1.45(0.32)&34.8(0.1)&2.8(0.3)&0.83&0.11&1.3(0.7)         \\
           &20, 20&0.58(0.08)&34.2(0.2)&3.7(0.4)&0.56&0.06&0.1(0.1)         \\
          &20, -20&0.97(0.23)&34.5(0.2)&3.5(0.4)&0.64&0.14&0.9(0.6)         \\
         &-20, -20&1.19(0.76)&35.5(0.4)&2.6(0.5)&0.35&0.10&4.3(0.4)         \\
         &-20, 20 &0.59(0.17)&36.2(0.2)&1.2(0.4)&0.39&0.14&0.1(0.1)         \\
\hline
\end{tabular}
\tablefoot{The reference position for each of the W33 sources is given in Table\,\ref{table:1}. The errors shown in parentheses are fitting uncertainties from the NH$_3$\,(1,1) fit in CLASS (see Sect.\,\ref{sect-2-2}).}
\label{table:A.1}
\end{table*}

%-------------------------- Table A.2
\begin{table*}[h]
\centering
\caption{Observed parameters of the NH$_3$\,(2,2) lines detected in W33.}
\begin{tabular}{l c c c c c c c c c }
\hline \hline
 Source & Offset & $\int$$T_{\rm MB}$d$v$ & $V_{\rm LSR}$ & $\Delta v$ & $T_{\rm MB}$ & rms &  \\
  & (\arcsec, \arcsec) & K\,km\,s$^{-1}$ & km\,s$^{-1}$ & km\,s$^{-1}$ & K & K &    \\
\hline
W33\,Main  & 0, 0 &-5.55(0.14)&33.8(0.1)&3.4(0.1)&-1.54& 0.05&       \\
           &      &-4.69(0.12)&38.7(0.1)&2.9(0.1)&-1.53& 0.05&       \\
           & 20, 0 &-6.81(0.36)&33.6(0.1)&3.3(0.2) & -1.93& 0.05&      \\
           &       &-7.15(0.36)&38.5(0.1)&3.8(0.2) & -1.77& 0.04&      \\
           & -20, 0 &7.35(0.16)&36.4(0.1)&3.9(0.1)&1.74& 0.14&       \\
           & 0, 20 & 4.32(0.12) &35.9(0.1)&2.4(0.1)&1.59& 0.04&     \\
           & 0, -20 &1.78(0.12)&36.7(0.1)&1.9(0.2)& 0.73& 0.09&       \\
           & 20, 20&4.11(0.13) &35.6(0.1)&2.5(0.1)&1.59& 0.04&      \\
           & -20, 20 &8.49(0.16)&36.1(0.1)&3.9(0.1)&2.04& 0.04&       \\
           & -20, -20 &8.27(0.18)&36.0(0.1)&4.2(0.1) & 1.75& 0.06&       \\
           &20, -20&0.43(0.10)&36.8(0.2)&1.0(0.3)&0.44& 0.15&      \\
           &$\vdots$&$\vdots$ &$\vdots$&$\vdots$&$\vdots$&$\vdots$         \\
W33\,A  &0, 0 &15.41(0.35)&37.6(0.1)&4.0(0.1)&3.62& 0.15&       \\
        &20, 0 & 11.64(0.23)&37.2(0.1)&4.1(0.1)&2.76& 0.14&      \\
        &-20, 0 &10.23(0.22)&37.8(0.1)&3.3(0.1)&3.05& 0.14&      \\
        &0, 20&11.48(0.24) &37.5(0.1)&3.7(0.1)&2.91& 0.14&     \\
        &0, -20 &7.37(0.16)&37.6(0.1)&3.4(0.1) & 2.04& 0.13&       \\
        & 20, 20 & 10.05(0.19) &37.0(0.1)&4.1(0.1)  &2.33& 0.06&       \\
        & -20, 20&10.13(0.22)&37.7(0.1) &2.9(0.1)&3.05& 0.03&      \\
        &-20, -20 &3.99(0.13)&37.7(0.1)&2.8(0.1)&1.31& 0.04&      \\
        &20, -20 &6.62(0.21)&37.1(0.1)&4.2(0.1)&1.45& 0.15&      \\
W33\,B  &0, 0 &7.52(0.18)&55.8(0.1)&3.1(0.1)&2.27& 0.11&       \\
        &20, 0 &3.82(0.12)&56.3(0.1)&2.6(0.1)&1.31& 0.06&      \\
        &-20, 0&7.35(0.19) &55.4(0.1)&3.2(0.2)&2.04& 0.14&      \\
        &0, 20 &5.49(0.15)&55.9(0.1)&2.8(0.1) & 1.59& 0.15&      \\
        &0, -20& 5.51(0.18) &56.1(0.1)&2.9(0.1)  &1.59& 0.14&      \\
        &20, 20&4.04(0.12)&56.6(0.1) &2.4(0.1)& 1.45& 0.06&      \\
        &-20, 20 &4.13(0.14)&55.5(0.1)&2.6(0.1)&1.31& 0.03&      \\
        &-20, -20 &6.75(0.21)&55.4(0.1)&3.0(0.1)&1.89& 0.04&       \\
        &20, -20 & 2.82(0.14)&56.5(0.1)&3.3(0.3)&0.73& 0.06&      \\
        &40, 0 &2.44(0.14)&56.6(0.1)&2.6(0.3)&0.73& 0.06&     \\
        &0, 40&2.90(0.14) &55.7(0.1)&2.6(0.2)&1.02& 0.06&     \\
           &-40, 0&3.93(0.31)&55.1(0.1)&2.8(0.3)&1.31& 0.05&         \\
           &0, -40&2.52(0.29)&56.8(0.2)&2.9(0.4)&0.82& 0.06&         \\
           &60, 0 &1.13(0.28)&56.1(0.4)&2.9(0.8)&0.37& 0.07&         \\
           &-60, 0&2.45(0.37)&54.2(0.4)&5.2(0.8)&0.44& 0.10&         \\
           &0, 60 &1.56(0.34)&56.3(0.7)&4.5(1.1)&0.33& 0.12&         \\
           &0, -60&1.34(0.21)&57.2(0.2)&2.0(0.3)&0.63& 0.08&         \\
W33\,Main1 & 0, 0 &6.03(0.11)&36.7(0.1)& 2.6(0.1)&2.21& 0.06&         \\
W33\,A1    & 0, 0 &5.12(0.14)&37.2(0.1)& 2.9(0.1)&1.63& 0.10&         \\
W33\,B1    & 0, 0 &3.27(0.15)&34.2(0.1)&4.6(0.3)&0.66& 0.06&         \\
           &20, 0 &0.99(0.28)&34.9(0.4)&2.6(0.9)&0.36& 0.04&         \\
           &-20, 0&1.29(0.34)&34.1(0.5)&3.8(0.9)&0.32& 0.08&         \\
           & 0, 20&2.79(0.32)&34.3(0.2)&3.9(0.5)&0.68& 0.08&         \\
           &0, -20&0.69(0.19)&33.1(0.2)&1.2(0.3)&0.54& 0.05&         \\
           &20, 20&2.74(0.38)&35.5(0.4)&6.0(0.8)&0.43& 0.14&         \\
          &20, -20&1.27(0.31)&35.2(0.3)&2.5(0.7)&0.48& 0.03&         \\
         &-20, -20&0.47(0.24)&35.6(0.3)&1.4(0.9)&0.31& 0.03&         \\
         &-20, 20 &0.92(0.22)&34.9(0.3)&2.2(0.5)&0.39& 0.08&         \\
\hline
\end{tabular}
\tablefoot{The reference position for each of the W33 sources is given in Table\,\ref{table:1}. The errors shown in parentheses are fitting uncertainties from the `GAUSS' fit in CLASS (see Sect.\,\ref{sect-2-2}).}
\label{table:A.2}
\end{table*}

%-----------------Figure A.1
\begin{figure}[h]
\vspace*{0.2mm}
\begin{center}
\includegraphics[width=1\textwidth]{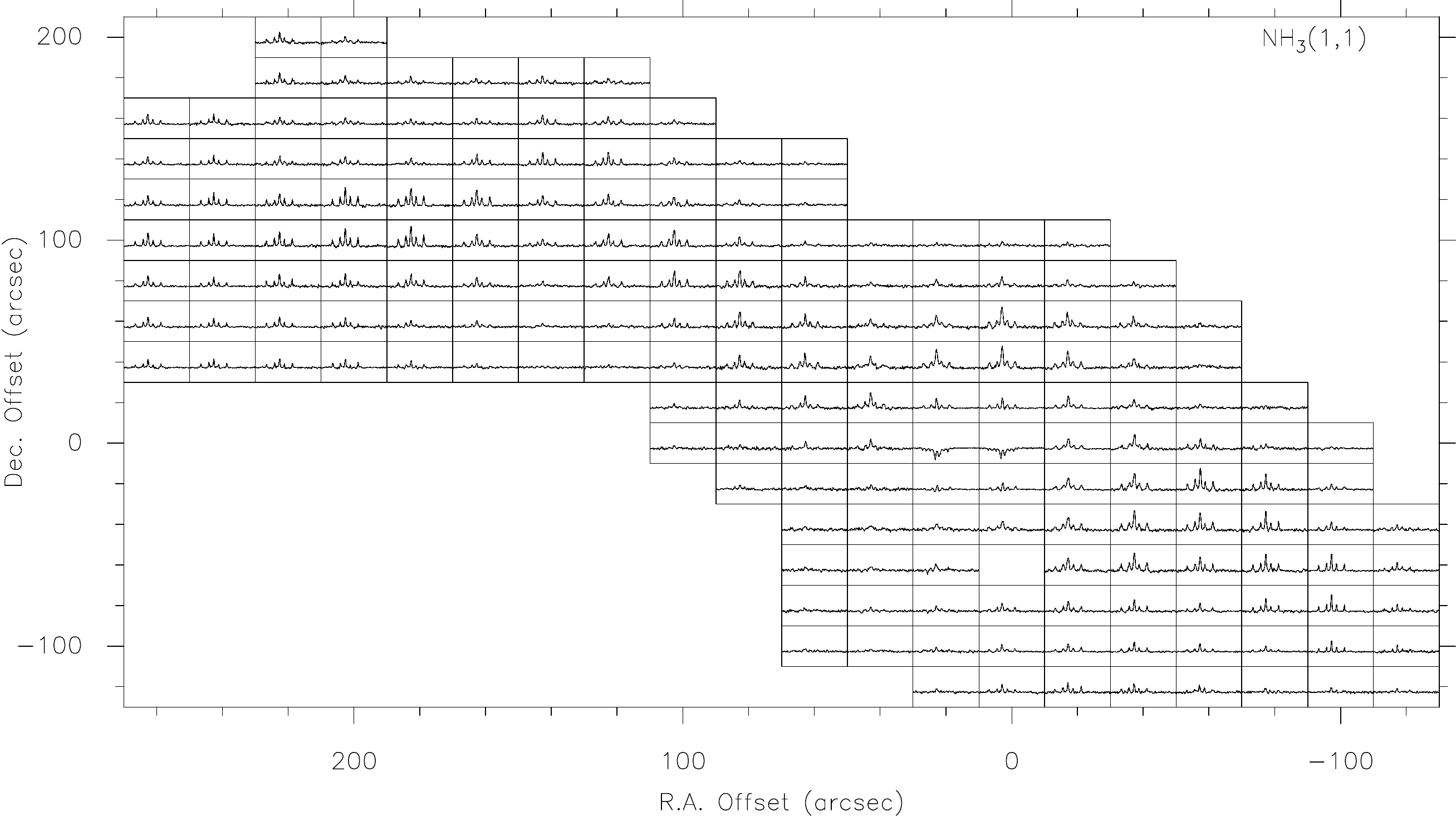}
\end{center}
\caption[]{The NH$_{3}$\,(1,1) line profiles of inversion transitions of W33\,Main and W33\,Main1. These are shown on $(\Delta \alpha, \Delta \delta)$ axes. The zero point is at the position of W33\,Main, R.A.\,: 18:14:13.50, DEC.\,: -17:55:47.0 (J2000). The X and Y axes indicate Right Ascension Offset (arcsec) and Declination Offset (arcsec), respectively. The main beam temperature scale for the NH$_{3}$ lines was obtained from continuum cross scans of 3C\,286 (see Sect. \ref{sect:Observation}). All radial velocities are on a $V_{\rm LSR}$ scale. At the assumed distance to the complex, $\sim$2.4\,kpc, 40$\arcsec$ is equivalent to 0.5\,pc. The individual spectra cover a velocity range of 32 to 38\,km\,s$^{-1}$ and the ordinate provides main beam brightness temperatures in the range --\,2.3 to 4.4\,K.}
\label{FgA.1}
\end{figure}

%-----------------Figure A.2
\begin{figure}[h]
\vspace*{0.2mm}
\begin{center}
\includegraphics[width=1\textwidth]{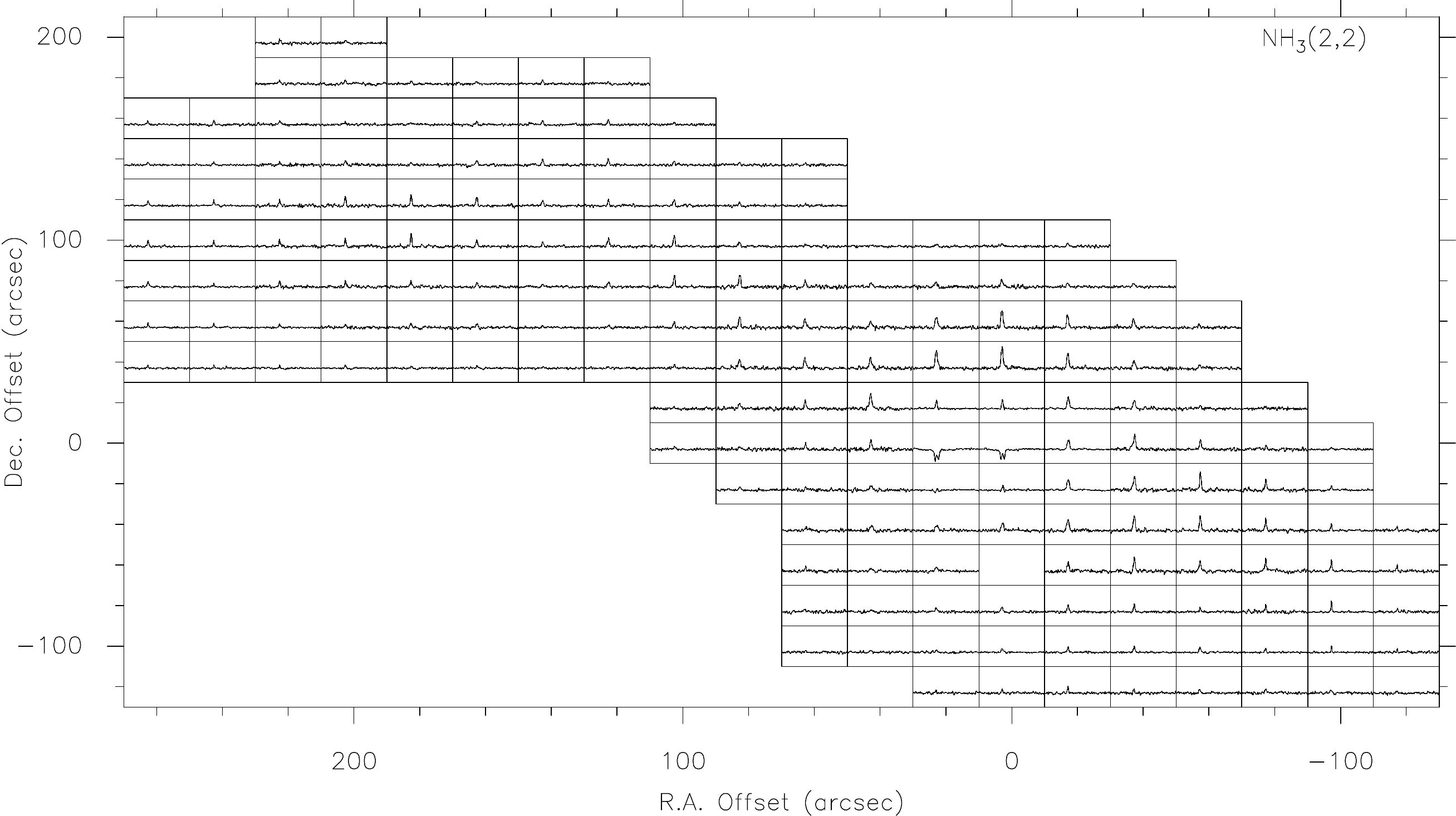}
\end{center}
\caption[]{The NH$_{3}$\,(2,2) line profiles of inversion transitions of W33\,Main and W33\,Main1. The individual spectra cover a velocity range of 32 to 38\,km\,s$^{-1}$ and the ordinate provides main beam brightness temperatures in the range --\,1.9 to 3.7\,K.}
\label{FgA.2}
\end{figure}

%-----------------Figure A.3
\begin{figure}[h]
\vspace*{0.2mm}
\begin{center}
\includegraphics[width=1\textwidth]{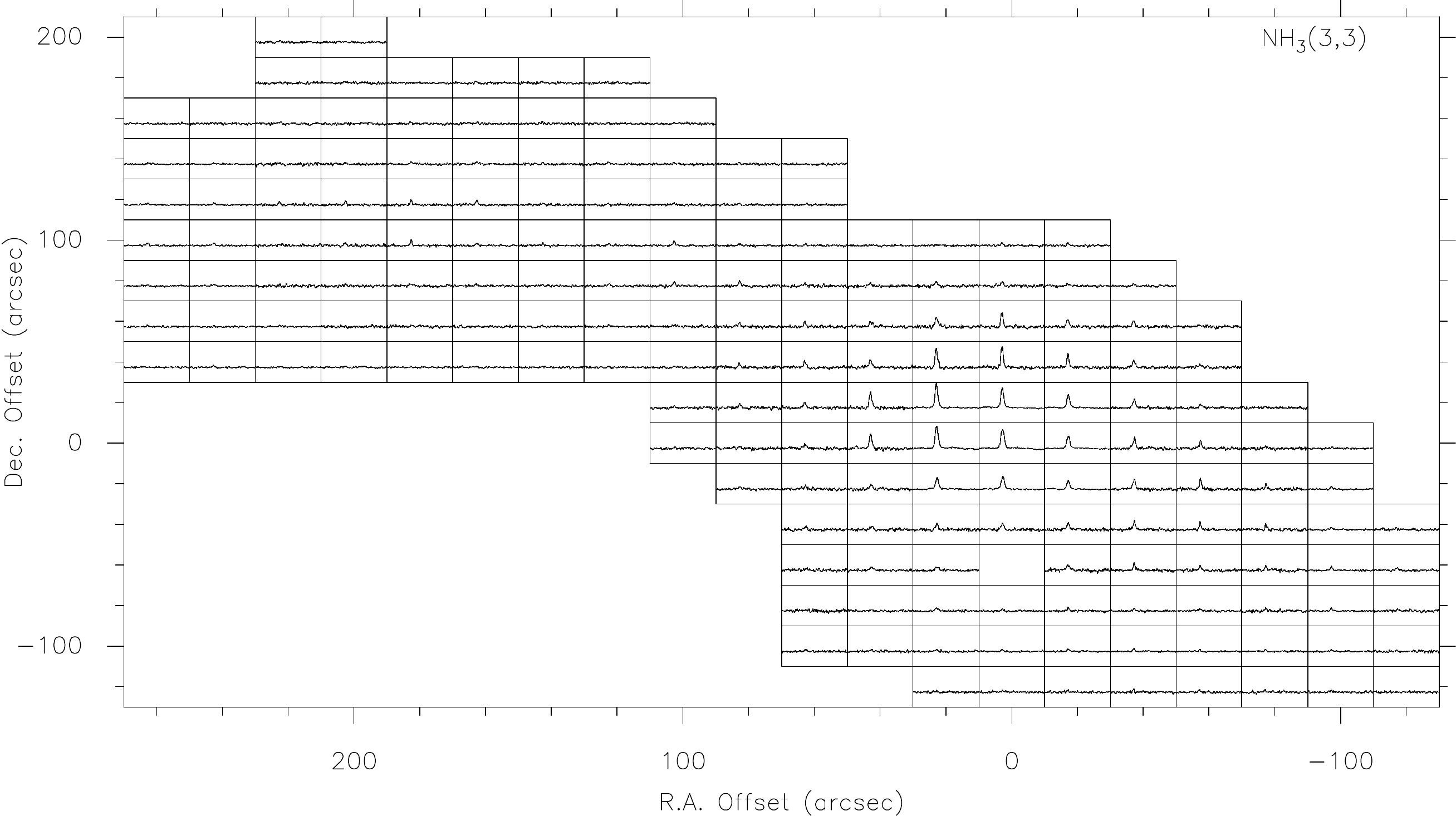}
\end{center}
\caption[]{The NH$_{3}$\,(3,3) line profiles of inversion transitions of W33\,Main and W33\,Main1. The individual spectra cover a velocity range of 32 to 38\,km\,s$^{-1}$ and the ordinate provides main beam brightness temperatures in the range 0.5 to 4.5\,K.}
\label{FgA.3}
\end{figure}

%-----------------Figure A.4
\begin{figure}[h]
\vspace*{0.2mm}
\begin{center}
\includegraphics[width=1\textwidth]{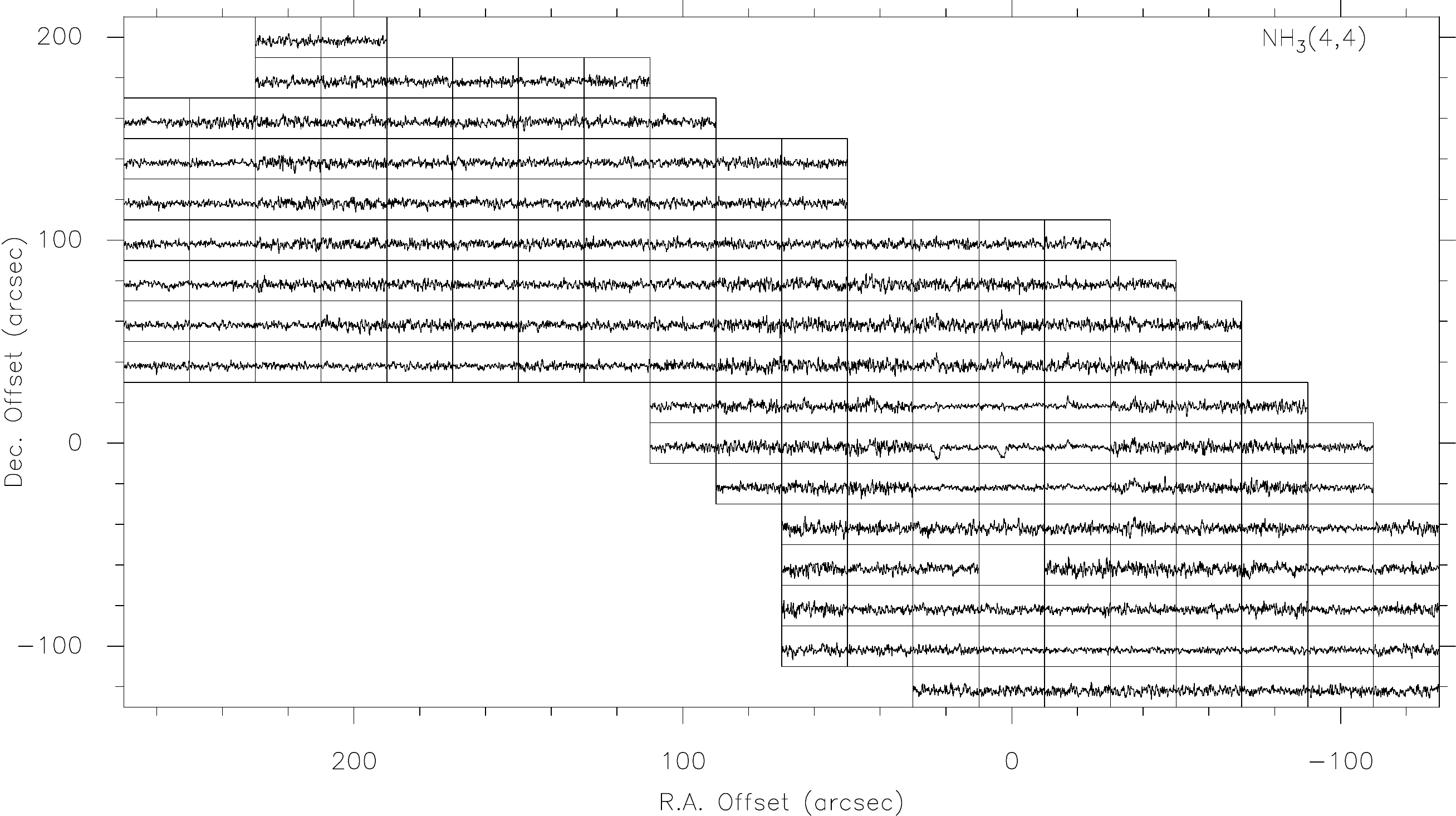}
\end{center}
\caption[]{The NH$_{3}$\,(4,4) line profiles of inversion transitions of W33\,Main and W33\,Main1. The individual spectra cover a velocity range of 32 to 38\,km\,s$^{-1}$ and the ordinate provides main beam brightness temperatures in the range --\,0.6 to 0.7\,K.}
\label{FgA.4}
\end{figure}

%-----------------Figure A.5
\begin{figure}[h]
\vspace*{0.2mm}
\begin{center}
\includegraphics[width=1\textwidth]{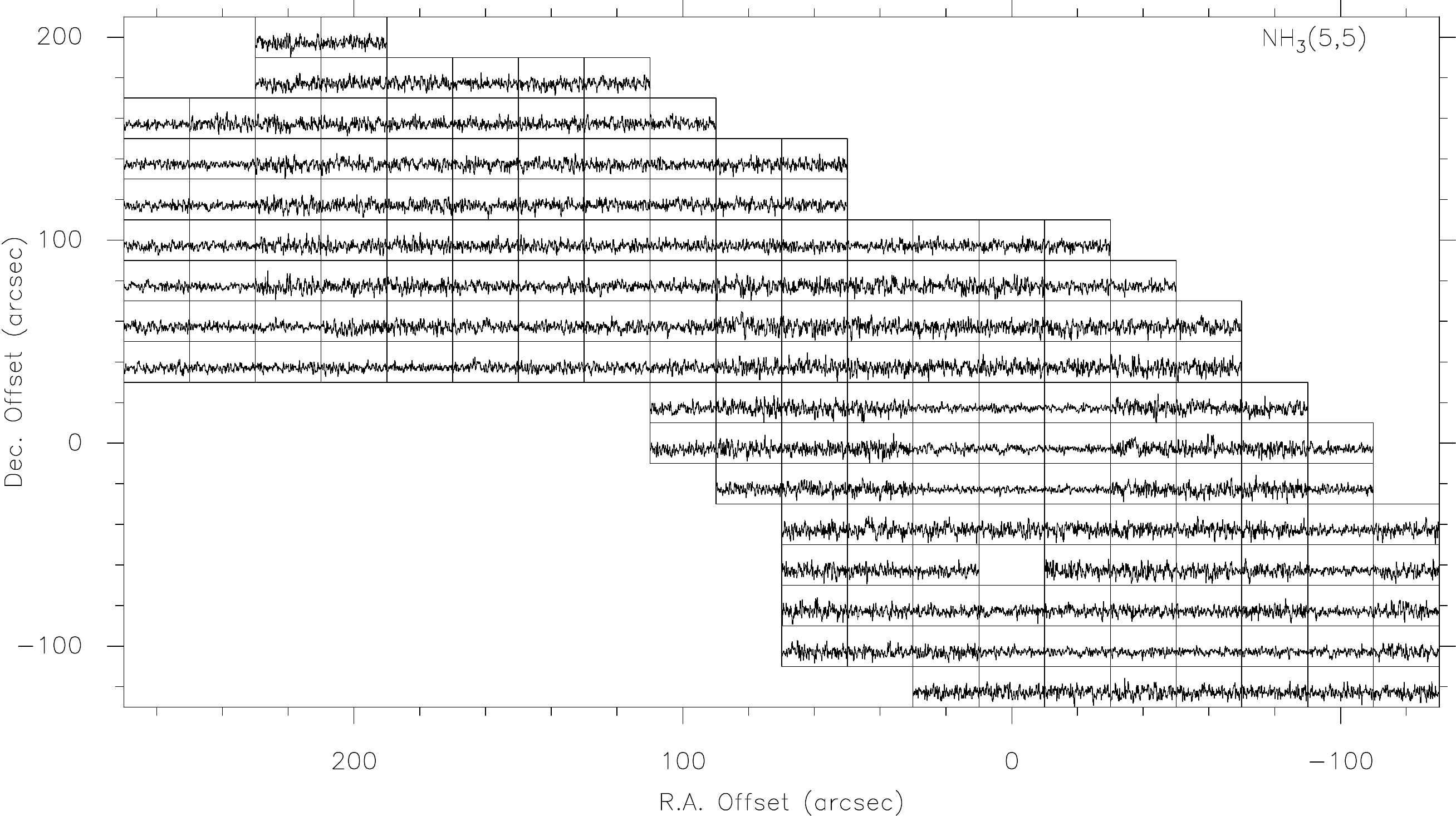}
\end{center}
\caption[]{The NH$_{3}$\,(5,5) line profiles of inversion transitions of W33\,Main and W33\,Main1. The individual spectra cover a velocity range of 32 to 38\,km\,s$^{-1}$ and the ordinate provides main beam brightness temperatures in the range --\,0.2 to 0.5\,K.}
\label{FgA.5}
\end{figure}

%-----------------Figure A.6
\begin{figure}[h]
\vspace*{0.2mm}
\begin{center}
\includegraphics[width=1\textwidth]{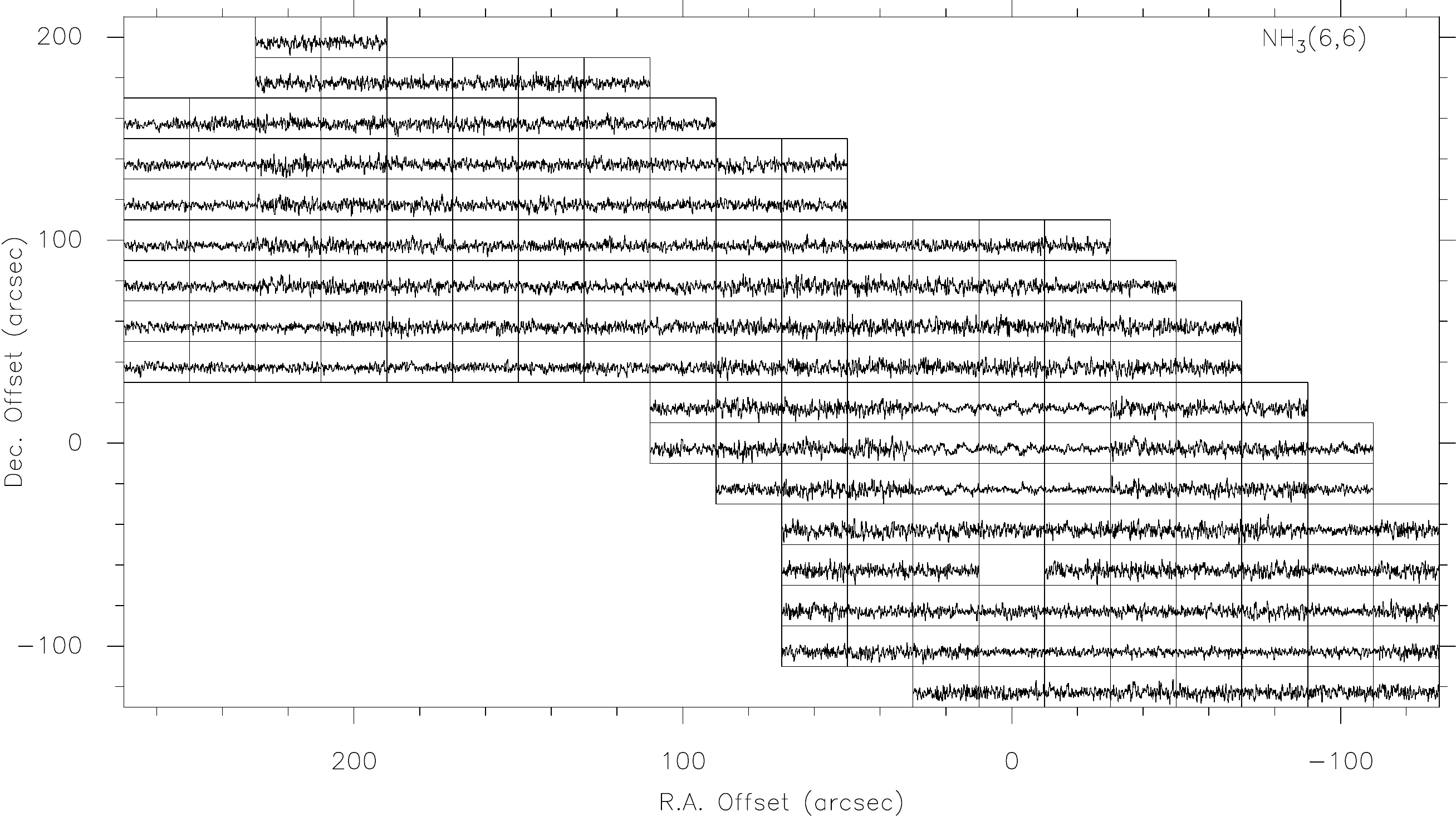}
\end{center}
\caption[]{The NH$_{3}$\,(6,6) line profiles of inversion transitions of W33\,Main and W33\,Main1. The individual spectra cover a velocity range of 32 to 38\,km\,s$^{-1}$ and the ordinate provides main beam brightness temperatures in the range --\,0.2 to 0.4\,K.}
\label{FgA.6}
\end{figure}

%-----------------Figure A.7
\begin{figure}[h]
\vspace*{0.2mm}
\begin{center}
\includegraphics[width=1\textwidth]{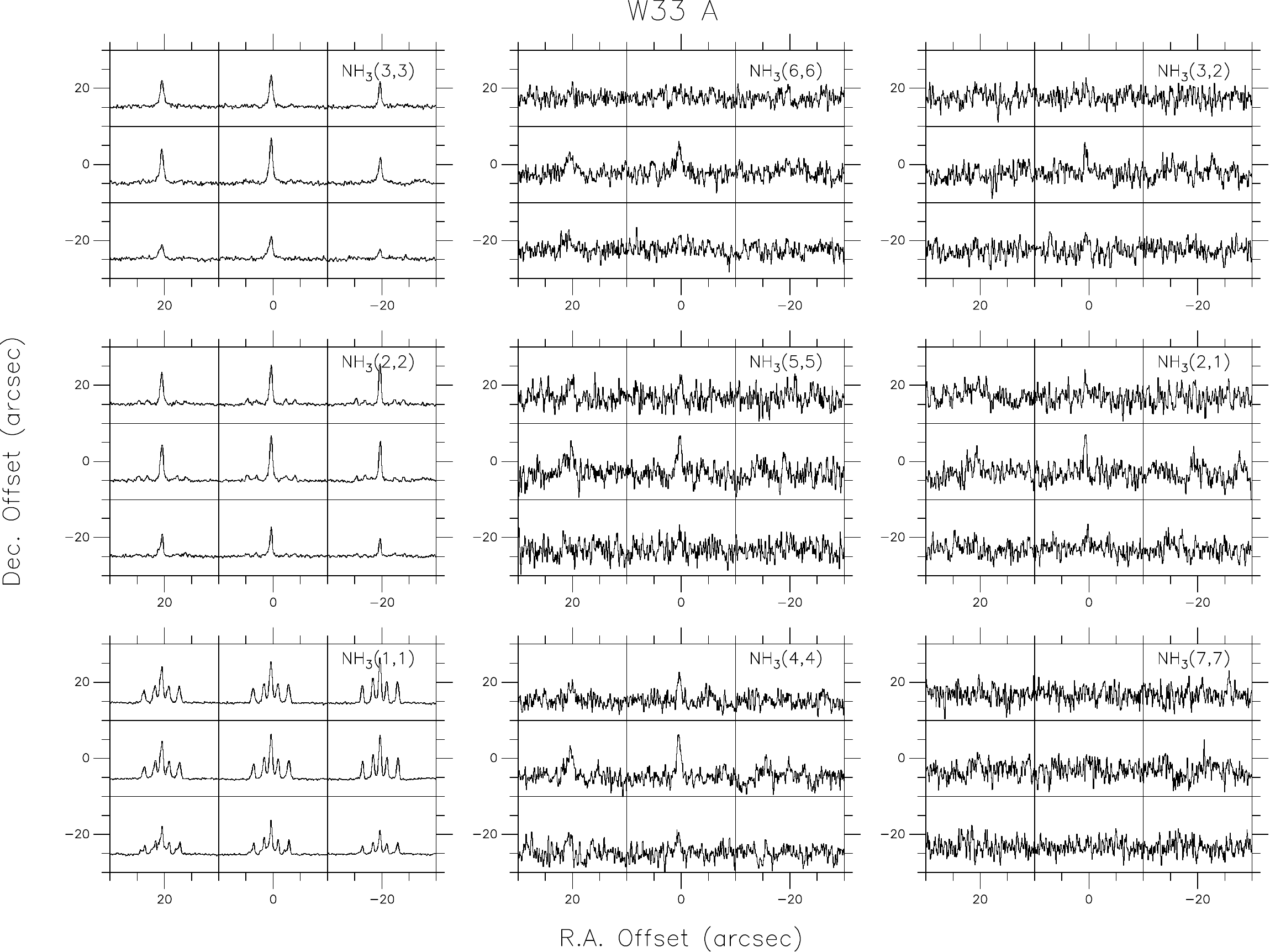}
\end{center}
\caption[]{Line profiles of inversion transitions of ammonia from W33\,A. These are shown on $(\Delta \alpha, \Delta \delta)$ axes. The zero point is at the position of W33\,A, R.A.\,: 18:14:39.10, DEC.\,: -17:52:03.0 (J2000). The X and Y axes indicate Right Ascension Offset (arcsec) and Declination Offset (arcsec), respectively. The main beam temperature scale for the NH$_{3}$ lines was obtained from continuum cross scans of 3C\,286 (see Sect. \ref{sect:Observation}). All radial velocities are on a $V_{\rm LSR}$ scale. At the assumed distance to the complex, $\sim$2.4\,kpc, 40$\arcsec$ is equivalent to 0.5\,pc. The individual spectra cover a velocity range of 32 to 38\,km\,s$^{-1}$ and the ordinate provides main beam brightness temperatures in the range --\,0.1 to 5.2\,K (\textit{left panel}), --\,0.1 to 0.5\,K (\textit{middle panel}), and --\,0.1 to 0.5\,K (\textit{right panel}).}
\label{FgA.7}
\end{figure}

%-----------------Figure A.8
\begin{figure}[h]
\vspace*{0.2mm}
\begin{center}
\includegraphics[width=1\textwidth]{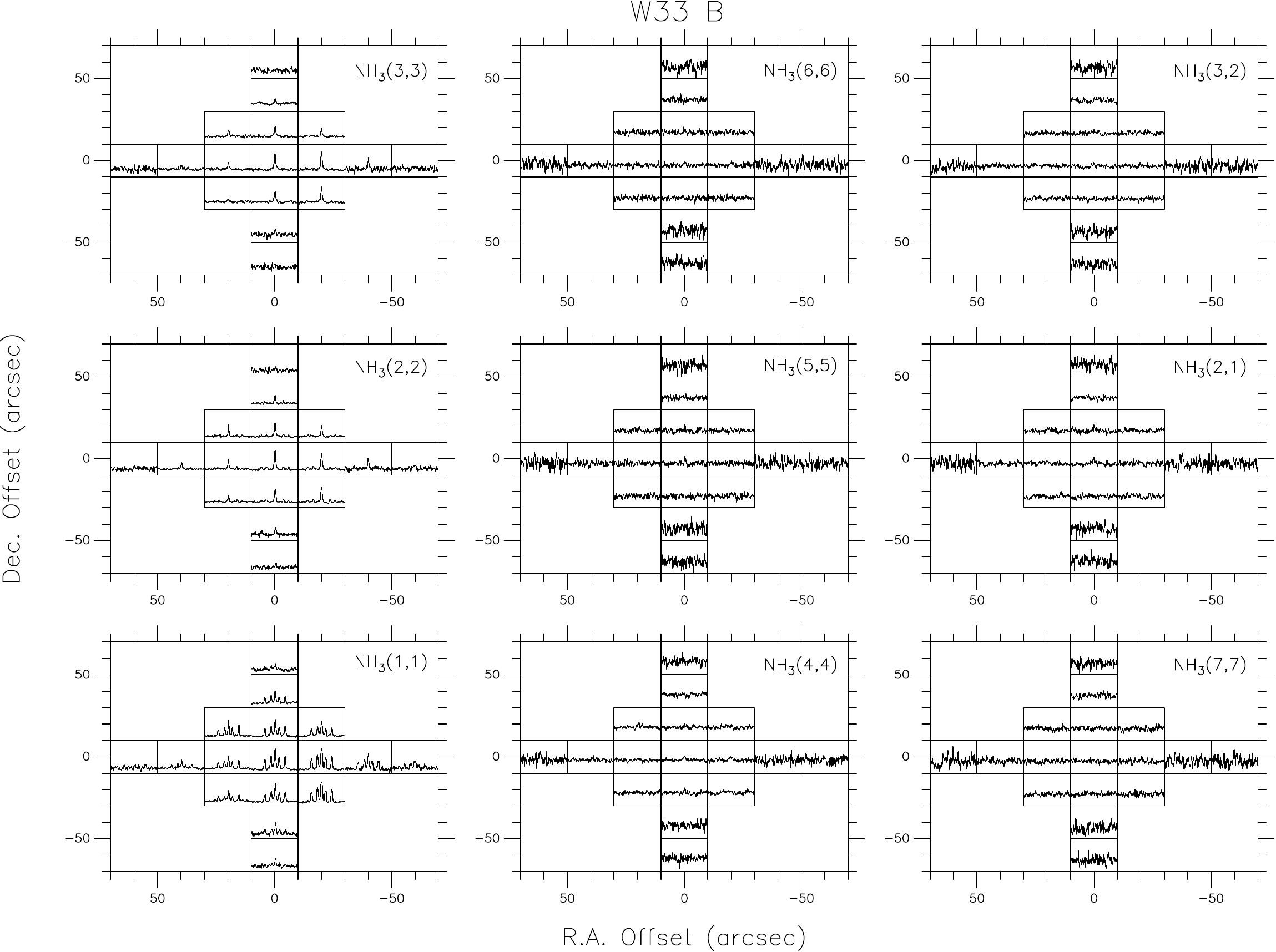}
\end{center}
\caption[]{Line profiles of inversion transitions of ammonia from W33\,B. These are shown on $(\Delta \alpha, \Delta \delta)$ axes. The zero point is at the position of W33\,B, R.A.\,: 18:13:54.40, DEC.\,: -18:01:52.0 (J2000). The X and Y axes indicate Right Ascension Offset (arcsec) and Declination Offset (arcsec), respectively. The main beam temperature scale for the NH$_{3}$ lines was obtained from continuum cross scans of 3C\,286  (see Sect. \ref{sect:Observation}). All radial velocities are on a $V_{\rm LSR}$ scale. At the assumed distance to the complex, $\sim$2.4\,kpc, 40$\arcsec$ is equivalent to 0.5\,pc. The individual spectra cover a velocity range of 54 to 62\,km\,s$^{-1}$ and the ordinate provides main beam brightness temperatures in the range --\,0.1 to 3.1\,K (\textit{left panel}), --\,0.1 to 0.3\,K (\textit{middle panel}), and --\,0.1 to 0.3\,K (\textit{right panel}).}
\label{FgA.8}
\end{figure}

%-----------------Figure A.9
\begin{figure}[h]
\vspace*{0.2mm}
\begin{center}
\includegraphics[width=1\textwidth]{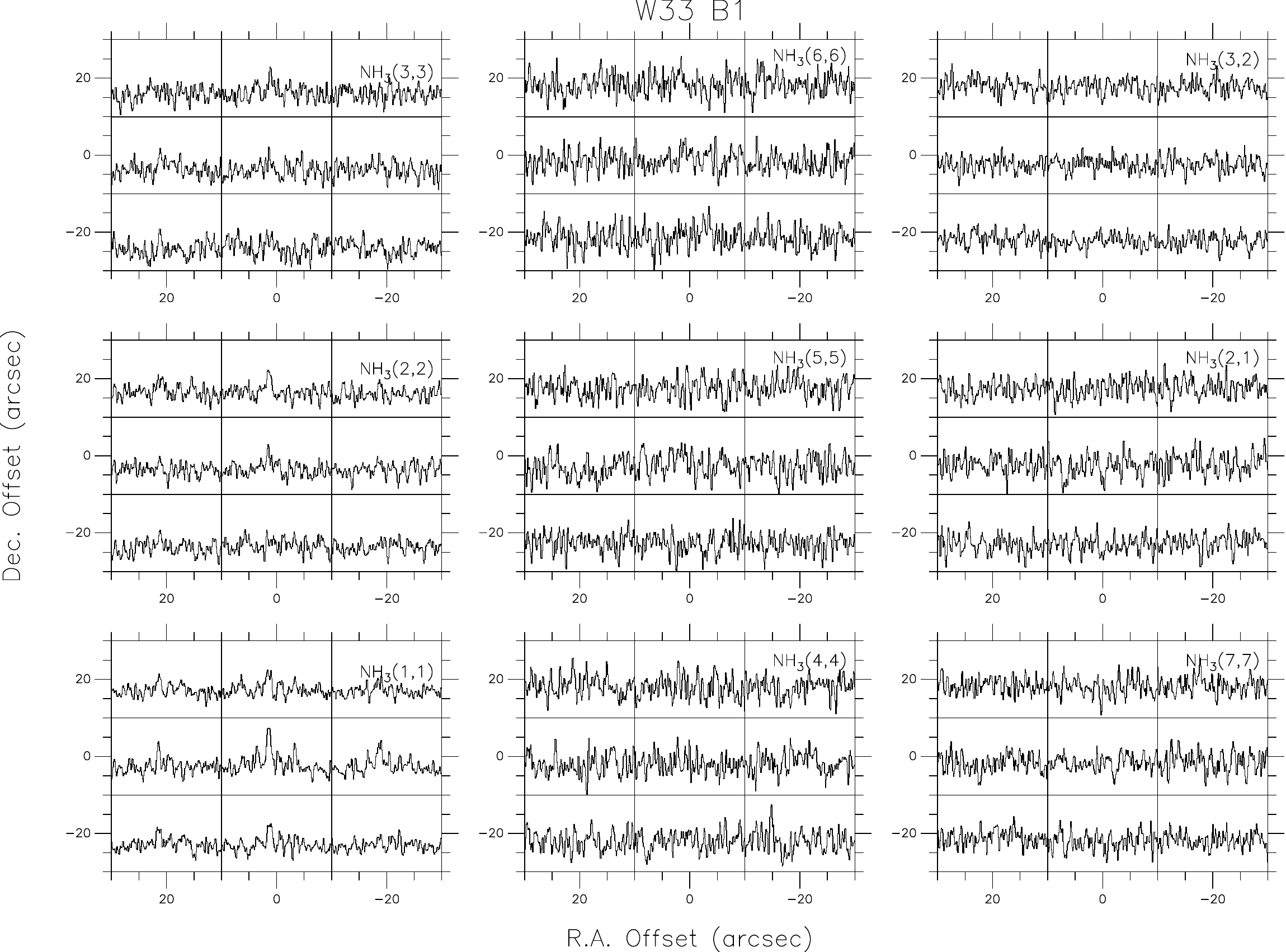}
\end{center}
\caption[]{Line profiles of inversion transitions of ammonia from W33\,B1. These are shown on $(\Delta \alpha, \Delta \delta)$ axes. The zero point is at the position of W33\,B1, R.A.\,: 18:14:07.10, DEC.\,: -18:00:45.0 (J2000). The X and Y axes indicate Right Ascension Offset (arcsec) and Declination Offset (arcsec), respectively. The main beam temperature scale for the NH$_{3}$ lines was obtained from continuum cross scans of 3C\,286  (see Sect. \ref{sect:Observation}). All radial velocities are on a $V_{\rm LSR}$ scale. At the assumed distance to the complex, $\sim$2.4\,kpc, 40$\arcsec$ is equivalent to 0.5\,pc. The individual spectra cover a velocity range of 32 to 38\,km\,s$^{-1}$ and the ordinate provides main beam brightness temperatures in the range --\,0.1 to 1.3\,K (\textit{left panel}), --\,0.1 to 0.3\,K (\textit{middle panel}), and --\,0.1 to 0.3\,K (\textit{right panel}).}
\label{FgA.9}
\end{figure}

%-------------------------- Table A.3
\begin{table*}[h]
\centering
\caption{Observed parameters of the NH$_3$\,(3,3) lines detected in W33.}
\begin{tabular}{c  c c c c c c c c }
\hline \hline
 Source & Offset & $\int$$T_{\rm MB}$d$v$ & $V_{\rm LSR}$ & $\Delta v$ & $T_{\rm MB}$ & rms   \\
  & (\arcsec, \arcsec) & K\,km\,s$^{-1}$ & km\,s$^{-1}$ & km\,s$^{-1}$ & K & K    \\
\hline
W33\,Main  & 0, 0 &23.03(0.09)&36.2(0.1)&5.8(0.1)&3.74&0.11     \\
           & 20, 0 &24.42(0.29)&35.9(0.1)&5.5(0.1) & 4.22&0.15       \\
           & -20, 0 &14.10(0.15)&36.3(0.1)&5.8(0.1)&2.33&0.09       \\
           & 0, 20 & 19.48(0.29) &35.8(0.1)&5.0(0.1)&3.63&0.13      \\
           & 0, -20 &14.10(0.15)&36.6(0.1)&5.8(0.1)& 2.33&0.09        \\
           & 20, 20&24.13(0.29) &35.5(0.1)&5.1(0.1)&4.51&0.15       \\
           & -20, 20 &13.38(0.14)&36.2(0.1)&5.3(0.1)&2.47&0.12       \\
           & -20, -20 &8.58(0.15)&36.2(0.1)&5.3(0.1) & 1.59&0.06        \\
           &20, -20&12.07(0.14)&36.5(0.1)&5.5(0.1)&2.04&0.09       \\
           &$\vdots$&$\vdots$ &$\vdots$&$\vdots$&$\vdots$&$\vdots$         \\
W33\,A  &0, 0 &12.26(0.17)&37.6(0.1)&4.6(0.1)&2.48&0.13        \\
        &20, 0 &9.45(0.15)&37.1(0.1)&4.9(0.1)&1.74&0.13       \\
        &-20, 0 &6.54(0.18)&37.9(0.1)&4.7(0.2)&1.31&0.10       \\
        &0, 20&8.43(0.14) &37.6(0.1)&4.6(0.1)&1.75&0.07     \\
        &0, -20 &7.12(0.15)&37.2(0.1)&5.9(0.2) & 1.16&0.13      \\
        & 20, 20 & 7.99(0.14) &37.3(0.1)&5.5(0.2)  &1.31&0.09       \\
        & -20, 20&4.94(1.31)& 37.7(0.1) &3.4(0.1)& 1.31&0.07       \\
        &-20, -20 &2.33(0.15)&37.5(0.2)&4.4(0.4)&0.58&0.04      \\
        &20, -20 &5.09(0.14)&36.4(0.1)&6.6(0.3)&0.73&0.07        \\
W33\,B  &0, 0 & 5.85(0.12)&55.7(0.1)&3.8(0.1)&1.43&0.12       \\
        &20, 0 &2.33(0.14)&56.2(0.1)&3.7(0.3)&0.58&0.04       \\
        &-20, 0&5.38(0.15) &55.5(0.1)&3.0(0.1)&1.74&0.07       \\
        &0, 20 &3.78(0.15)&56.0(0.1)&3.8(0.1) & 3.49&0.07       \\
        &0, -20& 3.78(0.15) &55.8(0.1)&3.9(0.2)  &0.87&0.33       \\
        &20, 20&2.33(0.14)& 56.6(0.1) &3.8(0.3)& 0.58&0.03     \\
        &-20, 20 &2.18(1.02)&55.4(0.1)&2.7(0.2)&0.73&0.06      \\
        &-20, -20 &4.51(0.15)&55.5(0.1)&3.1(0.1)&1.31&0.09        \\
        &20, -20 & 1.16(0.29)&55.9(0.4)&5.1(0.7)&0.29&0.04       \\
        &40, 0 &1.16(0.15)&56.2(0.2)&2.8(0.4)&0.44&0.03       \\
        &0, 40&1.31(0.14) &56.1(0.2)&3.1(0.4)&0.44&0.07    \\
           &-40, 0&3.22(0.34)&55.7(0.1)&3.1(0.5)&0.97&0.12&         \\
           &0, -40&1.75(0.26)&56.3(0.2)&2.9(0.5)&0.56&0.04&         \\
           &60, 0 &0.40(0.16)&55.5(0.2)&0.8(0.3)&0.48&0.03&         \\
           &-60, 0&1.65(0.49)&55.0(1.5)&9.6(2.8)&0.16&0.08&         \\
           &0, 60 &0.11(0.12)&56.2(0.2)&0.5(0.2)&0.21&0.01&         \\
           &0, -60&0.80(0.20)&55.2(0.3)&2.0(0.5)&0.37&0.04&         \\
W33\,Main1 & 0, 0 &2.79(0.11)&36.7(0.1)& 2.7(0.1)&0.97&0.06       \\
W33\,A1    & 0, 0 &2.80(0.21)&36.9(0.1)&4.0(0.4)&0.65&0.09         \\
W33\,B1    & 0, 0 &2.24(0.22)&34.8(0.3)&6.9(1.0)&0.30&0.06       \\
           &20, 0 &3.40(0.64)&40.2(1.5)&4.4(2.9)&0.22&0.14&         \\
           &-20, 0&0.40(0.20)&40.3(0.2)&1.0(0.4)&0.39&0.02&         \\
           & 0, 20&1.81(0.34)&35.7(0.3)&3.6(0.9)&0.46&0.11&         \\
           &0, -20&0.68(0.25)&36.4(0.3)&1.7(0.8)&0.38&0.05&         \\
           &20, 20&0.19(0.16)&36.3(0.5)&0.9(0.7)&0.19&0.01&         \\
          &20, -20&1.01(0.22)&35.2(0.2)&2.1(0.5)&0.46&0.04&         \\
         &-20, -20&1.34(0.31)&41.3(0.5)&4.1(1.0)&0.30&0.11&         \\
         &-20, 20 &0.68(0.19)&42.2(0.3)&1.9(0.5)&0.33&0.09&         \\
\hline
\end{tabular}
\tablefoot{For the reference position of each of the W33 sources, see Table\,\ref{table:1}. The errors shown in parentheses are fitting uncertainties from the `GAUSS' fit in CLASS (see Sect.\,\ref{sect-2-2}).}
\label{table:A.3}
\end{table*}

%-------------------------- Table A.4
\begin{table*}[h]
\centering
\caption{Observed parameters of the NH$_3$\,(4,4) lines detected in W33.}
\begin{tabular}{l c c c c c c c c }
\hline \hline
 Source & Offset & $\int$$T_{\rm MB}$d$v$ & $V_{\rm LSR}$ & $\Delta v$ & $T_{\rm MB}$ & rms   \\
  & (\arcsec, \arcsec) & K\,km\,s$^{-1}$ & km\,s$^{-1}$ & km\,s$^{-1}$ & K & K    \\
\hline
W33\,Main  & 0, 0 &-2.53(0.18)&33.7(0.3)&7.8(0.5)&-0.30&0.03       \\
           &      &-1.32(0.15)&38.7(0.1)&3.2(0.2)&-0.39&0.03     \\
           & 20, 0 &-4.23(0.41)&35.6(0.4)&7.4(0.6) & -0.55&0.16       \\
           &       &-0.75(0.35)&39.3(0.3)&2.3(0.8) & -0.32&0.09       \\
           & -20, 0 &1.80(0.16)&36.1(0.3)&5.9(0.8)&0.32&0.16       \\
           & 0, 20 & 0.33(0.16) &36.0(0.3)&2.0(0.6)&0.16&0.02       \\
           & 0, -20 &0.09(0.07)&35.5(0.3)&0.8(0.4)& 0.13&0.02        \\
           & 20, 20&0.49(0.16) &35.4(0.2)&2.1(0.5)&0.16&0.02      \\
           & -20, 20 & 1.96(0.16)&36.1(0.2)&4.3(0.6)&0.49&0.16      \\
           & -20, -20 & 1.30(0.16)&36.6(0.5)&6.6(1.6) & 0.16&0.03        \\
           &20, -20&0.33(0.16)&46.2(0.4)&1.8(0.7)&0.16&0.02       \\
           &$\vdots$&$\vdots$ &$\vdots$&$\vdots$&$\vdots$&$\vdots$         \\
W33\,A  &0, 0 &3.14(0.15)&37.7(0.1)&5.8(0.4)&0.50&0.06        \\
        &20, 0 &2.78(0.15)&37.6(0.4)&9.2(0.9)&0.33&0.05       \\
        &-20, 0 &0.82(0.16)&39.1(0.5)&4.3(1.3)&0.16&0.03       \\
        &0, 20&1.47(0.16) &38.1(0.2)&4.4(0.5)&0.33&0.03      \\
        &0, -20 &1.31(0.16)&36.7(0.4)&5.4(0.8) &0.16&0.05      \\
        & 20, 20 & 1.47(0.16) &38.9(0.5)&8.1(1.1) &0.16&0.05      \\
        & -20, 20&0.32(0.15)& 38.7(1.2) &4.0(1.9)& 0.16&0.05       \\
        &-20, -20 &0.49(0.16)&36.7( 1.7)&8.8(2.8)&0.07&0.05      \\
        &20, -20 &1.47(0.15)&35.9(0.5)&6.4(0.8)&0.16&0.07        \\
W33\,B  &0, 0 & 1.27(0.12)&55.4(0.2)&4.9(0.6)&0.24&0.05       \\
        &20, 0 &0.49(0.16)&57.5(0.9)&4.8(1.5)&0.08&0.05       \\
        &-20, 0&1.47(0.17) &53.9(0.5)&7.5(1.0)&0.16&0.03       \\
        &0, 20 &0.32(0.08)&56.1(0.2)&1.3(0.3) & 0.33&0.02       \\
        &0, -20& 0.82(0.16) &56.1(0.2)&2.5(0.4)  &0.33&0.05       \\
        &20, 20&0.82(0.16)& 57.4(0.2) &2.9(0.5)& 0.33&0.03      \\
        &-20, 20 &0.49(0.16)&55.7(0.6)&3.4(1.4)&0.16&0.05      \\
        &-20, -20 &1.64(0.33)&56.7(1.7)&6.8(2.3)&0.16&0.05       \\
        &20, -20 & 0.49(0.16)&55.1(0.9)&6.1(1.9)&0.16&0.05      \\
        &40, 0 &0.65(0.16)&58.9(0.5)&4.2(0.9)&0.16&0.07       \\
        &0, 40&0.32(0.16) &57.8(0.2)&4.3(0.4)&0.16&0.07      \\
W33\,Main\,1  &0, 0 & 0.50(0.07)&36.5(0.2)&2.9(0.5)&0.16&0.04       \\
W33\,A\,1     &0, 0 & 0.66(0.12)&38.2(0.4)&4.3(0.8)&0.14&0.04       \\
\hline
\end{tabular}
\tablefoot{For the reference position of each of the W33 sources, see Table\,\ref{table:1}. The errors shown in parentheses are fitting uncertainties from the `GAUSS' fit in CLASS (see Sect.\,\ref{sect-2-2}).}
\label{table:A.4}
\end{table*}

\clearpage
%-------------------------- Table A.5
\begin{table*}[h]
\centering
\caption{Calculated model parameter of NH$_3$\,(1,1) and NH$_3$\,(2,2) lines detected in W33.}
\begin{tabular}{l c c c c c  }
\hline \hline
Source & Offset& $T_{\rm ex}$ & $T_{\rm rot}$ & $T_{\rm kin}$ & $n_{\rm H_{2}}$/10$^{4}$ \\
 & (\arcsec, \arcsec) & K & K & K & cm$^{-3}$ \\
\hline
W33\,Main &0, 0   & 7.5 $\pm$ 1.1  & 23 $\pm$ 5 & 30 $\pm$ 7 & 1.2    \\
          &    & 7.6 $\pm$ 1.3  &   38 $\pm$ 6 & 45 $\pm$ 7 & 1.1    \\
          & 20, 0& 5.8 $\pm$ 2.6  &    27 $\pm$ 5& 39 $\pm$ 9 & 0.6     \\
          &      &  12.8 $\pm$ 5.3  &   29 $\pm$ 5& 44 $\pm$ 9 & 2.5     \\
          & -20, 0& 11.0 $\pm$ 3.0   &   28 $\pm$ 1 &42 $\pm$ 1  & 2.0   \\
          &0, 20&  5.9 $\pm$ 0.1  &   15 $\pm$ 1& 17 $\pm$ 1   &  0.9   \\
          &0, -20& 4.7 $\pm$ 0.1  &   12 $\pm$ 1 &13 $\pm$ 1  & 0.6    \\
          & 20, 20 & 6.7 $\pm$ 0.3   &   25 $\pm$ 1&35 $\pm$ 1  & 0.9    \\
          &-20, 20& 18.4 $\pm$ 1.0  &    28 $\pm$ 1 &42 $\pm$ 2  & 4.9  \\
          &-20, -20& 7.9 $\pm$ 0.4   &   28 $\pm$ 2 &42 $\pm$ 3  & 1.1    \\
          &20, -20& 3.9 $\pm$ 0.1  &   9 $\pm$ 1 &10$\pm$ 1 & 0.4  \\
          &$\vdots$&$\vdots$ &$\vdots$&$\vdots$&$\vdots$                   \\
W33\,A &0, 0& 8.9 $\pm$ 0.3  &    15 $\pm$ 1 &16 $\pm$ 1   & 2.6  \\
       &20, 0& 7.8 $\pm$ 0.1  &   14 $\pm$ 1&16 $\pm$ 1    & 1.8  \\
       &-20, 0& 8.1 $\pm$ 0.2  &   13 $\pm$ 1 &14 $\pm$ 1    & 2.4   \\
       &0, 20& 8.2 $\pm$ 0.1  &    14 $\pm$ 1 &16 $\pm$ 1    & 2.1  \\
       &0, -20& 7.7 $\pm$ 0.1  &   14 $\pm$ 1 &16 $\pm$ 1     & 1.7  \\
       & 20, 20 & 7.5 $\pm$ 0.1  &   14 $\pm$ 1&16 $\pm$ 1    & 1.6  \\
       & -20, 20& 8.4 $\pm$ 0.2  &   13 $\pm$ 1 &14 $\pm$ 1 & 2.7   \\
       &-20, -20& 5.9 $\pm$ 0.3  &   12 $\pm$ 1&13 $\pm$ 1     & 1.1  \\
       &20, -20&  5.5 $\pm$ 0.6  &   27 $\pm$ 3&39 $\pm$ 4    &  0.6  \\
W33\,B &0, 0 &  5.9 $\pm$ 0.1  &   11 $\pm$ 1 & 13 $\pm$ 1    & 1.1  \\
       &20, 0& 4.5 $\pm$ 0.1  &   11$\pm$ 1 &12 $\pm$ 1    & 0.5  \\
       &-20, 0& 5.7 $\pm$ 0.1  &    12 $\pm$ 1 &13 $\pm$ 1    & 0.9  \\
       &0, 20& 5.1 $\pm$ 0.1  &    12 $\pm$ 1 &13 $\pm$ 1    &  0.7 \\
       &0, -20& 5.5 $\pm$ 0.1   &   12 $\pm$ 1&13 $\pm$ 1     &  0.9 \\
       &20, 20& 5.0 $\pm$ 0.1   &   12 $\pm$ 1 &13 $\pm$ 1   &  0.7 \\
       &-20, 20& 4.9 $\pm$ 0.2  &   12 $\pm$ 1&13 $\pm$ 1   &  0.7 \\
       &-20, -20& 5.6 $\pm$ 0.2   &   11 $\pm$1&12 $\pm$ 1   &  1.1 \\
       &20, -20& 4.3 $\pm$ 0.1   &   12 $\pm$ 1&13 $\pm$ 1   &  0.4 \\
       &40, 0& 3.9 $\pm$ 0.1  &   12 $\pm$ 1 &13 $\pm$ 1    &  0.3 \\
       &0, 40& 4.4 $\pm$ 0.1   &   11 $\pm$ 1 &12 $\pm$ 1   &  0.6 \\
           &-40, 0& 4.8 $\pm$ 0.1   &   11 $\pm$ 1 &12 $\pm$ 1   &  0.7 \\
           &0, -40& 5.4 $\pm$ 0.6   &   20 $\pm$ 1 &25 $\pm$ 2   &  0.6 \\
           &60, 0 &  3.1 $\pm$ 0.1   &   8 $\pm$ 1 &9 $\pm$ 1   &  0.1 \\
           &-60, 0& 3.9 $\pm$ 0.1   &   11 $\pm$ 1 &12 $\pm$ 1   &  0.3 \\
           &0, 60 & 3.9 $\pm$ 0.8   &   27 $\pm$ 3 &39 $\pm$ 4   &  0.2 \\
           &0, -60& 14.6 $\pm$ 4.6  &   22 $\pm$ 2 &28 $\pm$ 4   &  4.4 \\
W33\,Main1 & 0, 0 & 7.2 $\pm$ 0.1   &   12 $\pm$ 1 &13 $\pm$ 1   &  1.8 \\
W33\,A1    & 0, 0 & 5.9 $\pm$ 0.1   &   11 $\pm$ 1 &13 $\pm$ 1   &  1.1 \\
W33\,B1    & 0, 0 & 4.3 $\pm$ 0.1   &   11 $\pm$ 2 &13 $\pm$ 2   &  0.5 \\
           &20, 0 & 6.0 $\pm$ 4.8   &   17 $\pm$ 3 &20 $\pm$ 3   &  0.8 \\
           &-20, 0& 4.9 $\pm$ 1.2   &   17 $\pm$ 2 &20 $\pm$ 3   &  0.5 \\
           & 0, 20& 3.4 $\pm$ 0.1   &   10 $\pm$ 1 &11 $\pm$ 1   &  0.2 \\
           &0, -20& 3.8 $\pm$ 0.1   &   9 $\pm$ 1 & 10 $\pm$ 1   &  0.4 \\
           &20, 20& 8.6 $\pm$ 0.7   &   27 $\pm$ 2 &38 $\pm$ 3   &  1.3 \\
          &20, -20& 3.8 $\pm$ 0.2   &   25 $\pm$ 3 &35 $\pm$ 4   &  0.2 \\
         &-20, -20& 3.1 $\pm$ 0.1   &   7 $\pm$ 2 &8 $\pm$ 1   &  0.1 \\
         &-20, 20 & 6.9 $\pm$ 1.8   &   27 $\pm$ 1 &39 $\pm$ 4   &  0.9 \\
\hline
\end{tabular}
\tablefoot{The reference positions for each of the W33 sources are given in Table\,\ref{table:1}. The errors shown in parentheses are calculated uncertainties (see Appendix\,C of \citealt{2020A&A...643A.178T}).}
\label{table:A.5}
\end{table*}

\newpage
%----------------------------------Appendix B
%\Online
\twocolumn
\section{Calibration stability and NH$_{3}$\,(3,3) line variations towards the peak position of W33\,Main}
\label{Appendix B}
%---------------------------------------------------Figure B1
\begin{figure}[h]
\includegraphics[width=0.48\textwidth]{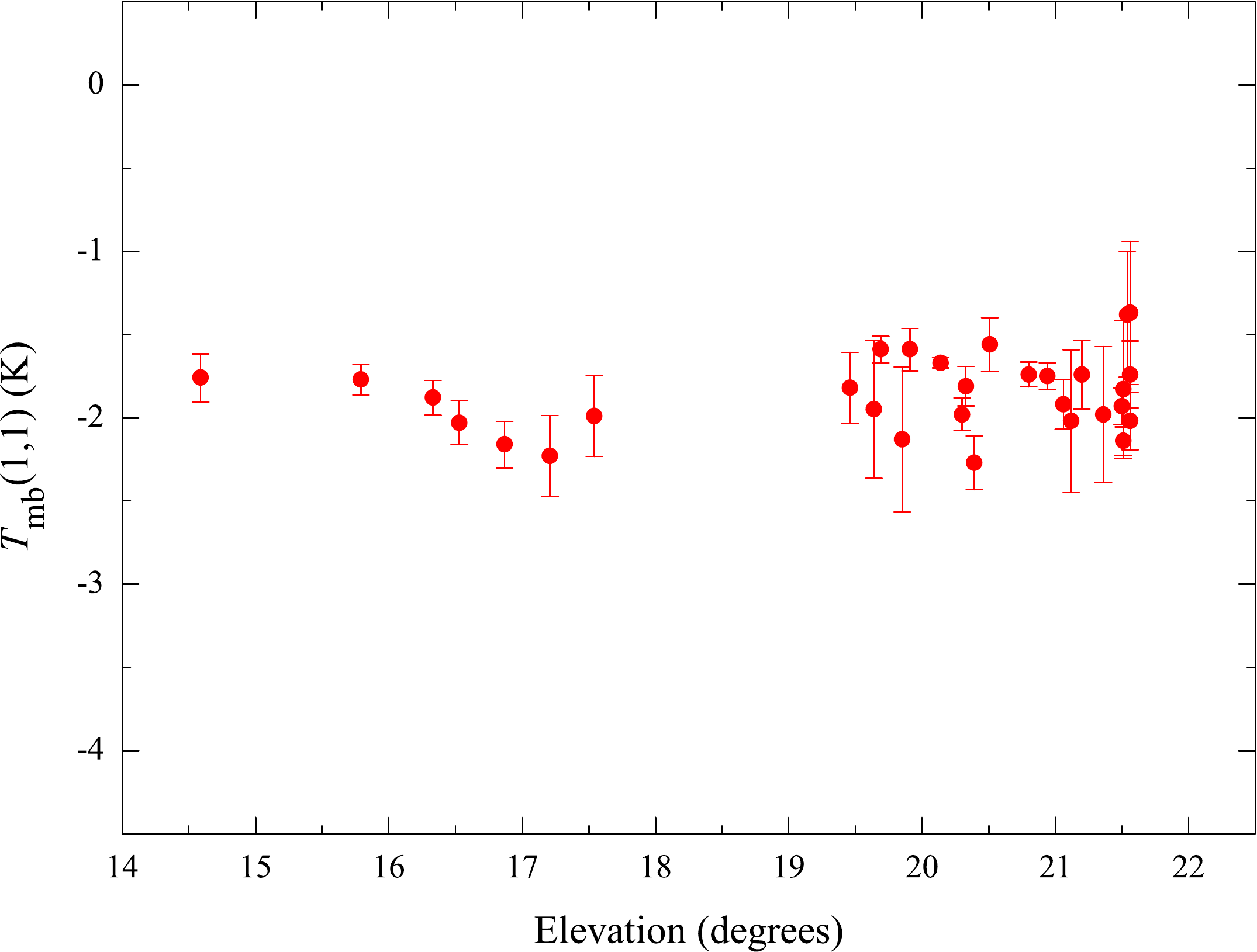}
\caption[]{Uncorrected NH$_3$\,(1,1) main line peak intensities against elevation of repeated observations
toward the reference position (see Table\,\ref{table:1}) of the W33\,Main central position. The standard deviations of the mean of the flux is about $4.2\%$.}
\label{FgB.1}
\end{figure}

The W33\,Main peak position was observed many times to check the stability of the system. The reference position is R.A.:\,18:14:13.50, DEC.:\,-17:55:47.0. To present the peak distribution against elevation (Fig.\,\ref{FgB.1}), we fitted the NH$_3$\,(1,1) main lines (the central group of NH$_3$\,(1,1) hyperfine components). From Fig.\,\ref{FgB.1}, we clearly see that there is no significant systematic variation. The standard deviations of the mean of the peak intensities is about $4.2\%$, thus the observational system of the Effelsberg telescope is stable.

\begin{figure}[h]
\vspace*{0.2mm}
\centering
\includegraphics[width=0.469\textwidth]{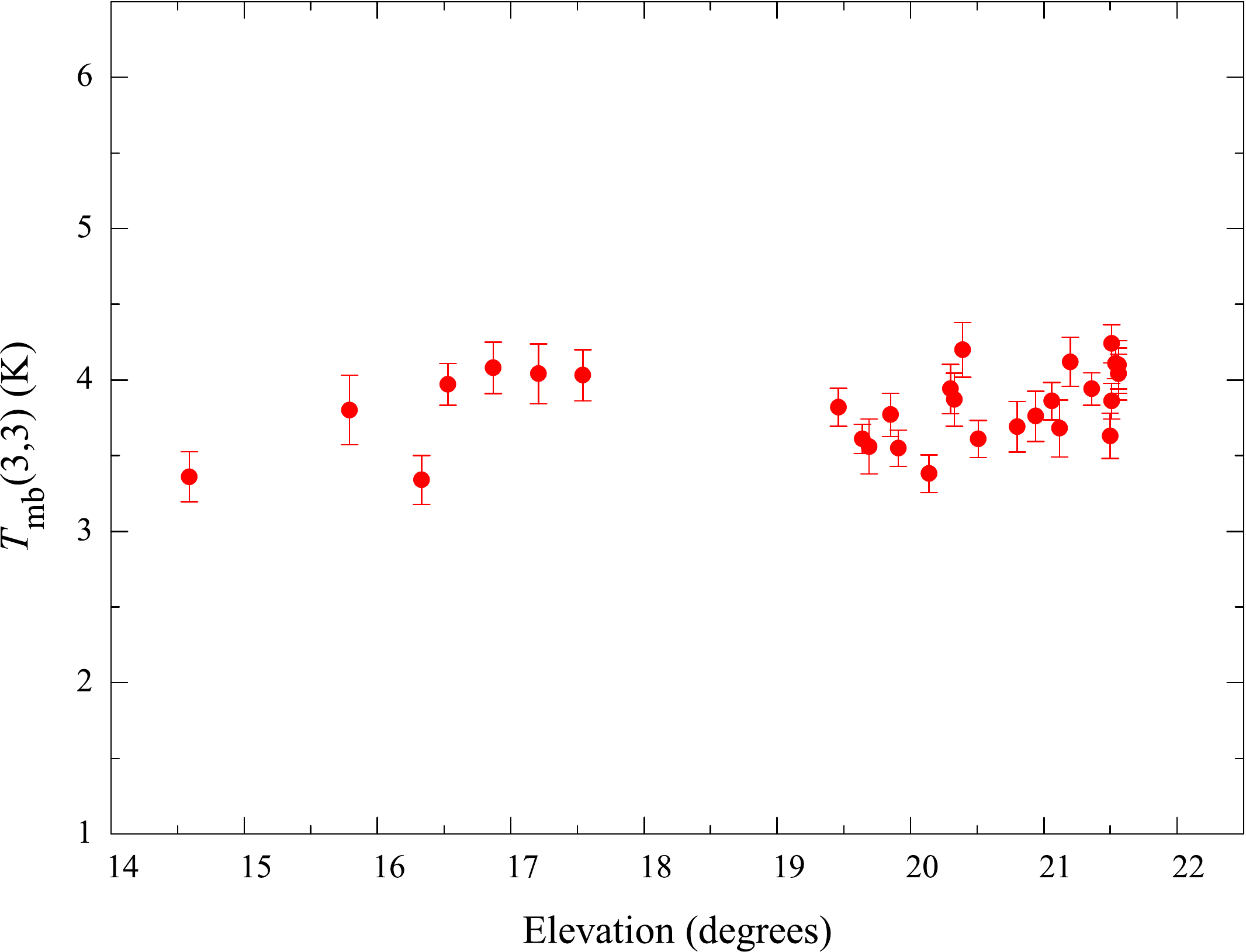}
\hspace{0.25cm}
\includegraphics[width=0.5\textwidth]{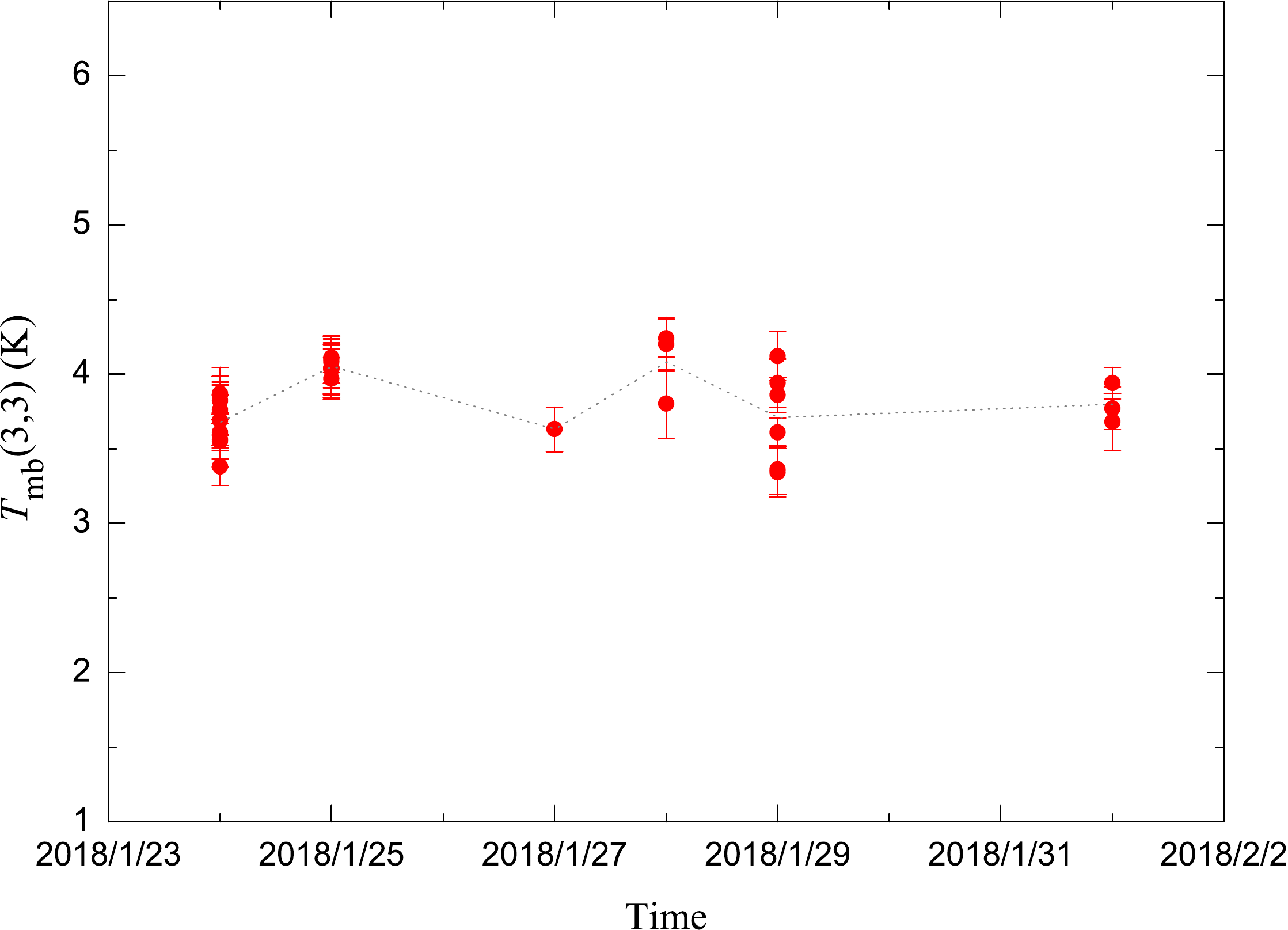}
\caption[]{\textit{Top panel}: NH$_3$\,(3,3) main line peak intensities against elevation of repeated observations
toward the reference position of W33\,Main (see Table\,\ref{table:1}). The standard deviations of the mean of the flux density is about $4.4\%$. \textit{Bottom panel}: NH$_3$\,(3,3) main line peak intensities against the epoch of the observation, also toward the  W33\,Main peak position. The gray dotted line connects the average $T_{\rm mb}$\,(3,3) values of each day.}
\label{FgB.2}
\end{figure}

We also studied the variation of the NH$_{3}$\,(3,3) line, because it is the strongest one, and as a maser, also possibly variable on short time scales (see Fig.\,\ref{W33Main,A} left panel and Sect.\,\ref{sect-3-1}). Therefore, we present the NH$_3$\,(3,3) main line peak intensities against elevation (Fig.\,\ref{FgB.2} top panel) and NH$_3$\,(3,3) main line peak intensities against the epoch of the observation (Fig.\,\ref{FgB.2} bottom panel) toward the W33\,Main peak position. The gray dotted line in Fig.\,\ref{FgB.2}., bottom panel, connects the average $T_{\rm mb}$\,(3,3)
values of each day. From the top panel of Fig.\,\ref{FgB.2}, we obtain that the standard deviations of the mean of the peak intensities
is about $4.4\%$, which can be even more clearly seen in Fig.\,\ref{FgB.2}., bottom panel. Again the main beam brightness temperature variations of this NH$_{3}$\,(3,3) line are not large enough to indicate significant variations during the week covered by our Effelsberg observations.

%-----------------Figure B2

%----------------------------------Appendix C

\section{Results of the NH$_{3}$ line observations}
\label{Appendix C}

%-----------------Figure C.1
\begin{figure*}
\centerline{\hbox{
\includegraphics[width=6.5cm,height=6.2cm,angle=0]{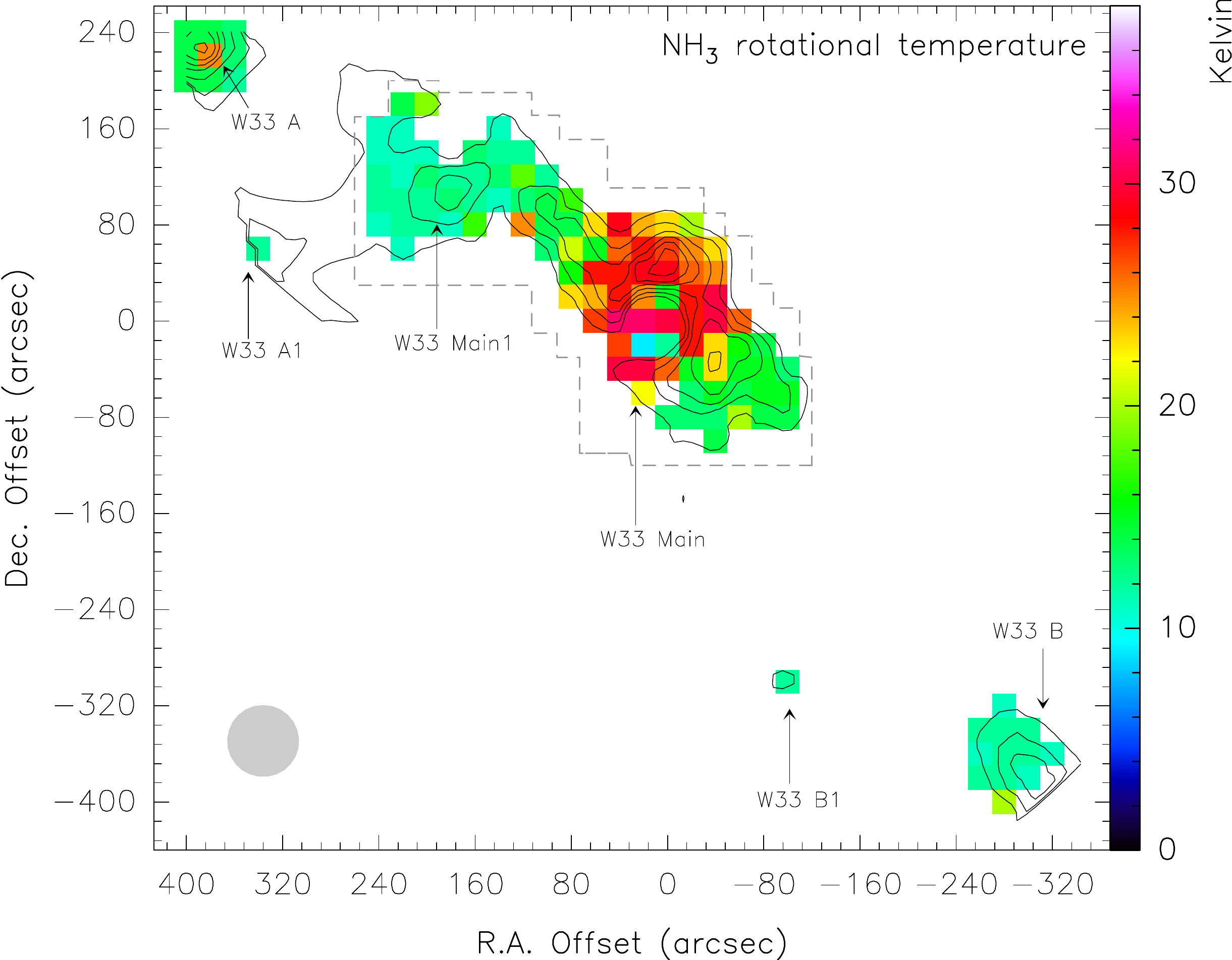}
\hspace{0.08cm}
\includegraphics[width=6.3cm,height=6.2cm,angle=0]{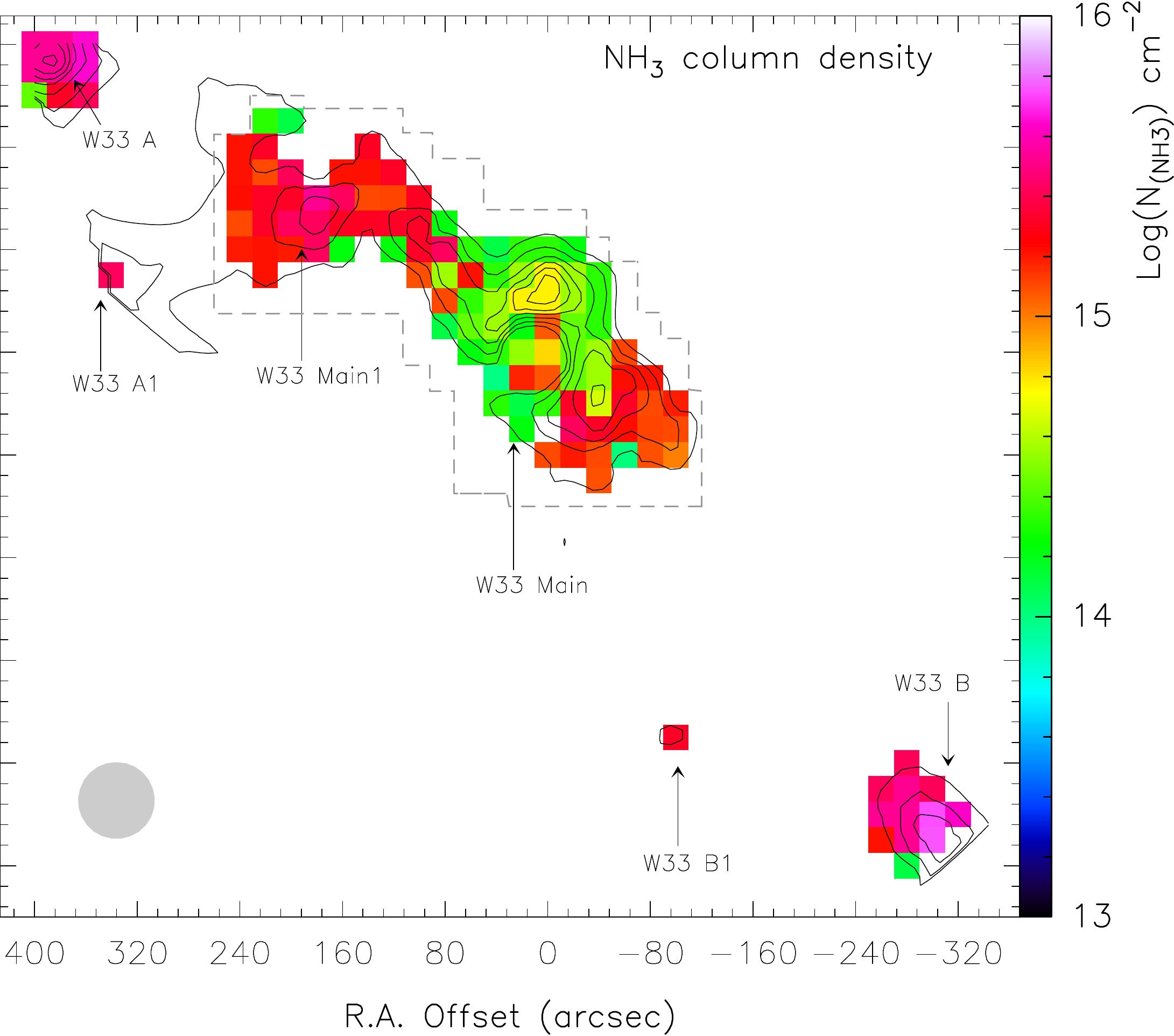}
\hspace{0.08cm}
\includegraphics[width=6.3cm,height=6.2cm,angle=0]{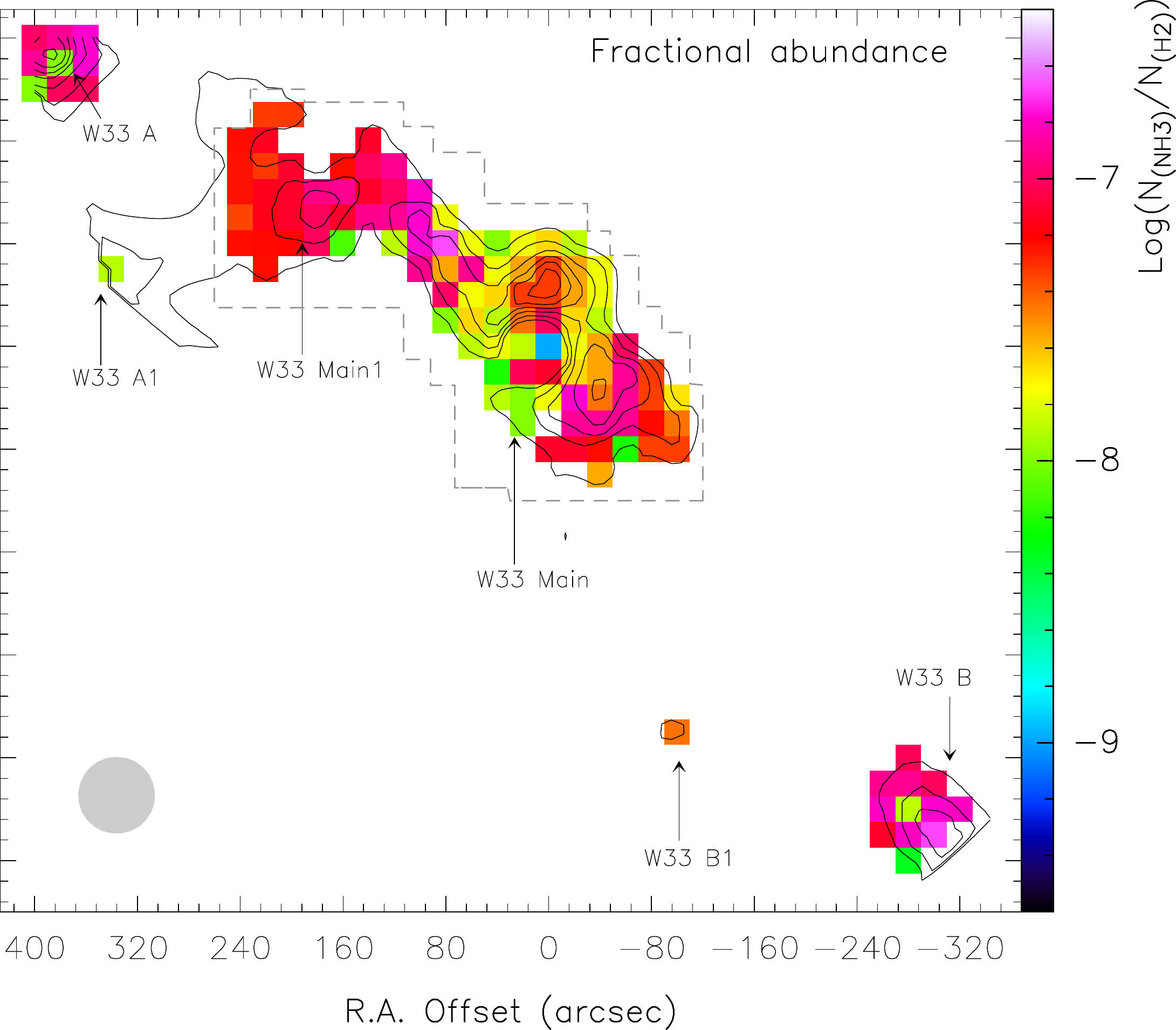}}}
\caption[]{Maps of NH$_3$ rotational temperature in units of Kelvin (\textit{left}), the logarithm of the
total-NH$_3$ column density in units of cm$^{-2}$(\textit{middle}), and the corresponding logarithm of the
fractional abundance (\textit{right}). The reference position is R.A.\,: 18:14:13.50, DEC.\,: -17:55:47.0 (J2000). The integration range is 32 to 40 \,km\,s$^{-1}$. Contours are the same as in the left panel of Fig.\,\ref{integrated-intencity}. The limits of the mapped region over significant parts of our map are indicated by gray dashed lines. The half-power beam width is illustrated as a gray filled circle in the lower left corner of each image.}
\label{Fig:C.1}
\end{figure*}

%-----------------Figure C.2
\begin{figure*}[t]
\centering
\includegraphics[width=0.486\textwidth]{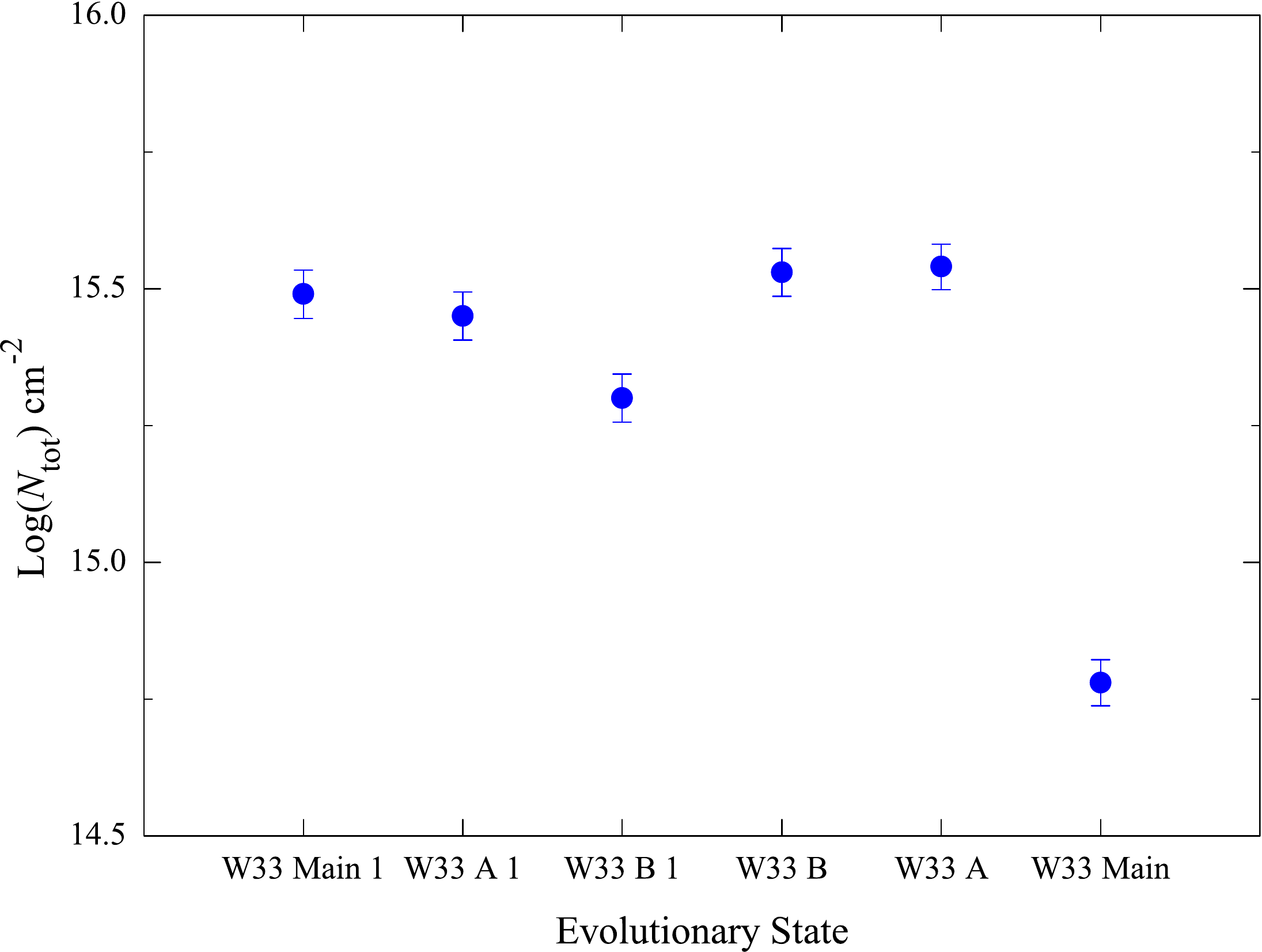}
\hspace{0.2cm}
\includegraphics[width=0.486\textwidth]{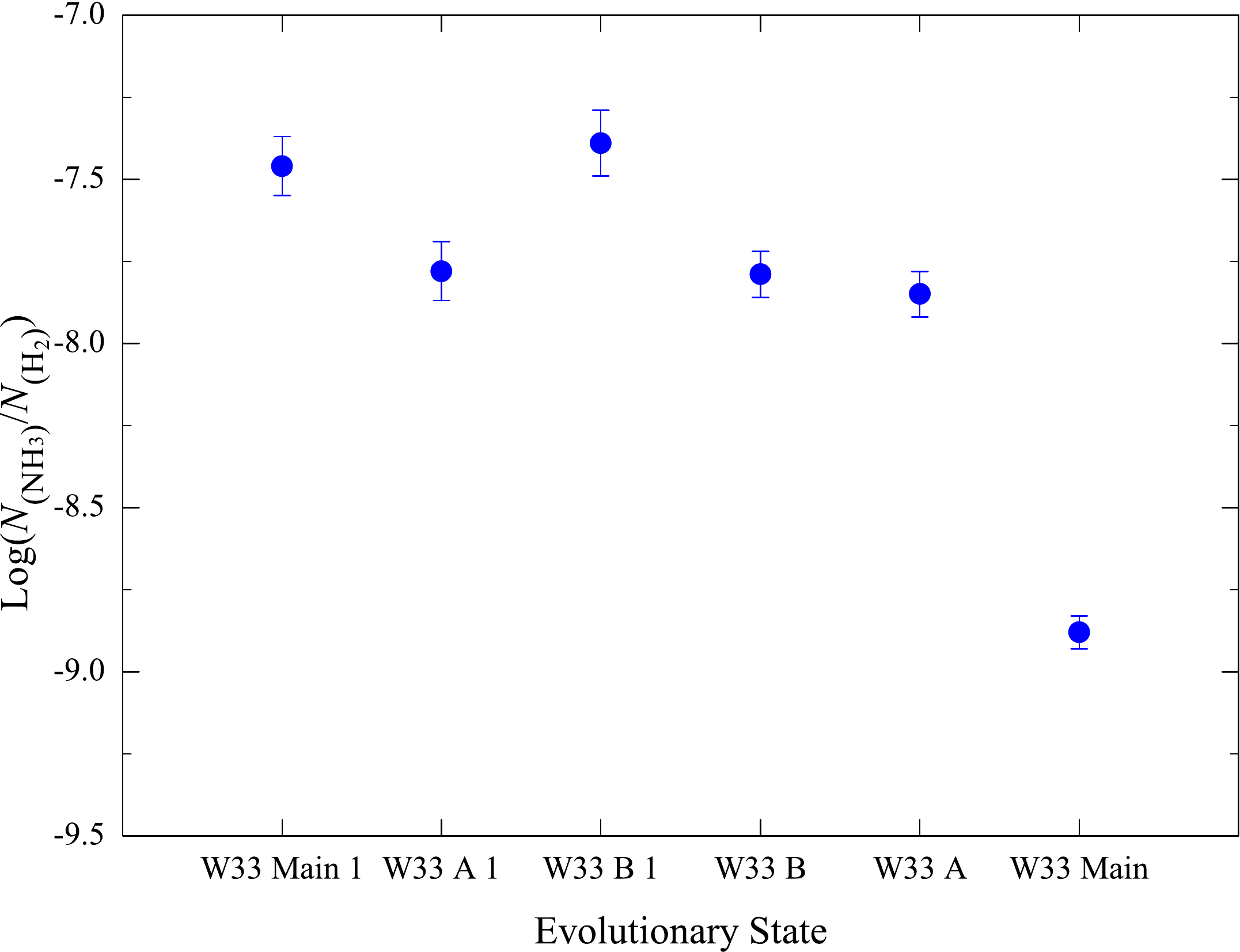}
%\hspace{0.2cm}
\includegraphics[width=0.486\textwidth]{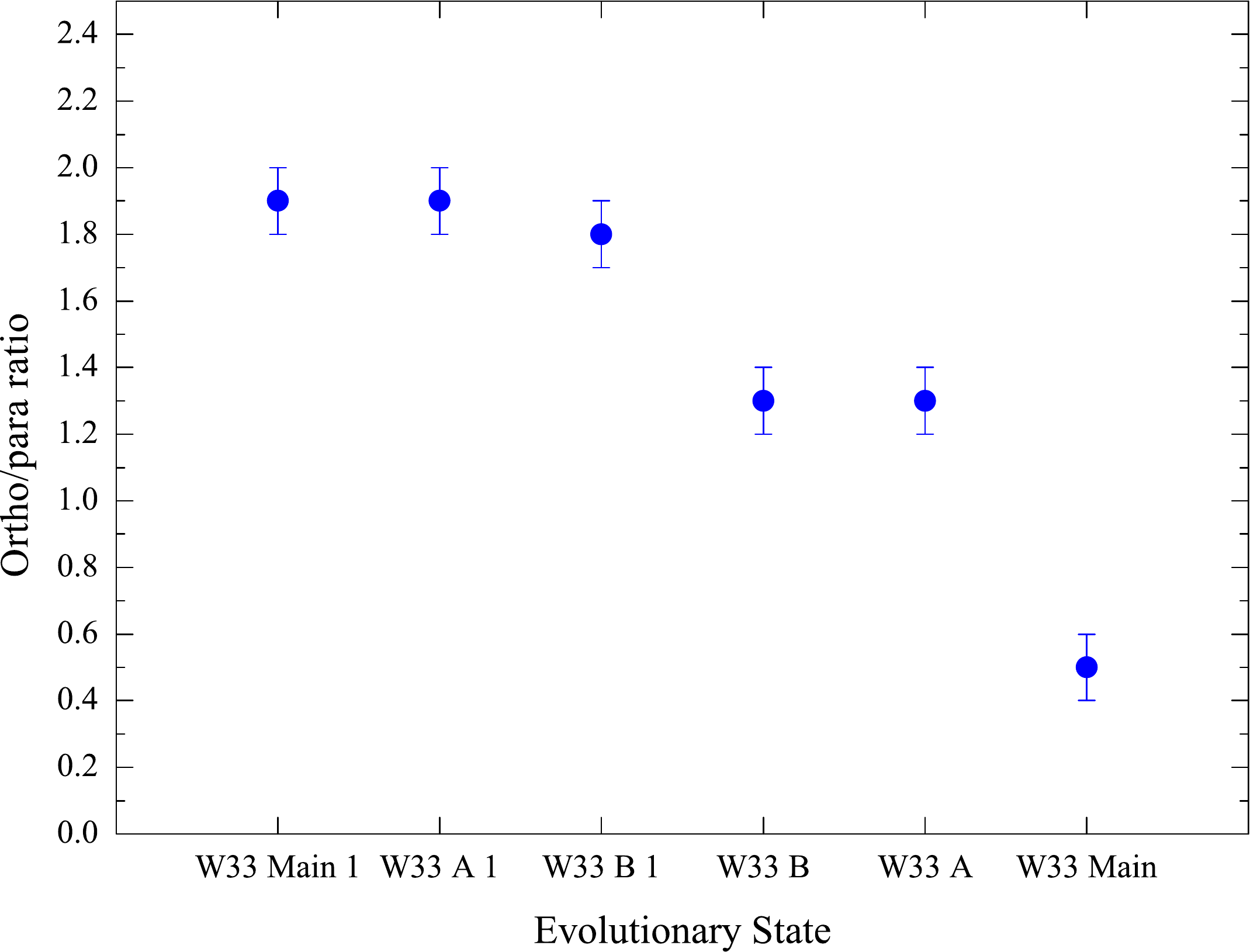}
\hspace{0.2cm}
\includegraphics[width=0.486\textwidth]{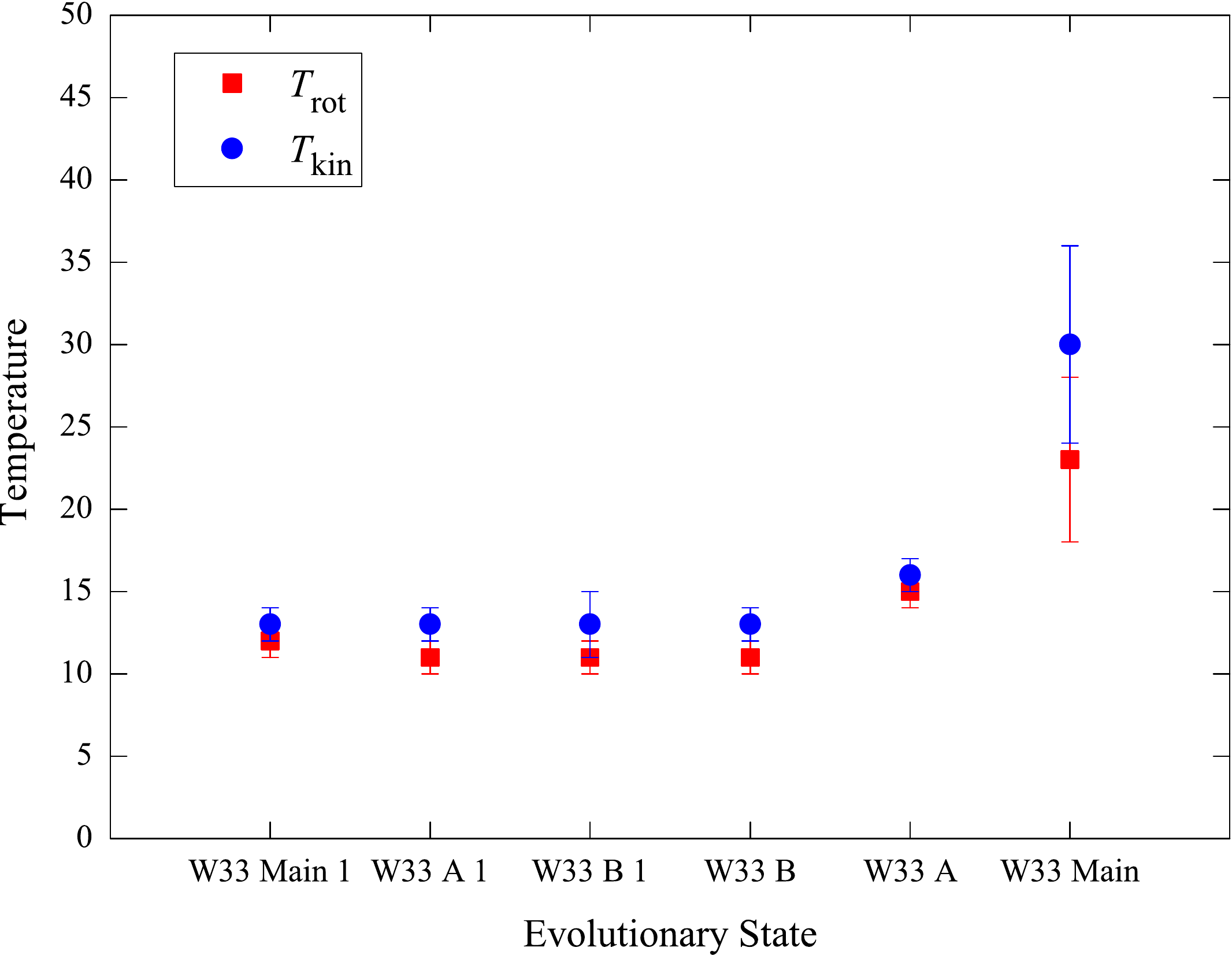}
\caption[]{Column densities derived from total-$N$(NH$_3$) vs. the evolutionary sequence of the six W33 source (\textit{top left}), total fractional NH$_3$ abundance, $N$(total-NH$_3$)/$N$(H$_{2}$), vs. the evolutionary sequence (\textit{top right}), ortho-to-para ratio of NH$_3$ vs. the evolutionary sequence (\textit{bottom left}), and gas temperature derived from NH$_3$ vs. the evolutionary sequence (\textit{bottom right}). For the chosen positions, see Table\,\ref{table:1}. Note that the ortho-to-para ratio of W33\,Main may be affected by systematic errors which cannot be quantified (see Sect.\,\ref{maser-line}).}
\label{Fig:C.2}
\end{figure*}

\end{appendix}
\end{document}